\documentclass[10pt,a4paper]{article}

\usepackage[a4paper, left=2.5cm, right=2.5cm]{geometry}

\usepackage{amsmath,amssymb}
\usepackage{amsfonts}
\usepackage[dvipsnames]{xcolor}
\usepackage{graphicx}
\usepackage{dcolumn}
\usepackage{comment}
\usepackage{authblk}

\usepackage{epstopdf}
\usepackage{color}
\usepackage[utf8]{inputenc}
\usepackage{amsthm}
\usepackage{fontenc}

\usepackage{soul}

\usepackage{bm}
\RequirePackage{slashed}\usepackage{cancel}
\usepackage[normalem]{ulem}
\newcommand{\delete}[1]{{\blue{\sout{#1}}}}
\newcommand{\replace}[2]{\red{\sout{#1}}\blue{#2}}

\def\sumint{\int \! \!\ \! \! \! \! \!\ \! \! \!\! \!\sum}
\def\be{\begin{eqnarray} &&}
\def\nonu{\nonumber \\ &&}
\def\ee{\end{eqnarray}}
\usepackage{array,booktabs}
\newcolumntype{C}{>{$\displaystyle}c<{$}}
\newcommand*{\blue}{\textcolor{blue}}
\newcommand*{\red}{\textcolor{Red}}
\newcommand*{\green}{\color{OliveGreen}}

\def\be{\begin{eqnarray} &&}
\def\nonu{\nonumber \\ &&}
\def\ee{\end{eqnarray}}

\def\sumint{\int \! \!\ \! \! \! \! \!\ \! \! \!\! \!\sum}
\newcommand{\blf}[1]{\bf  {\tilde #1}}
\newcommand{\bfm}[1] {\mbox{\boldmath{$#1$}}}
\newcommand{\bfi}{\begin{figure}}
\newcommand{\efi}{\end{figure}}

\usepackage{multirow}
\usepackage{hhline}
\usepackage{tikz}
\usepackage{makecell}
\usepackage{comment}

\definecolor{tcA}{rgb}{0,0,0}
\usepackage{array}
\newcolumntype{P}[1]{>{\centering\arraybackslash}p{#1}}

\DeclareMathOperator*{\SumInt}{%
\mathchoice%
  {\ooalign{$\displaystyle\sum$\cr\hidewidth$\displaystyle\int$\hidewidth\cr}}
  {\ooalign{\raisebox{.14\height}{\scalebox{.7}{$\textstyle\sum$}}\cr\hidewidth$\textstyle\int$\hidewidth\cr}}
  {\ooalign{\raisebox{.2\height}{\scalebox{.6}{$\scriptstyle\sum$}}\cr$\scriptstyle\int$\cr}}
  {\ooalign{\raisebox{.2\height}{\scalebox{.6}{$\scriptstyle\sum$}}\cr$\scriptstyle\int$\cr}}
}

\newcommand{\lightbrown}[1]{{\textcolor[rgb]{0.71, 0.4, 0.11}{#1}}}

\newcommand{\pink}[1]{{\textcolor[rgb]{1,0,0.5}{#1}}}

\newcommand{\gianni}[1]{{ \green{{\bf Gianni}:\ {#1}}}}

\newcommand{\matteo}[1]{{ \pink{{\bf Matteo}:\ {#1}}}}
\newcommand{\emanuele}[1]{{ \lightbrown{{\bf Emanuele}:\ {#1}}}}
\usepackage{xcolor}

\begin{document}
\title{\huge Theoretical insights on nuclear double parton distributions}

 \author[1]{Federico Alberto Ceccopieri}

\author[2]{Filippo Fornetti}

\author[3]{Emanuele Pace}

\author[4]{Matteo Rinaldi\thanks{Corresponding author email: matteo.rinaldi@pg.infn.it}}

\author[5]{Giovanni Salm\`e}

\author[6]{Nicholas Iles}

\affil[1]{\rule[25pt]{0pt}{0pt}  Universit\'e de Liège, B4000, Liège, Belgium.}

\affil[2]{\rule[25pt]{0pt}{0pt} Dipartimento di Fisica e Geologia, Universit\`a degli Studi di Perugia and INFN Sezione di Perugia, Italy.}

\affil[3]{\rule[25pt]{0pt}{0pt} Universit\`a di Roma Tor Vergata, Via della Ricerca Scientifica 1, 00133 Rome, Italy.}

\affil[4]{\rule[25pt]{0pt}{0pt} Istituto Nazionale di Fisica Nucleare, Sezione di Perugia, Via A. Pascoli, Perugia, Italy.}

\affil[5]{\rule[25pt]{0pt}{0pt} Istituto Nazionale di Fisica Nucleare, Sezione di Roma, Piazzale A. Moro 2, 00185 Rome, Italy.}

\affil[6]{\rule[25pt]{0pt}{0pt} Dipartimento di Fisica e Geologia, Universit\`a degli Studi di Perugia, Italy.}

\maketitle

\begin{abstract}
In this {{paper}}, we address double parton scattering  (DPS)
in $pA$ collisions. {Within} the Light-Front  
{approach}, {we {formally}  derive the two contributions to the nuclear double parton distribution (DPD), namely: DPS1, involving two partons from the same nucleon, and DPS2, where the two partons belong to different  parent nucleons.}
 We then generalize the sum rule for {hadron} DPDs  to the nuclear case and analytically show how all contributions combine to give the expected results. {In addition} partial sum rules for the DPDs related to  DPS1 and DPS2 mechanisms {are discussed} 
 {for the first time}.
 The deuteron system {{is}} considered for the first calculation  of the nuclear DPD
 {by using {{a}} realistic wave function obtained from the {very refined nucleon-nucleon} AV18 potential, {embedded in a rigorous Poincar\'e covariant formalism}}.
 {Results {{are}} used to test} sum rules  and properly verify that DPS1 contribution compares with the DPS2 one, although smaller. {We also introduce EMC-like ratios involving nuclear and free DPDs to address the potential role of DPS in understanding {in depth} {the EMC} effect.}

\end{abstract}

\section{Introduction}
{Achieving} higher and  higher center-of-mass energies in particle collisions has opened new frontiers in physics, bringing multi-parton interactions (MPI) to the forefront of hadronic collision research. As accelerator capabilities have advanced, these interactions—once considered negligible background effects—have emerged as crucial phenomena that demand thorough investigations.

Among the various MPI processes, double parton scattering (DPS) represents the simplest and most accessible case for study, serving as a vital testing ground for our understanding of multiple particle interactions. In this case, two partons of one hadron simultaneously interact with  two partons of the other colliding hadron.
 Fundamental {theoretical} studies on DPS can be found in, e.g., Refs. \cite{Goebel:1979mi,Paver:1982yp,Mekhfi:1983az,Gaunt:2009re,Diehl:2011yj,book,Blok:2011bu,Diehl:2011tt,Manohar:2012pe,Diehl:2023jje} and in, e.g., Refs. \cite{Bali:2018nde,Bali:2020mij,Bali:2021gel,Reitinger:2024ulw} for lattice investigations.
Experimental evidence for DPS has been accumulating through various analyses \cite{AxialFieldSpectrometer:1986dfj,S2,CDF:1993sbj,CDF:1997lmq,data2,D0:2009apj,ATLAS:2016ydt,LHCb:2012aiv}, with a particularly compelling demonstration emerging from the same-sign WW production at the LHC \cite{CMS:2022pio,ATLAS:2025bcb}.

This distinctive signature, where two W bosons with identical charge states are produced simultaneously, provides one of the cleanest experimental window into double parton scattering processes,
{ since the corresponding SPS background can  be suppressed by appropriate cuts}.
 Let us also mention
that recent experimental and phenomenological studies on triple parton scattering can be found in, e.g., Refs. \cite{CMS:2021qsn,dEnterria:2016ids}.  {Notably},
DPS {also} provides unique insights into hadronic structure \cite{noiprl,rapid,jhepc,Cotogno:2020iio,noiPLB,Diehl:2021wvd} by revealing novel properties of the parton dynamics, including double parton correlations and the
mean transverse separation between quarks and gluons.
These fundamental information are encoded in double parton distributions (DPDs), entering the DPS cross-section.

In this scenario, nuclear targets offer unique advantages for studying DPS, significantly enhancing the detection of { signal production} rates that are typically sub-leading w.r.t. single parton scattering (SPS) signals {backgrounds}.
Moreover, DPS {investigation} of proton-nucleus ($pA$) and nucleus-nucleus ($AA$) collisions provide crucial insights also into MPI  in proton-proton ($pp$) collisions {relevant for data-analyses at LHC}. Recent studies have demonstrated substantial progresses in understanding nuclear DPS mechanisms across various final states
\cite{dEnterria:2017yhd,Cattaruzza:2005nu,dEnterria:2012jam,Salvini:2013xpz,Cazaroto:2016nmu,Huayra:2019iun,Blok:2019fgg,Blok:2020oce,Blok:2020ckm} and implemented in PYTHIA Monte Carlo simulations \cite{Fedkevych:2019ofc}. {Let us also mention some experimental analyses of nuclear DPS, e.g., those in Refs. \cite{LHCb:2020jse,Maneyro:2024twb,CMS:2024wgu}.}
We would like to mention that some experimental effort has been devoted
to DPS measurements in $pA$ collisions, whose analyses and results are reported in Refs. \cite{LHCb:2020jse,CMS:2024wgu}.

The theoretical framework for DPS in $pA$ collisions reveals two distinct mechanisms: $i)$ DPS1: a process analogous to proton-proton DPS, where two partons from the incident nucleon interact with two partons within a single target nucleon in the nucleus; $ii)$ DPS2: a novel mechanism unique to nuclear targets, where two partons from the incoming nucleon interact with partons from different nucleons in the nucleus.
This two-mechanism approach, first established in Refs. \cite{Strikman:2001gz,Blok:2012jr}, provides a comprehensive description of double parton scattering reaction mechanism in nuclear environments.
In nuclear collisions, accurate characterization of both DPS1 and DPS2 mechanisms is crucial for providing reliable DPS cross-section predictions.
 The first {ingredient}, {to be considered for DPS1,} is the nucleon light-cone momentum distribution {(LCMD)}, which can be  {reliably} calculated for light nuclei \cite{Fornetti:2023gvf} and reasonably approximated for heavy nuclei. The second {one} is the free nucleon DPD \cite{Blok:2012jr}. On the contrary,
the DPS2 contribution presents greater computational complexity, requiring the evaluation of off-diagonal  {two-body nuclear densities}. While this calculation can be realistically {addressed} for light nuclei, it becomes particularly challenging for heavy ions and approximate methods are required. {{For heavy nuclei}}, the DPS1 cross-section {scales}  as $~A ~\sigma^{pp}_{DPS}$, whereas the DPS2 contribution scales approximately as $A^{1/3}\sigma_{DPS1}=A^{4/3} \sigma^{pp}_{DPS}$ \cite{Blok:2012jr}. This scaling behavior indicates that the DPS2 mechanism dominates in nuclear collisions, providing a crucial insight for experimental observations and theoretical predictions.

{Since  current investigations address the short-range correlations between nucleons, and hence large kinetic energies are involved, a reliable relativistic framework for strongly bound systems is necessary. In view of this,} our approach, based on the Poincar\'e covariant light-front (LF) formalism
{for nuclei (see, e.g., Refs. \cite{KP,Lev:1993pfz} for generalities,   Ref. \cite{Lev:2000vm} for the deuteron and \cite{DelDotto:2016vkh} for the A=3 mirror nuclei)}, 
establishes a rigorous  framework for characterizing DPS1 and DPS2 contributions in nuclear DPDs.
{  It is worthwhile to recall that, within the LF Hamiltonian dynamics, systems are composed by a fixed number of on-mass-shell particles. This means that in a nucleus one has only $A$ nucleons and that the square of the four-momentum of each nucleon is equal to its mass: $p^2 = M^2$.}
{As  initial case study, nuclear double-parton distributions  are calculated  for the deuteron {by} using realistic wave-functions derived from the phenomenological {nucleon-nucleon} AV18 potential \cite{Wiringa:1994wb}, but formally embedded in the LF framework. In general, the choice of light nuclei takes advantage  of the fact that  reliable wave-functions are available, so as to provide an ideal, but challenging, approach for computing these unknown distributions  (results beyond the deuteron will be presented elsewhere).}
{{Before} evaluating deuteron DPS1 and DPS2},
novel partial sum rules for baryon number and momentum conservation in DPDs  are introduced for the first time.
These {novel}  sum-rules, obtained by generalizing the integral properties of nucleon  DPDs \cite{Gaunt:2009re}, yield a quantitative foundation for comparing the relative magnitudes of DPS1 and DPS2 mechanisms.
The nuclear DPD sum-rules have been numerically verified and the results point to the conclusion that the DPS1 contribution is not negligible and compares with the DPS2 one.
The results here obtained have broader   implications, particularly when studies of double-parton scattering in electromagnetic processes {are being planned} at  facilities such as Jefferson Lab and  upcoming  Electron-Ion Colliders
\cite{Ceccopieri:2021luf,preparation,Blok:2025kvs}.   Moreover, many-body nuclear form factors, which can be evaluated from the off-forward two-body nuclear density, {e.g. the one}  relevant to the DPS2 mechanism,  play a significant role in other processes, such as
the $J/\psi$ diffractive photoproduction \cite{Guzey:2022jtv}. In conclusion, future experimental analyses will benefit of theoretical studies of nuclear DPS reactions and, in turn, they  will provide   valuable insights into those fundamental quantities.

The paper is structured as follows. In Sect. II we formally generalize the  definition   of the hadron DPD to the nuclear case and the expressions for  DPS1 and DPS2 contributions are properly obtained. In Sect. III the basic DPD sum rules {are} extended to the nuclear case, in particular the partial sum rules for the two mechanisms are introduced. {In  Sect. IV, the deuteron densities obtained within the Poincar\'e-covariant LF framework are introduced, and  calculations of the nuclear contributions to DPD are presented. }In Sect. V, the results of the {whole} calculations of the deuteron DPDs are showed and discussed. In Sect. VI { we draw our} conclusions.

\section{ Nuclear DPS}
In this  Section, {the general formalism for describing the nuclear DPS is introduced.} We prove that the Light-Cone (LC) correlator, which  defines the DPDs in  QCD, when acting on a nuclear system, results in two distinct distributions, corresponding to  DPS1 and DPS2 mechanisms, as discussed in Refs. \cite{Strikman:2001gz,Blok:2012jr}.
 Let us first recall the DPDs for the nucleon case (see, e.g., Ref. \cite{Diehl:2011yj}):

\begin{align}
    \tilde D^\tau_{ij}(x_1,x_2, \mathbf{y}_\perp) = 2 p^+ \int \dfrac{d z_1^-}{2 \pi} \dfrac{d z_2^-}{2 \pi} \int d y^- ~e^{i x_1 p^+ z_{1}^-} e^{i x_2 p^+ z_{2}^-}  \langle p;~ \tau | \theta_i(y,z_1) \theta_j(0,z_2) |\tau;~p\rangle  \Big|_{z_1^+=z_2^+=y^+=0}^{\mathbf z_{\perp,1}= \mathbf z_{\perp,2}=\mathbf 0}~,
    \label{Eq:nucl_DPD}
\end{align}
where  {$\{ij\}$ are the flavors of the involved quarks, } $|\tau {;p} \rangle$ is the nucleon state {with intrinsic dofs}, {$p^\mu\equiv\{p^\pm, {\bf p}_\perp\}$  the nucleon four-momentum with}
$p^\pm=(p^0\pm p^3)/\sqrt{2}$  \ and $x_\ell =q^+_\ell/p^+$ the longitudinal momentum fraction carried by  the $\ell$-th   parton w.r.t. the nucleon momentum, { being} $q_\ell^\mu$  the momentum of the $\ell$-parton.  Moreover, ${\bf y}_\perp$ represents the transverse distance between the two partons.
{Clearly,} one must have $x_1 + x_2 \le 1 $.
The bi-linear operators appearing in the above expression read:
 \begin{align}
    \theta_i(y,z) = \bar q_i \left( y -\dfrac{1}{2}z \right)\dfrac{1}{2}\gamma^+ q_i \left(y+ \dfrac{1}{2}z \right) ~,
    \label{Eq:correlatorDPD}
\end{align}
 where   $q_i(y\pm z/2)$ is the quark field operator  for a parton of flavor $i$.
 The generalization to a nucleus $A$   is:
\begin{align}
\label{Eq:nucl_DPD_gen}
    \tilde D_{ij}^{A}(x_1^B,x_2^B,\mathbf y_\perp) = 2 P^+  \int \dfrac{d z_1^-}{2 \pi} \dfrac{d z_2^-}{2 \pi} \int d y^- ~e^{i x_1^B P^+ z_1^-} e^{i x_2^B P^+ z_2^-}  \langle A |\theta_i(y,z_1) \theta_j(0,z_2) |A \rangle \Big|_{z_1^+=z_2^+=y^+=0}^{\mathbf z_{\perp,1}= \mathbf z_{\perp,2}=\mathbf 0}~.
\end{align}
Now $P^+$ is the plus component of the nucleus four-momentum in the  nucleus rest frame, whereas $|A\rangle$ denotes the nuclear state.

In order to correctly evaluate the above expression, it is essential to define the nuclear state $|A\rangle$ in a suitable framework.
To this aim, {we} consider a {LF} approach \cite{KP,Brodsky:1997de,Diehl:2000xz}, {since it remarkably allows us to easily separate the centre of mass motion from the intrinsic one (as in the non relativistic case), both at the level of the nucleus and the nucleon, given the subgroup properties of the LF-boosts}.
Let us define the conventions for the momenta of partons, nucleons and nuclei, respectively: i)
{{ $q^\mu_\ell$ are the components the four-momentum of the $\ell-$th parton  in the nucleus rest frame},} ii) $p^\mu_r~$
{{are the components of} the four-momentum of the $r-$th  nucleon {in the nucleus rest
frame,} {with $r=1,2, \dots,A$} and $M$ the nucleon mass {{(as noticed in the Introduction}} $p^2_r = M^2$ in LF formalism),  iii) {
{$M_A$ is the nucleus mass (recall that $P^\mu$ = \{$P^\pm=M_A,{\bf P}_\perp={\bf 0}_\perp$\} in the rest frame of the nucleus)}}. }
The longitudinal momentum fractions carried by the $\ell$-th parton w.r.t. the nucleus can be introduced:

 \begin{align}
     x_\ell^B = \dfrac{q_\ell^+}{P^+}~.
     \label{eq:xell_def}
 \end{align}

One can also define the longitudinal-momentum fractions carried by the $r$-th nucleon w.r.t. the parent nucleus:
\begin{align}
     \xi_r = \dfrac{p_r^+}{P^+}~.
 \end{align}

With these definitions, the nucleon momentum vector can be expressed as follows {in terms of the intrinsic LF coordinates \{$\xi_r, {\bf k}_{\perp,r}$\}}:

\begin{align}
    p_r^+ = \xi_r P^+;~~~{\bf p}_{\perp,r}=\xi_r {\bf P}_\perp+{\bf k}_{\perp,r}~,
\end{align}
 and the following constraints from  the four-momentum conservation are found:

 \begin{align}
 \label{Eq:cons1}
     \sum_r^A p_r^+ = P^+  &\Rightarrow \sum_r^A \xi_r=1~,
     \\
     \sum_r {\bf p}_{\perp,r} = {\bf P}_\perp~ &\Rightarrow  \sum_{r} {\bf k}_{\perp,r} = {\bf 0}~.
     \label{Eq:cons2}
 \end{align}
In the present analysis, we {assume a factorized form of} the nuclear state $|A\rangle$, {as the Cartesian product of i) the nuclear wave function (wf) that takes into account} only nucleonic dofs and ii)  an intrinsic part, with partonic dofs.  In this framework, one gets:

\begin{align}
\label{Eq:A_state}
    |A \rangle &= \sum_{\tau_1,..,\tau_A=n,p} \sum_{\lambda_1,..,\lambda_A} \dfrac{1}{[{2 (2 \pi)^{3}}]^{(A-1)/2}} \int \left[\prod_{r=1}^A \dfrac{d\xi_r d^2 k_{\perp,r}}{\sqrt{\xi_r}}\right]  \delta \left(1-\sum_{l=1}^A \xi_l \right) \delta \left(
 \sum_{l=1}^A \mathbf k_{\perp,l} \right)
  \\
 \nonumber
 &\times
 \psi(\xi_1,..,\xi_A, \mathbf k_{1,\perp},...,\mathbf k_{\perp,A},\tau_1,...,\tau_A,\lambda_1,...,\lambda_A)
|n_1 \rangle \cdot \cdot \cdot |n_A\rangle~,
\end{align}
where, $\psi$ is the  LF nuclear wf that describes the dynamics of A nucleons (i.e. the collection of {centres of masses of QCD singlet-states}) {in the frame where ${\bf P}_\perp=0$,} and $\lambda_r$ is the projection of the spin of $r$-th nucleon along the $z$ axis.
{Of course, the {nuclear} wf is completely antisymmetric under the exchange of two {nucleons}.}

{
In analogy with the {Fock} decomposition of the hadronic states in terms of free parton states  {(see, e.g., Ref. \cite{Diehl:2000xz})},  the above independent nucleons state, {$|n_i\rangle$,} {should consist only of} plane-waves {describing}  the centre-of-mass (CM) motion of each nucleon, {as whole}. {{However}}, in order to completely describe the nuclear state, one should also include the intrinsic partonic part, {once we adopt the insights of Ref. \cite{Strikman:2001gz}, where   the introduction of a  large
scale, i.e. the nucleus radius, suggests separating the effects of nucleonic and partonic dofs, obtaining a workable approximation}. In particular, we simply assume that:}

\begin{align}
    |n_r \rangle = \underbrace{| \xi_r P^+,\xi_r {\bf P_\perp}+\mathbf k_{\perp,r}, \tau_r, \lambda_r\rangle}_{CM~ plane-wave}~ \otimes  \underbrace{|\phi_r \rangle}_{intrinsic}~.
\end{align}
Since our calculations will parametrize the intrinsic dependence {of the nucleon state} entirely through parton distribution functions, as discussed {in what follows}, we will henceforth omit explicit reference to {such a dependence}.
Here, $\tau_r$ represents the nucleon isospin.
{{Once the intrinsic part is properly normalized,} the orthonormalization rule of the nucleon states is}:

\begin{align}
\label{Eq:norm}
    \langle n_r | n'_k \rangle &=
 2 (2 \pi)^3 p_r^+ \delta(p_r^+ -p^{\prime +}_k)\delta^{(2)}({\bf p}_{\perp,r}- {\bf p}'_{\perp,k})\delta_{\tau_r,\tau^\prime_k} \delta_{\lambda_r,\lambda^\prime_k}  {\delta_{r,k}}
    \\
    \nonumber
    &=2 (2 \pi)^3  \xi_r \delta \left(\xi_r- \frac{{P'}^+}{P^+}
    \xi'_k\right) \delta^{(2)}(\xi_r \mathbf{P}_\perp + \mathbf k_{\perp r}- \xi'_k \mathbf{P}'_\perp - \mathbf k'_{\perp k}) \delta_{\tau_r,\tau^\prime_k} \delta_{\lambda_r,\lambda^\prime_k}  {\delta_{r,k}}~,
\end{align}
{Where we denote with $\delta_{r,k}$ the orthogonality over all the intrinsic quantum numbers.}
The above expression can be generalized to the nuclear case, viz.:

\begin{align}
   \langle A |A' \rangle &=2(2 \pi)^3 P^+ \delta(P^+ - P'^+) \delta^{(2)} (\mathbf{P}_\perp -\mathbf{P}'_\perp)~,
\end{align}
therefore, {after inserting Eq. \eqref{Eq:A_state},} the orthonormalization rule of the nuclear wfs {reads}:

{
\begin{align}
   \langle A |A' \rangle &=  \dfrac{1}{[2 (2\pi^3)]^{A-1}}  \sum_{\tau_1,...,\tau_A=n,p}   \int \left[\prod_{r=1}^{A} \dfrac{d\xi_r d^2 k_{\perp,r}}{\sqrt{\xi_r}}\right] \left[\prod_{r=1}^{A} \sum_{\lambda_r}\right]
   \\
   \nonumber
    & \times
   \psi^\dagger(\xi_1,..,\xi_A, \mathbf k_{1,\perp},...,\mathbf k_{\perp,A},\tau_1,...,\tau_A,\lambda_1,...,\lambda_A) \delta \left(1-\sum_{k=1}^A \xi_k \right) \delta \left(
 \sum_{l=1}^A \mathbf k_{\perp,l} \right)
 \\
 \nonumber
    &\times \Biggl\{ \sum_{\tau'_1,...,\tau'_A=n,p}
 \int\left[\prod_{r=1}^{A} \dfrac{d\xi'_r d^2 k'_{\perp,r}}{\sqrt{\xi_r'}}\right] \left[\prod_{r=1}^{A} \sum_{\lambda'_r}\right]
    ~
    \langle n_1 | n_1' \rangle  \langle n_2 | n_2' \rangle \dots  \langle n_A | n_A' \rangle
    \\
 \nonumber
 & \times \psi(\xi'_1,..,\xi'_A, \mathbf k'_{\perp,1},...,\mathbf k'_{\perp,A},\tau'_1,...,\tau'_A,\lambda'_1,...,\lambda'_A)
 \delta \left(1-\sum_{k=1}^A \xi'_k \right) \delta \left(
 \sum_{l=1}^A \mathbf k'_{\perp,l} \right)\Biggr\}~.
\end{align}
}

 {By inserting the result of   Eq. \eqref{Eq:norm}, the quantity  in the curly brackets becomes}:
\begin{align}
\nonumber
    & \int \left[\prod_{r=1}^{A} \dfrac{d\xi'_r d^2 k'_{\perp,r}}{\sqrt{\xi_r'}}\sum_{\lambda'_r,\tau'_r} \right]
    \left[\prod_r^A   \langle n_r | n_r' \rangle \right]
    \\
    \nonumber
    &\times
    \psi(\xi'_1,..,\xi'_A, \mathbf k'_{\perp,1},...,\mathbf k'_{\perp,A},\tau'_1,...,\tau'_A,\lambda'_1,...\lambda'_A)
 \delta \left(1-\sum_{k=1}^A \xi'_k \right) \delta \left(
 \sum_{l=1}^A \mathbf k'_{\perp,l} \right) = \\
 \nonumber
 & =  [2 (2 \pi^3)]^A \left[ \prod_{r=1}^{A} \sum_{\lambda'_r,\tau'_r} \sqrt{\xi_r} \right] P^+  \delta \left( P^+ - P'^+\right) \delta \left( \mathbf{P}_\perp -\mathbf{P}'_\perp \right)
 \\
 \nonumber
 &\times
 \psi(\xi_1,..,\xi_A, \mathbf k_{1,\perp},...,\mathbf k_{\perp,A},\tau'_1,...,\tau'_A,\lambda'_1,...,\lambda'_A)
 \\
 &\times
 \prod_{i=r}^A \delta(\tau_r-\tau'_r)\delta(\lambda'_r-\lambda_r)~,
\end{align}
and therefore:
\begin{align}
    \langle A |A' \rangle &=  2 (2 \pi^3) P^+ \delta \left( P^+ - P'^+\right) \delta \left( \mathbf{P}_\perp -\mathbf{P}'_\perp \right) \int \left[\prod_{r=1}^{A} d\xi_r d^2 k_{\perp,r} \sum_{\lambda_r} \sum_{\tau_r=n,p}\right]  \\
    \nonumber
    &\times |\psi(\xi_1,..,\xi_A, \mathbf k_{1,\perp},...,\mathbf k_{\perp,A},\tau_1,...,\tau_A,\lambda_1,...,\lambda_A)|^2
    \delta \left(1-\sum_{k=1}^A \xi_k \right) \delta \left(
 \sum_{l=1}^A \mathbf k_{\perp,l} \right)~.
\end{align}
Thus the normalization of the nuclear LF wf reads:

\begin{align}
\label{Eq:norm_wf}
 &\sum_{\tau_1,...,\tau_A=n,p}   \int \left[\prod_{r=1}^{A} d\xi_r d^2 k_{\perp,r} \sum_{\lambda_r}\right]\delta \left(1-\sum_{k=1}^A \xi_k \right) \delta \left(
 \sum_{l=1}^A \mathbf k_{\perp,l} \right)
 \\
 \nonumber
 &\times
 |\psi(\xi_1,..,\xi_A, \mathbf k_{1,\perp},...,\mathbf k_{\perp,A},\tau_1,...,\tau_A,\lambda_1,..,\lambda_A)|^2
  =1~.
\end{align}
Once the nuclear state has been properly defined in terms of its constituents and the normalization of the wave function is established,  the expression given in Eq. (\ref{Eq:A_state}) can be used in the nuclear DPD in Eq. (\ref{Eq:nucl_DPD_gen}).

\subsection{Nuclear DPDs in impulse approximation}
In order to properly allow the bilinear operators in Eq. (\ref{Eq:nucl_DPD_gen}) to act on the intrinsic part of the nucleonic states which describe the {{nucleon internal structure}}, we adopt the decomposition of the nuclear {{operators}} in impulse approximation {(see also Ref. \cite{Strikman:2001gz})}. In this framework, each operator in Eq. (\ref{Eq:nucl_DPD_gen}) is given, {for each quark flavor,} by an incoherent sum over operators acting on  nucleon states:

\begin{align}
\label{Eq:operator}
     \theta_i(y,z_1) \theta_j(0,z_2)  \Longrightarrow \Theta_i(y,z_1) \Theta_j(0,z_2)
     &=\sum_{l,r}^A \theta^l_i\left(y,{z_1} \right)
         \theta^r_j \left(0,{z_2} \right)
         \\
         \nonumber
        & {= \sum_{l=1}^A \theta^l_i\left(y,{z_1} \right)
         \theta^l_j \left(0,{z_2} \right)+ \sum_{l\ne r}^A \theta^l_i\left(y,{z_1} \right)
         \theta^r_j \left(0,{z_2} \right)}~,
\end{align}
{where the {superscript} $l(r)$ in $\theta$ specifies that the operator acts only on the partons pertaining to the nucleon $l(r)$.}
From now on, taking care of antisymmetry {in the nuclear wf},  {the operator  acts on  i) a single state $|n_1\rangle$ when $r=l$, obtaining an overall factor $A$ (DPS1),  and on the pair $|n_1\rangle |n_2 \rangle$, when $l\ne r$, so that a factor $A(A-1)$ has to be introduced} (DPS2):

\begin{align}
\label{Eq:operator_final}
     \Theta_i(y, z_1) \Theta_j(0, z_2)  = A \left[ \theta^1_i(y, z_1)  \right] \left[ \theta^1_j(0, z_2)  \right] +A(A-1)\left[ \theta^1_i(y, z_1)  \right] \left[ \theta^2_j(0, z_2)  \right]~,
\end{align}
(see also Ref.  \cite{Blok:2012jr}). In conclusion,
 within this approach  DPS1 and DPS2 mechanisms directly arise from the decomposition of the above bilinear operators.

 Once the above operator is considered in Eq. (\ref{Eq:nucl_DPD_gen}) {and} the nuclear expansion of Eq. (\ref{Eq:A_state}) for the state $|A\rangle$ is taken into account, one gets the following expression for the DPD for the nucleus:

\begin{align}
\nonumber
    \tilde D_{ij}^{A}(x_1^B,x_2^B,\mathbf y_\perp) &= \dfrac{2 P^+}{[2(2 \pi)^3]}   \sum_{\tau_1,\tau_2 = n,p}  \sum_{\lambda_1,\lambda_2} \sum_{\lambda'_1,\lambda'_2} \int \dfrac{d z_1^-}{2 \pi} \dfrac{d z_2^-}{2 \pi} d y^-
     \dfrac{ d\xi_1 d\xi_2 d  \xi'_1}{\sqrt{\xi_1 \xi_2  \xi'_1 (\xi_1+\xi_2- \xi'_1)}}
    d^2 k_{1,\perp} d^2k_{2,\perp} d^2  k'_{\perp,1}
    \\
    \nonumber
    &\times  e^{i x_1^B P^+ z_1^-} e^{i x_2^B P^+ z_2^-} \psi(\xi_1,\xi_2, \mathbf k_{1,\perp},\mathbf k_{2,\perp},\tau_1,\tau_2,\lambda_1,\lambda_2)\psi^{\dagger}( \xi'_1, \xi'_2, { \mathbf{ k}'}_{\perp,1}, {\mathbf {k}'}_{\perp,2},\tau_1,\tau_2,\lambda'_1,\lambda'_2)
    \\
    &\times
    \langle  n'_1 | \langle  n'_2 |\Theta_i \left(y,{z_1} \right) \Theta_j \left(0,{z_2} \right) |n_1 \rangle | n_2 \rangle \Big|_{z_1^+=z_2^+=y^+=0}^{\mathbf z_{\perp,1}= \mathbf z_{\perp,2}=\mathbf 0}~,
        \label{Eq:gen}
\end{align}

where
 $\xi'_2=\xi_1+\xi_2-\xi'_1$,   $ \mathbf k'_{\perp,2}=  \mathbf k_{1,\perp}+\mathbf k_{2,\perp}- \mathbf k'_{\perp,1}$
and

\begin{align}
 \label{den1}
    &
    \psi(\xi_1,\xi_2, \mathbf k_{1,\perp},\mathbf k_{2,\perp},\tau_1,\tau_2,\lambda_1,\lambda_2)\psi^{\dagger}( \xi'_1, \xi'_2, { \mathbf{ k}'}_{\perp,1}, {\mathbf {k}'}_{\perp,2},\tau_1,\tau_2,\lambda'_1,\lambda'_2)
    \\
    \nonumber
    &
     \equiv
\sum_{\tau_3,..,\tau_A=n,p} \sum_{\lambda_3,..,\lambda_A}  \int \left[\prod_{r=3}^{A-1} {d\xi_r d^2 k_{\perp,r}}\right]
\psi(\xi_1,..,\xi_A, \mathbf k_{1,\perp},...,\mathbf k_{\perp,A},\tau_1,...,\tau_A,\lambda_1,...,\lambda_A)
  \\
 \nonumber
 &\times
 \psi^{\dagger}(\xi'_1,\xi'_2,\xi_3,..,\xi_A, \mathbf {k}'_{\perp,1},\mathbf {k}'_{\perp,2},\mathbf {k}_{\perp,3},...,\mathbf k_{\perp,A},\tau_1,...,\tau_A,\lambda'_1,\lambda'_2,\lambda_3,...,\lambda_A).
\end{align}
In Eq. (\ref{den1}), one has
$ \xi_A= 1 - \sum_{r=1}^{A-1}\xi_r $ and $ \mathbf k_{\perp,A}= -\sum_{r=1}^{A-1} \mathbf k_{\perp,r}$.
The normalization of the above quantity reads:

\begin{align}
\label{den2}
\sum_{\tau_1,\tau_2}\sum_{\lambda_1,\lambda_2} \int d\xi_1 d\xi_2 d^2 k_{1,\perp}d^ 2k_{2,\perp}
    |\psi(\xi_1,\xi_2, \mathbf k_{1,\perp},\mathbf k_{2,\perp},\tau_1,\tau_2,\lambda_1,\lambda_2)|^2 =1~.
\end{align}

{Notice that in order to compare the results here presented to those discussed in Ref. \cite{Blok:2012jr}  and to the usual nuclear PDFs described in impulse approximation, see e.g. Ref. \cite{Pace:2022qoj}, it is necessary to evaluate DPDs and PDFs as functions of: }
\begin{align}
\label{Eq:xrisc}
    x_{\ell} \equiv \dfrac{q_\ell^+}{P^+} \dfrac{M_A}{M} =  \dfrac{q_\ell^+}{P^+}
    \dfrac{1}{ \raisebox{-0.06cm}{$\bar \xi$}} =   \dfrac{x_\ell^B}{ \raisebox{-0.06cm}{$\bar \xi$}}
 \end{align}
{with}:
\begin{align}
\bar \xi = \frac{M}{M_A} \sim \frac{1}{A}~.
\end{align}
{
{From now on, all distributions here studied  will be evaluated as functions of $x_\ell$ so that}
$\sum_\ell x_\ell= M_A/M \sim A$. This amounts to have an average nucleon plus-component given by $\sim P^+_A/A$}.
One should notice that:
\begin{align}
   x_1 + x_2  \le ( q^+_1 / P_A^+) (M_A/M) + [(P_A^+ - q^+_1) / P_A^+] (M_A/M) =   M_A / M~.
   \label{eq:x1_x2}
\end{align}

{Let us discuss  separately the two possibilities where the  operators act on partons belonging to i) the same nucleon, i.e.  the DPS1 mechanism,  or ii) two different nucleons, i.e. the DPS2 mechanism.}

\subsection{The DPS1 contribution}
In this case one can exploit the inner product $\langle n'_2|n_2\rangle$ in Eq. (\ref{Eq:gen}).
Thus, {by using } $x_l$ of Eq. \eqref{Eq:xrisc}, one can simplify Eq. (\ref{Eq:gen}):

\begin{align}
\label{eq:dps1}
    \tilde D_{ij}^{A,1}(x_1,x_2,\mathbf y_\perp) &= A~ 2 P^+ \sum_{\tau, \tau_2 = n,p}
    \int \dfrac{d z_1^-}{2 \pi}  \int\dfrac{d z_2^-}{2 \pi}  \int d y^-
     \int \dfrac{ d\xi_1 }{\xi_1 }  \int d\xi_2
    \int d^2 k_{1,\perp} \int d^2k_{2,\perp}  \bar \xi^2
    \\
    \nonumber
    & \times
    e^{i x_1 P^+ z_1^- /\bar \xi} e^{i x_2 P^+ z_2^-/\bar \xi}
    |\psi(\xi_1,\xi_2, \mathbf k_{1,\perp},\mathbf k_{2,\perp},\tau,\tau_2)|^2~
    \\
    \nonumber
    &\times
    \langle  n_1 | \theta_i\left(y,{z_1}\right) \theta_j \left(0,{z_2} \right) |n_1 \rangle  \Big|_{z_1^+=z_2^+=y^+=0}^{\mathbf z_{\perp,1}= \mathbf z_{\perp,2}=\mathbf 0}~.
\end{align}
where {the $\bar \xi^2$} is needed to preserve the normalization of the distribution and one defines:

\begin{align}
     |\psi(\xi_1,\xi_2, \mathbf k_{1,\perp},\mathbf k_{2,\perp},\tau,\tau_2)|^2 \equiv \sum_{\lambda_1,\lambda_2}
      |\psi(\xi_1,\xi_2, \mathbf k_{1,\perp},\mathbf k_{2,\perp},\tau,\tau_2,\lambda_1,\lambda_2)|^2~.
      \label{psi12}
\end{align}

{Recalling that the plus component of the
$i$-th nucleon  momentum is $p^+_i = \xi_i P^+$, one recasts Eq. \eqref{eq:dps1} as follows}:

\begin{align}
    \tilde D_{ij}^{A,1}(x_1,x_2,\mathbf y_\perp) &= A \sum_{\tau, \tau_2 = n,p}
    \int \dfrac{d z_1^-}{2 \pi} \int \dfrac{d z_2^-}{2 \pi}  \int d y^-
    \int  \dfrac{ d\xi_1 }{\xi_1 } \int d\xi_2
   \int  d^2 k_{1,\perp} \int d^2k_{2,\perp}~  2 \dfrac{p_1^+}{\xi_1}\bar \xi^2
    \\
    \nonumber
    &\times
     e^{i x_1/\xi_1 p_1^+ z_1^-/\bar \xi} e^{i x_2/\xi_1 p_1^+ z_2^-/\bar \xi}
    |\psi(\xi_1,\xi_2, \mathbf k_{1,\perp},\mathbf k_{2,\perp},\tau,\tau_2)|^2
    \\
    \nonumber
    &\times
    \langle  n_1 | \theta_i(y,z_1) \theta_j(0,z_2) |n_1 \rangle  \Big|_{z_1^+=z_2^+=y^+=0}^{\mathbf z_{\perp,1}= \mathbf z_{\perp,2}=\mathbf 0}~,
\end{align}

and one gets:
\begin{align}
     \tilde D_{ij}^{A,1}(x_1,x_2,\mathbf y_\perp) &=  A \sum_{\tau = n,p}
\int d\xi_1 ~ \dfrac{ \bar \xi^2  }{\xi_1^2 }~\rho_\tau^A (\xi_1) ~\tilde D_{ij}^{\tau}\left(x_1\dfrac{\bar \xi}{\xi_1}, x_2\dfrac{\bar \xi}{\xi_1},\mathbf y_\perp \right) ~,
    \label{DA1}
\end{align}
where $\rho_\tau^A (\xi_1)$ is the {one-body} LCMD {of a nucleus with $A$ nucleons,} given by
\begin{align}
  \rho_\tau^A (\xi_1)  \equiv \sum_{\tau_2 = n,p} \int d\xi_2 \int d^2 k_{1,\perp} \int d^2k_{2,\perp} |\psi(\xi_1,\xi_2, \mathbf k_{1,\perp},\mathbf k_{2,\perp},\tau,\tau_2)|^2
    \label{Eq:density1}
\end{align}
{and normalized as follows
\begin{align}
  \sum_\tau  \int d\xi ~ \rho_\tau^A(\xi) = 1 ~.
  \label{eq:normrho}
\end{align}}
 It is immediate to recognize that
in the DPS1 case one must have
$\bar \xi x_1/\xi_1+\bar \xi x_2/\xi_1 \leq 1$ {(recall that $x_l$ is defined in Eq. \eqref{Eq:xrisc})}.
{In Eqs. {\eqref{DA1}} and \eqref{Eq:DPD1_momentum}, the effective range of integration  is  $\xi_{min} =  (x_1 + x_2)\bar \xi \leq \xi \leq 1$  since the above DPD  vanishes for $x_1 \bar \xi/\xi +x_1 \bar \xi/\xi \leq 1$. }
Moreover, in order to compare our findings with that of Ref. \cite{Blok:2012jr}, {one should recall  that}
a LCMD normalized to the number of nucleons is {adopted there:}

\begin{align}
\label{bar}
    \bar \rho_\tau (\xi) \equiv A \rho_\tau(\xi)~.
\end{align}

{ The main result of this subsection is Eq. (\ref{DA1}) which expresses
the nuclear double distributions as a convolution of the LCMD of the nucleus $A$ with nucleonic double parton distributions. As it is well known,
the nucleonic DPDs present a singular behaviour as $y_\perp\rightarrow 0$, induced by the socalled splitting term, and necessitate a proper ultraviolet regularisation. Once this is provided, the $y$-integral can be performed and this leads to finite Fourier transform of DPDs depending on  $\bf {k}_\perp$, conjugated to $\bf{y}_\perp$ and allows the computation of DPDs sum rules which are defined at  $\bf {k}_\perp=0$. The nuclear DPDs inherit this issue from their nucleonic counterpart
since the convolution over LCMD involves
only longitudinal variables while the { internal structure}  remains unchanged.
Therefore, we refer the interested reader to Refs. \cite{Diehl:2018kgr,Diehl:2019rdh,Diehl:2020xyg} for the detailed treatment of these renormalization effects. In a nutshell, it amounts to introduce in the Fourier transform {of the nucleon DPDs} {a multiplicative regularizing function $\Phi$, viz.}}
\begin{align}
    D^\tau_{ij}(x_1,x_2, {\bf k}_\perp) \equiv  \int d^2y_\perp ~ e^{-i {\bf k}_\perp \cdot {\bf y}_\perp}  \Phi(y_\perp \nu) \tilde D^\tau_{ij}(x_1,x_2,{\bf y}_\perp)
\end{align}
{ The $\Phi$ functional form is constructed such as to tame the $y_\perp \to 0$ singularity with an appropriate choice of the regulating scale $\nu$.
{Therefore, the regularization procedure, applied to the nuclear DPD corresponding to the DPS1 mechanism, still leads
{to the Fourier-transformed version of
Eq. (\ref{DA1}) but in momentum space:}}}

\begin{align}
    D_{ij}^{A,1}(x_1,x_2,\mathbf k_\perp)
= A \sum_{\tau = n,p} \int d\xi ~ \dfrac{  \bar \xi^2 }{\xi^2 } ~
    \rho_\tau^A(\xi) ~  D_{ij}^{\tau}\left(x_1\dfrac{\bar \xi}{\xi},x_2 \dfrac{\bar \xi}{\xi},\mathbf k_\perp \right)~.
    \label{Eq:DPD1_momentum}
\end{align}

{ As a result of the impulse approximation, we anticipate that, within the DPS2 mechanism to be discussed in the next Section,  the two interacting partons belong to two distinct nucleons in the nucleus,
and, as such, their distribution is regular as
$y_\perp \to 0$ allowing the use of ordinary Fourier transform {since in this case the splitting effect is not relevant for the two partons involved in the reaction}.}

\subsection{The DPS2 contribution}
Let us now consider the case where the operator in
 Eq. (\ref{Eq:gen}) acts on two different nucleons and the overall factor $A(A-1)$ is included (cf. Eq. \eqref{Eq:operator_final}).
{Still using} $x_\ell$ {defined in}  Eq. \eqref{Eq:xrisc},
   the DPS2 contribution reads:
\begin{align}
    \label{Eq:gen2}
    \tilde D_{ij}^{A,2}(x_1,x_2,\mathbf y_\perp) &=A(A-1)\sum_{\tiny \begin{aligned}
        & \lambda_1,\lambda_2, \\
        & \lambda'_1,\lambda'_2
    \end{aligned}} \sum_{\tau_1, \tau_2 = n,p}
    \int \dfrac{d z_1^-}{2 \pi} \int \dfrac{d z_2^-}{2 \pi}
    \int d y^-
    \int  \dfrac{ d\xi_1}{\sqrt{\xi_1 }} \int  \dfrac{d\xi_2} {\sqrt{\xi_ 2}} \int  \dfrac{d  \xi'_1}{\sqrt{ \xi'_1 (\xi_1+\xi_2- \xi'_1)}}
    \\
    \nonumber
    &\times  \int   d^2 k_{1,\perp}  \int  d^2k_{2,\perp}  \int  d^2  k'_{\perp,1}\dfrac{2 P^+}{[2(2 \pi)^3]} e^{i x_1 P^+ z_1^-/\bar \xi} e^{i x_2 P^+ z_2^-/\bar \xi} ~\bar \xi^2
    \\
    \nonumber
    &\times
    \psi(\xi_1,\xi_2, \mathbf k_{1,\perp},\mathbf k_{2,\perp},\tau_1,\tau_2,\lambda_1,\lambda_2)
    \psi^{\dagger}( \xi'_1, \xi'_2,  \mathbf k'_{\perp,1},\mathbf k'_{\perp,2},\tau_1,\tau_2,\lambda'_1,\lambda'_2)
    \\
    \nonumber
    &\times
    \langle  n'_1 | \theta_i(y,z_1) |n_1 \rangle
    \langle  n'_2 |\theta_j(0,z_2)  | n_2 \rangle \Big|_{z_1^+=z_2^+=y^+=0}^{\mathbf z_{\perp,1}= \mathbf z_{\perp,2}=\mathbf 0}~.
\end{align}

{One notices  that the rightmost matrix element in Eq. \eqref{Eq:gen2}, i.e. $$ \langle  n'_2 |\theta_j(0,z_2)  | n_2 \rangle \Big|_{z_2^+=0}^{\mathbf z_{\perp,2}=\mathbf 0}$$  is the same appearing  in the GPDs  correlator \cite{Diehl:2003ny}, viz.
\begin{align}
\label{Eq:LCGPD}
    \Phi^{\tau,i}_{\lambda,\lambda'}(x,\zeta,\mathbf \Delta_\perp) = \int \frac{d z^-}{2 \pi}e^{i x p^+ z^-} \langle\tau'| \bar q_i \left(-\frac{z}{2}\right) \frac{\gamma^+}{2} q_i\left(\frac{z}{2}\right)|\tau \rangle \big |_{z^+ = {\bf z_\perp}=0}
\end{align}
where $|\tau~(\tau')\rangle$ is the nucleon initial (final) state with initial (final) spin polarization $\lambda~(\lambda')$.   Defining
the transverse and longitudinal momentum transfer, respectively:

\begin{align}
 \begin{cases}
 \mathbf \Delta_{\perp,r} = \mathbf  k'_{\perp,r}- \mathbf k_{\perp,r}
\\
 ~
 \\
  \Delta_r^+ =  P^+ (\xi'_r - \xi_r)~,
\end{cases}
\end{align} one has \begin{align}
\mathbf \Delta_{\perp}=\mathbf \Delta_{\perp,1} = -\mathbf \Delta_{\perp,2}~,
\end{align}
since the inner product of the spectator states leads to $\mathbf k_{1,\perp}+\mathbf k_{2,\perp} = \mathbf k'_{\perp,1}+ \mathbf k'_{\perp,2}$. Moreover, in analogy with GPDs \cite{Diehl:2003ny}, one can introduce
 the skewness variable:

\begin{align}
    \zeta_r = -\dfrac{\Delta_r^+}{2 \bar p_r^+} = - \dfrac{\xi'_r-\xi_r}{\xi'_r+\xi_r}~,
\end{align}
where $\bar p_r^{~+}$ is
the mean momentum of the $r$-th nucleon, i.e.:
\begin{align}
    \bar p_r^{~+} = \dfrac{p_r^+ + p'^{+}_r}{2}= \dfrac{P^+}{2}(\xi_r+ \xi'_r)~.
\end{align}
}

In addition, also the {leftmost} matrix element in Eq. (\ref{Eq:gen2})
 can be manipulated  to  {eventually} get the GPDs {correlator, as follows}
\begin{align}
    \langle  n'_1 | \theta_i(y,z_1) |n_1 \rangle &= \langle  n'_1 | \bar q_i \left(y- \dfrac{1}{2}z_1\right) \frac{\gamma^+}{2}  q_i \left(y+ \dfrac{1}{2}z_1\right) |n_1 \rangle
    \\
    \nonumber
    & = \langle  n'_1 |e^{-i \hat p  y} \bar q_i \left(- \dfrac{1}{2}z_1\right) \frac{\gamma^+}{2}   q_i \left( \dfrac{1}{2}z_1\right) e^{i \hat p  y} |n_1 \rangle= e^{-i y \cdot ( p'_1-p_1)} \langle  n'_1 | \theta_i(0,z_1) |n_1 \rangle~.
\end{align}
{Notably, the dependence on the four-vector $y$ is factorized out, and recalling that i) $y^+=0$ and  ii) $a^\mu b_\mu = {a^+ b^- + a^- b^+}- \mathbf a_{\perp} \cdot \mathbf b_{\perp}$, one has:}

\begin{align}
\label{Eq:expo}
   e^{-i y ( p'_1-p_1)} &=
   e^{-i y^- ( p'^{+}_1-p^+_1)} e^{-i \mathbf y_\perp \cdot (\mathbf  p_{\perp,1}-\mathbf p'_{\perp,1})}=e^{-i y^- ( p'^{+}_1-p^+_1)} e^{-i \mathbf y_\perp \cdot (\mathbf  k_{1,\perp}-\mathbf k'_{\perp,1})} e^{-i\mathbf y_\perp \cdot \mathbf P_\perp (\xi_1-\xi'_1) }
   \\
   \nonumber
   & = e^{-i y^- ( p'^{+}_1-p^+_1)} e^{i \mathbf y_\perp \cdot \mathbf \Delta_\perp}  e^{-i\mathbf y_\perp \cdot \mathbf P_\perp (\xi_1-\xi'_1) } =e^{-i y^-P^+(\xi'_1-\xi_1)  }e^{i \mathbf y_\perp \cdot \mathbf \Delta_\perp} e^{-i\mathbf y_\perp \cdot \mathbf P_\perp (\xi_1-\xi'_1) }
\end{align}
{Then, the integration} over $y^-$ {in Eq. \eqref{Eq:gen2} can be safely performed, getting}
\begin{align}
    \int dy^-~ e^{i y^-P^+(\xi'_1-\xi_1)}  =  \dfrac{2 \pi}{P^+} \delta(\xi'_1-\xi_1) ~.
\end{align}
 Once the above delta function is considered, {there is no longitudinal-momentum transfer and} the term in the exponent proportional to ${\bf P}_\perp$ in Eq. (\ref{Eq:expo}) is zero.
{In addition the integration over $z^-_{1(2)}$  leads to \begin{align}
\label{Eq:LCGPD2}
    \Phi^{\tau,i}_{\lambda,\lambda'}(x_i{\bar \xi\over \xi_r},0,\mathbf \Delta_\perp) = \int \frac{d z^-}{2 \pi}~e^{i (x_i \bar\xi/\xi_r) p^+_rz^-_r}  ~\langle  n'_r | \theta_i(0,z_r) |n_r \rangle \big |_{z^+ = {\bf z_\perp}=0}~.
\end{align}}
Finally, one obtains
\begin{align}
\label{Eq:gen15}
    \tilde D_{ij}^{A,2}(x_1,x_2,\mathbf y_\perp) &=  \dfrac{A(A-1)}{(2 \pi)^2} \sum_{\tau_1, \tau_2 = n,p} \sum_{\lambda_1,\lambda_2} \sum_{\lambda'_1,\lambda'_2} \int
      d\xi_1~\dfrac{\bar \xi}{\xi_1} \int d\xi_2~\dfrac{\bar \xi } {\xi_2}
    \int d^2 k_{1,\perp} \int d^2k_{2,\perp} \int d^2  k'_{\perp,1} ~e^{i \mathbf y_\perp \cdot \mathbf \Delta_\perp}
    \\
    \nonumber
    &\times     \psi(\xi_1,\xi_2, \mathbf k_{1,\perp},\mathbf k_{2,\perp},\tau_1,\tau_2,\lambda_1,\lambda_2)\psi^{\dagger}( \xi_1, \xi_2,  \mathbf k'_{\perp,1},\mathbf k'_{\perp,2},\tau_1,\tau_2,\lambda'_1,\lambda'_2)
    \\
    \nonumber
    &\times
    \Phi^{\tau_1,i}_{\lambda_1,\lambda'_1}\left(x_1 \dfrac{\bar \xi}{\xi_1},0, \mathbf\Delta_\perp \right) \Phi^{\tau_2,j}_{\lambda_2,\lambda'_2}\left(x_2 \dfrac{\bar \xi}{\xi_2},0, -\mathbf \Delta_\perp \right)~,
    \end{align}
where the LC  correlator for the nucleon $\tau$,  {$\Phi^\tau(x,\zeta,\mathbf \Delta_\perp)$ (see Eq. \eqref{Eq:LCGPD}) is } parametrized through
GPDs \cite{Diehl:2003ny} { at leading twist  as follows:  }

\be
 \label{Eq:correlator}
    \Phi^{\tau,i}_{\lambda,\lambda'} (x,\zeta, {\bf \Delta}_\perp) = \dfrac{1}{2 p^+}
 \left[ H^\tau_i(x,\zeta, t) \bar u(p',\lambda') \gamma^+ u(p,\lambda)+\dfrac{E^\tau_i(x,\zeta,t)}{2 M} \bar u(p',\lambda') i \sigma^{+ \mu}\Delta_\mu u(p,\lambda) \right]~.
\ee
In Eq. (\ref{Eq:correlator}) $u$ is the usual Dirac spinor,
$t = \Delta^2$,
$H$ and $E$ are the chiral-even (helicity conserving) and chiral-odd (helicity flip) GPDs of flavor $i$ and for the nucleon $\tau$.  In our case, $\zeta=0$, ${\bf\Delta}_{i\perp}={\bf k}_{i\perp}$  and $p^\mu$ is the parent-nucleon momentum.

Also in this case, one can define the DPD in momentum space:

\be
    \label{Eq:DPD2full}
  D_{ij}^{A,2}(x_1,x_2,\mathbf k_\perp)  =  A(A-1)
  \sum_{\tau_1, \tau_2 = n,p}  \sum_{\lambda_1,\lambda_2} \sum_{\lambda'_1,\lambda'_2}
  \int
      d\xi_1~\dfrac{\bar \xi}{\xi_1} \int d\xi_2~\dfrac{\bar \xi } {\xi_2} ~ \rho_{\tau_1 \tau_2}^A (\xi_1,\xi_2, {\bf k_\perp},\lambda_1,\lambda_2,\lambda'_1,\lambda'_2)
%
    \nonu\times ~
\Phi^{\tau_1,i}_{\lambda_1,\lambda'_1}\left(x_1 \dfrac{\bar \xi}{\xi_1},0, \mathbf k_\perp \right) \Phi^{\tau_2,j}_{\lambda_2,\lambda'_2}\left(x_2 \dfrac{\bar \xi}{\xi_2},0, -\mathbf k_\perp \right)~,
\ee
{where $\rho_{\tau_1 \tau_2}^A (\xi_1,\xi_2, {\bf k_\perp},\lambda_1,\lambda_2,\lambda'_1,\lambda'_2)$  is a two-body {spin-dependent} LCMD of a nucleus, with $A$ nucleons},  given by
\begin{align}
\label{Eq:density2}
    &\rho_{\tau_1 \tau_2}^A (\xi_1,\xi_2, {\bf k_\perp},\lambda_1,\lambda_2,\lambda'_1,\lambda'_2) \equiv  \int  d^2 k_{1,\perp} \int d^2k_{2,\perp}
    \\
    \nonumber
    &\times
    \psi(\xi_1,\xi_2, \mathbf k_{1,\perp},\mathbf k_{2,\perp},\tau,\tau_2,\lambda_1,\lambda_2) \psi^\dagger(\xi_1,\xi_2, \mathbf k_{1,\perp}+{\bf k_\perp},\mathbf k_{2,\perp}-{\bf k_\perp},\tau,\tau_2,\lambda'_1,\lambda'_2)~,
\end{align}
with the normalization that follows from Eq. {\eqref{den2}.}
{The expression in Eq. \eqref{Eq:DPD2full}} closely parallels the findings in Ref. \cite{Blok:2012jr}, {but
terms beyond the leading {{spin conserving}} one are found with  a different {{normalization.}} {{We recall that usually unpolarized distributions are dominant w.r.t. those related to spin flip.}}
It demonstrates that the nuclear DPD, associated with the DPS2 mechanism, is dependent on nucleon GPDs.  The key distinction in our approach lies in the {identification of} distinct contributions:
$i)$ the contribution with spin conservation between initial and final states, i.e.  when $\lambda_i = \lambda'_i$ for $i=1,2$;
$ii)$ the one spin-flip contribution, i.e. when
 when $\lambda_{1(2)} \neq \lambda'_{1(2)}$ and $iii)$  two spin-flips contribution, i.e. when $\lambda_{i} \neq \lambda'_{i}$ for $i=1,2$.

 In principle, the nuclear DPD depends on both the leading-twist GPDs, $H$ and $E$. However, for low $k_\perp$ values, the spin flip contributions are suppressed by factors $\mathcal{O}(k_\perp/(2M))$ and $\mathcal{O}(k^2_\perp/(4M^2)$ and therefore, as a first approximation, the  leading term is the one related to the spin conservation. Numerical details on this issue will be provided { in a dedicated Section}.
{Let us point out that, although the nuclear DPD associated with the DPS2 mechanism depends on the product of GPDs, the overall distribution nevertheless vanishes when $x_1+x_2 > M_A / M$
{(cf. Eq. \eqref{eq:x1_x2})}.

 In the case of spin conservation,
similarly to Eq. (\ref{Eq:DPD1_momentum}), one can relate the nuclear DPD to an off-forward LCMD:
\begin{align}
\label{Eq:gen6}
  D_{ij}^{A,2}(x_1,x_2,\mathbf k_\perp)  &= A(A-1)
  \sum_{\tau_1, \tau_2 = n,p}  \int
     \dfrac{ d\xi_1}{\xi_1} \int
     \dfrac{d\xi_2 }{ \xi_2}
     \rho_{\tau_1 \tau_2}^A (\xi_1,\xi_2, {\bf k_\perp})~
 \bar{\xi}^2
 \\
 \nonumber
 &\times
    H^{\tau_1}_i\left(x_1 \dfrac{\bar \xi}{\xi_1},0, \mathbf k_\perp \right) H^{\tau_2}_j\left(x_2 \dfrac{\bar \xi}{\xi_2},0, -\mathbf k_\perp \right)~,
\end{align}
where {the off-forward LCMD, depending on the non-diagonal product of nuclear wave functions,  is}:
\begin{align}
 \label{Eq:density2b}
   & \rho_{\tau_1 \tau_2}^A (\xi_1,\xi_2, {\bf k_\perp}) \equiv
  \\
  \nonumber
 &  \sum_{\lambda_1,\lambda_2}\int  d^2 k_{1,\perp} d^2k_{2,\perp} \psi(\xi_1,\xi_2, \mathbf k_{1,\perp},\mathbf k_{2,\perp},\tau,\tau_2,\lambda_1,\lambda_2) \psi^\dagger(\xi_1,\xi_2, \mathbf k_{1,\perp}+{\bf k_\perp},\mathbf k_{2,\perp}-{\bf k_\perp},\tau,\tau_2,\lambda_1,\lambda_2)
\\
\nonumber
    &= \sum_{\lambda_1,\lambda_2}
    \rho_{\tau_1 \tau_2}^A (\xi_1,\xi_2, {\bf k_\perp},\lambda_1,\lambda_2,\lambda_1,\lambda_2)
    ~.
\end{align}
 In the following, the nuclear distribution
 normalized to the number of permutations $A(A-1)$ is used:
\be
 \label{Eq:density2bb}
    \bar \rho_{\tau_1 \tau_2}^A (\xi_1,\xi_2, {\bf k_\perp}) = A (A-1) \rho_{\tau_1 \tau_2}^A (\xi_1,\xi_2, {\bf k_\perp}) ~,
\ee
 in analogy with $\bar \rho_\tau^A$ normalized to $A$.}
{As already mentioned,} the two-body LCMD depends on the non-diagonal product of nuclear wave functions {and this feature} introduces significant computational complexity when a numerical evaluation of the corresponding DPD with realistic nuclear wave-functions is carried out.
However, since
various parametrizations for the nucleon GPDs {can be found in literature and  calculations for  light-nuclei  LF wfs are already available \cite{Kievsky:1992um,Wiringa:1994wb,Pudliner:1995wk}}, reliable predictions for the nuclear DPD corresponding to the DPS2 mechanism can be realistically evaluated in future studies.
Following the line of Ref. \cite{Blok:2012jr}, one can notice that the overall integrand is peaked around $\xi_i \sim 1/A$ and, therefore, {for}  the leading term one has:
\be
    D^{A,2}_{ij}(x_1,x_2,{\bf k_\perp}) \simeq  \sum_{\tau_1, \tau_2=n,p} H^{\tau_1}_i(x_1,0, {\bf k_\perp})H^{\tau_2}_j(x_2,0,{\bf -k_\perp}) F_{A,\tau_1,\tau_2}^{double}({\bf k_\perp})
    \label{Eq:blo}
\ee
where
\begin{align}
    F_{A,\tau_1,\tau_2}^{double}( {\bf k_\perp}) &= \int \dfrac{ d\xi_1}{\xi_1} \int \dfrac{d\xi_2 }{ \xi_2} ~\bar \xi^2
  \bar  \rho^A_{\tau_1,\tau_2}(\xi_1,\xi_2, {\bf k_\perp})
\end{align}
{In particular, Eq. (\ref{Eq:blo}) coincides with Eq. (9)  in Ref. \cite{Blok:2012jr} as expected.}

A similar quantity has been previously calculated for $^3$He and $^4$He nuclei to predict the cross-section for $J/\psi$ electroproduction in $eA$ collisions at the future Electron-Ion Collider \cite{AbdulKhalek:2021gbh}, in order to move beyond the conventional impulse approximation \cite{Guzey:2022jtv}. In forthcoming studies, nuclear DPDs entering the DPS cross-section  will be correlated with observables from {various} experimental processes, {paving the way} for studying these novel quantities, ultimately deepening our fundamental understanding of the nuclear partonic structure.
 Moreover, in potential DPS events at {{lepton-hadron}} colliders \cite{Chekanov:2007ab,Ceccopieri:2021luf,preparation} with {virtual and quasi-virtual} photon  probes, the DPS2 contribution to the cross-section will not involve any unknown nucleon DPDs. Consequently, the DPS2 mechanism would represent an assessable {contribution to the nuclear DPS mechanism} for, {e.g.,} light nuclei, especially given the availability of both nucleon GPDs information and realistic nuclear wave-functions.

{Before closing this Section, {we recall that} in the case of the DPS2 mechanism, the $\xi_1$ and $\xi_2$ integration {extrema} in Eq. \eqref{Eq:gen6} are from $\xi_{i,min} = x_i \bar \xi$ to 1}.
\section{ Sum rules for nuclear DPDs}

It is noteworthy that the nuclear DPDs, corresponding to  DPS1 and DPS2 mechanisms, simultaneously {affect} the nuclear DPS cross-section. Given the limited constraints on nuclear wave-functions for heavy systems, in Ref. \cite{Blok:2012jr} the authors argued that, under certain approximations, the DPS1 cross-section scales as $\sigma_{DPS1} \sim A\sigma_{DPS}^{pp}$ and  $\sigma_{DPS2} \sim A^{1/3}\sigma_{DPS1}$. Here, in order to better understand the relative weights  of DPS1 and DPS2,  we {extend}  the so-called Gaunt-Stirling (GS) sum rules \cite{Gaunt:2009re} originally introduced for  {nucleonic } DPDs {to the nuclear case}.
{In particular, we introduce partial sum rules (PSR), i.e. the contributions to the GS sum rules for nuclear DPDs corresponding to the DPS1 and DPS2 mechanisms, separately}.
In this way, we are able to  provide a rigorous method  to estimate the magnitude of DPDs corresponding to the two mechanisms, without relying on additional approximations, but  only on the strong constraint  given by the  GS sume rules.
Quite interestingly, the GS  sum rules {in the nuclear case} are satisfied only when both mechanisms are considered simultaneously.  Once the two contributions are calculated we can provide an overall  estimate of the relative suppression of the DPS1   w.r.t. the DPS2 contribution, for any nuclear system without  approximations. In what follows, actual numerical calculations are presented for the deuteron, leaving the  A=3 and 4 nuclei for future studies.

Let us first introduce the \textit{number sum rule} (NSR) for a proton.
For a quark {{or antiquark}} of flavor $i$ and a valence quark $j_v$  we have {(see Ref. \cite{Gaunt:2009re})}
\begin{align}
 \int_{0}^{1-x_{1}} d x_{2}~ D_{i j_v}\left(x_{1}, x_{2}, k_{\perp}=0\right)=
 \begin{cases} N_{j_{v}} d_i(x_1) & \text { for } i \neq j \\
\left(N_{j_{v}}-1\right) d_i(x_1) &\text { for } i=j
\\
\left(N_{j_{v}}+1\right) d_i(x_1) &\text { for } i=\bar j
\end{cases}
\label{Eq.NSRnucle}
\end{align}
where $N_{j_v}$ is the number of valence quarks of flavor $j_v$ inside the proton and $d_i(x)$ is the usual PDF for a quark of flavor $i$.
 The extension to the nuclear target, {where we recall that $x_l$ is defined in Eq. (\ref{Eq:xrisc})}, reads

\begin{equation}
    \begin{aligned}
\int_{0}^{1/\bar{\xi}-x_{1}} d x_{2}~ D^A_{i j_{{v}}}\left(x_{1}, x_{2}, k_{\perp}=0\right)=
 \begin{cases} N_{j_{{v}}}^A  d_{i}^A(x_1)& \text { for } i\neq j \\
\left(N_{j_{{v}}}^A-1\right) d_{i}^A(x_1) &\text { for } i=j
\\
\left(N_{j_{{v}}}^A+1\right) d_{i}^A(x_1) &\text { for } i=\bar j
\end{cases},
\end{aligned}
\label{SR_nuclear1}
\end{equation}
where $N_{j_{{v}}}^A$ is the total number of the valence quarks $j_{{v}}$ in the nucleus and $d^A_i(x)$ the nuclear PDF for the quark of flavor $i$, viz.
\begin{align}
    d^A_i(x)= \sum_{\tau=p,n} d^{A, \tau}_i(x) =\sum_{\tau=p,n}  \int_{0}^{1} d \xi~  \bar \rho^A_{\tau} (\xi) \dfrac{\bar \xi}{\xi} d^\tau_i \left(x \dfrac{\bar \xi}{\xi}\right) ~,
    \label{Eq:nuclear_PDF}
\end{align}
{where $d^{A, \tau}_i(x)$ is the nucleon $\tau$ PDF contribution to the nuclear PDF {and $\bar \rho(\xi) $ is the nuclear LCMD normalized to the number of nucleons.}}

{ Moreover, for model calculation predictions, it can be useful to introduce the overall normalization of the nucleon DPDs. Since in the present analysis we are mainly focused on the nuclear DPDs in the valence region, we introduce relations which hold only at the low hadronic scale of a model, where there are only valence quark distributions, {  i.e., $d_{\bar q}(x)=0$ and $d_q(x)=d_{q_v}(x)$}.
Assuming that also the second parton is a valence one and integrating Eq.  \eqref{Eq.NSRnucle} over $x_1$  one obtains: }

\begin{align}
\label{Eq:norm_proton}
 \int_{0}^{1} d x_{1} \int_{0}^{1-x_{1}} d x_{2} D_{i_{v} j_{v}}\left(x_{1}, x_{2}, k_{\perp}=0\right)=
 \begin{cases}N_{i_{v}} N_{j_{v}} & \text { for } i \neq j \\
\left(N_{i_{v}}-1\right) N_{j_{v}} &\text { for } i=j\end{cases}
\end{align}
{In the nuclear case one has}
\begin{equation}
    \begin{aligned}
 \int_{0}^{1/\bar{\xi}} d x_{1} \int_{0}^{1/\bar{\xi}-x_{1}} d x_{2} D^A_{i_{{v}} j_{{v}}}\left(x_{1}, x_{2}, k_{\perp}=0\right)=
 \begin{cases}N_{i_{{v}}}^A N_{j_{{v}}}^A & \text { for } i \neq j \\
\left(N_{i_{{v}}}^A-1\right) N_{j_{{v}}}^A &\text { for } i=j\end{cases}.
\end{aligned}
\label{SR_nuclear_norm}
\end{equation}

Finally, let us introduce the \textit{momentum sum rule} (MSR). For the nucleon case \cite{Gaunt:2009re} one has
\begin{align}
  \sum_j \int_0^{1-x_1} dx_2~x_2 D_{ij}(x_1,x_2,0)=(1-x_1)d_i(x_1)
  \label{MSRn}.
\end{align}
Therefore, the generalization to the nuclear case reads as follows
\begin{align}
  \sum_j \int_0^{1/\bar \xi-x_1} dx_2~x_2 D^A_{ij}(x_1,x_2,0)=\left( \frac{1}{\bar \xi}-x_1\right)d^A_i(x_1)\sim(A-x_1)d^A_i(x_1)~.
  \label{Eq:MSR}
\end{align}
Let us remark that  the above sum rules are preserved in the pQCD evolution of DPDs \cite{Gaunt:2009re,Ceccopieri:2014ufa} { with the exceptions of the normalization rules Eqs. (\ref{Eq:norm_proton})-(\ref{SR_nuclear_norm})}. The above properties are fulfilled when both DPS1 and DPS2 mechanisms are taken into account. However, one can define
  partial sum rules  for the two DPDs in order to study the relative impact of the two contributions.
{In what follows use has been made of one- and two-body LC momentum number densities, i.e. $\bar \rho_\tau(\xi)$ and $\bar \rho_{\tau_1 \tau_2}(\xi_1,\xi_2,{\bf k_{\perp}=0}) = A (A-1) \rho_{\tau_1 \tau_2}(\xi_1,\xi_2,{\bf k_{\perp}=0})$. In fact, as discussed in details in Appendix \ref{app:1bd}-\ref{app:2bd}, these quantities satisfy the same PDF and DPD sum rules, respectively.}

\subsection{PSR for DPS1}

In this section, the PSR
for the DPS1 mechanism, Eq. (\ref{Eq:DPD1_momentum}), are introduced. Let us first start with the NSR:

\begin{equation}
\int_{0}^{1/\bar{\xi}-x_{1}} d x_{2} D_{i j_{{v}}}^{A,1}\left(x_{1}, x_{2}, k_\perp=0\right)=\sum_{\tau=n, p} \int_{0}^{1} d \xi~ \frac{\bar \xi^2}{\xi^{2}} \bar\rho^A_{\tau}(\xi) \int_{0}^{1/\bar{\xi}-x_{1}} d x_{2}  D_{i j_{{v}}}^{\tau}\left(\frac{x_{1}}{\xi}  \bar{\xi}, \frac{x_{2}}{{\xi}} {\bar{\xi}}, 0\right),
\label{APSR1}
\end{equation}
{where the one-body light-cone momentum number density, {$\bar\rho^A_{\tau}(\xi)$, is defined in Eq. (\ref{bar})} }
{(see Appendix \ref{app:1bd}). }
By changing variable, $y_2 = x_2 ~\bar{\xi}/
{\xi}\in [0,1]$,
one gets:
\begin{equation}
\int_{0}^{1/\bar{\xi}-x_{1}} d x_{2} D_{i j_{{v}}}^{A,1}\left(x_{1}, x_{2}, k_\perp=0\right)=\sum_{\tau=n,p} \int_{0}^{1} d \xi ~ \bar \rho^A_{\tau} (\xi) \dfrac{\bar \xi}{\xi} \int_{0}^{(1/\bar{\xi} - x_1)\bar{\xi}/\xi} d y_{2}  D_{i j_{{v}}}^{\tau}\left(x_1 \dfrac{\bar \xi}{\xi}, y_2, 0\right) ~,
\label{APSR2}
\end{equation}
{Indeed, in Eq. \eqref{APSR2} the upper limit of the  integral on  $y_2$  is larger than what is actually needed, since} the nucleon DPD  vanishes for
$y_2 > 1-x_1 \bar{\xi}/\xi$  
{and}
\begin{align}
\label{Eq:condition}
    \dfrac{1/\bar{\xi} -x_1}{ \xi/\bar{\xi}} = \dfrac{1}{\xi}-\dfrac{x_1 \bar{\xi}}{ \xi} \geq 1-\dfrac{x_1 \bar{\xi}}{\xi}~.
\end{align}
Then, one can use the NSR in Eq. (\ref{SR_nuclear1}) {that reads}
\begin{equation}
\int_{0}^{1-x_{1} \bar \xi/\xi} d y_{2} D_{i j_{{v}}}^{\tau}\left(x_1 \dfrac{\bar \xi}{\xi}, y_{2}, 0\right)  = \begin{cases}\left(N^\tau_{j_{{v}}}-1\right) d_{i}^{\tau}(x_1 \bar \xi/\xi) & i=j \\
N_{j_{{v}}}^{\tau} d_{i}^{\tau}(x_1 \bar \xi/\xi) & i \neq j
\\
\left(N^\tau_{j_{{v}}}+1\right) d_{i}^{\tau}(x_1 \bar \xi/\xi) & i=\bar j
\end{cases}.
\label{SR_nucleon}
\end{equation}
where $N^\tau_{j_{{v}}}$ is the number of the valence quarks $j$ in the nucleon $\tau$.
By inserting the above relation in  Eq. \eqref{APSR2}, we get the final expression for the PSR {(see Eq. (\ref{Eq:nuclear_PDF}) for the definition of the contribution of the nucleon $\tau$ to the nuclear PDF)}
\begin{align}
    \int_{0}^{1/\bar{\xi}-x_{1}} d x_{2} D_{i j_{{v}}}^{A,1}\left(x_{1}, x_{2}, 0\right) =   \sum_{\tau=n,p} \begin{cases}\left(N^\tau_{j_{{v}}}-1\right) d_{i}^{A,\tau}(x_1) & i=j \\
N_{j_{{v}}}^{\tau} d_{i}^{A,\tau}(x_1) & i \neq j
\\
\left(N^\tau_{j_{{v}}}+1\right) d_{i}^{A,\tau}(x_1) & i=\bar j
\end{cases},
\label{NSR_DPS1}
\end{align}
For the MSR we get:

\begin{align}
\sum_j  \int_0^{1/\bar{\xi}-x_1} dx_2~x_2 D^{A,1}_{ij}(x_1,x_2,0) &=  \sum_j  \sum_{\tau=n,p} \int d\xi~ \bar \rho_\tau^A(\xi) \dfrac{\bar \xi^2}{\xi^2} \int_0^{{\xi/\bar \xi - x_1}} dx_2~x_2 D^\tau_{ij}\left(x_1 \dfrac{\bar \xi}{\xi},x_2 \dfrac{\bar \xi}{\xi},0 \right)~,
\end{align}
and by defining $y_2 = x_2 \bar \xi/\xi$ we have:

\begin{align}
  \sum_j    \int_0^{1/\bar{\xi}-x_1} dx_2~x_2 D^{A,1}_{ij}(x_1,x_2,0) &= \sum_j  \sum_{\tau=n,p} \int d\xi~ \bar \rho_\tau^A(\xi)  \int_0^{1-x_1 \bar \xi/\xi } dy_2~y_2 D^\tau_{ij}\left(x_1 \dfrac{\bar \xi}{\xi},y_2 ,0 \right)~.
\end{align}
{The nucleon MSR, Eq. (\ref{MSRn}), can be used to evaluate the integral over $y_2$ to obtain}:

\begin{align}
  \sum_j   \int_0^{1/\bar{\xi}-x_1} dx_2~x_2 D^{A,1}_{ij}(x_1,x_2,0) &= \sum_{\tau=n,p} \int d\xi~ \bar \rho_\tau^A(\xi) \left(1-x_1\dfrac{\bar \xi}{\xi} \right)d^{\tau}_i\left(x_1 \dfrac{\bar \xi}{\xi}\right).
  \label{MSR_DPS1}
\end{align}

\subsection{PSR for DPS2}

In order to evaluate  both NSR and MSR for the nuclear DPD corresponding to the DPS2 mechanism,  {DPDs are evaluated by taking} the expression in Eq. (\ref{Eq:gen6}) in the limit ${\bf k_\perp}=0$, {viz.}
\begin{equation}
    D_{ij}^{A,2}(x_1,x_2,{\bf k_\perp=0}) =  \sum_{\tau_1, \tau_2 = p,n} \int d\xi_1~{\bar \xi\over \xi_1} \int d \xi_2~{\bar \xi \over \xi_2} ~\bar\rho_{\tau_1 \tau_2}^{A}(\xi_1, \xi_2) ~d_i^{\tau_1}\left(\frac{\bar{\xi} x_1}{\xi_1}\right)d_j^{\tau_2}\left(\frac{\bar{\xi} x_2}{\xi_2}\right) ~,
\end{equation}
where  $\xi_1 + \xi_2 \le 1$  {and the double LC momentum number density   is defined by
\begin{align}
\label{Eq:density_bar_2}
    \bar \rho_{\tau_1 \tau_2}^A (\xi_1,\xi_2) \equiv \bar \rho_{\tau_1 \tau_2}^A (\xi_1,\xi_2, {0})=A (A-1) \int  d^2 k_{1,\perp} d^2k_{2,\perp} |\psi(\xi_1,\xi_2, \mathbf k_{1,\perp},\mathbf k_{2,\perp},\tau_1,\tau_2)|^2 ~.
\end{align}}
{The main properties of these quantities are discussed in details in appendix \ref{app:2bd}.}
The NSR {reads}
\begin{align}
\nonumber
    \int_{0}^{1/\bar{\xi}-x_{1}} d x_{2}~ D_{ij_v}^{A,2}\left(x_{1}, x_{2}, k_\perp=0\right) &=
       \sum_{\tau_1, \tau_2 = p,n} \int d\xi_1~{\bar \xi\over \xi_1} \int d \xi_2~{\bar \xi \over \xi_2} ~
      \bar \rho_{\tau_1\tau_2}^A (\xi_1,\xi_2)~ d_i^{\tau_1}\left(\frac{\bar{\xi} x_1}{\xi_1}\right)
      \\
      &\times
\int_{0}^{(\xi_2/\bar{\xi})} d x_{2}~d_{j_v}^{\tau_2}\left(\frac{\bar{\xi} x_2}{\xi_2}\right)
~,\label{eq:DA2}
\end{align}
since $d_{j_v}^{\tau_2}\left({\bar{\xi} x_2}/{\xi_2}\right)$ vanishes for $x_2 \geq \xi_2/\bar{\xi}$ {and $1/\bar{\xi}-x_{1} \geq \xi_2/\bar{\xi}$}.
Also in this case, let us  change the integration variable from $x_2$ to $y_2 = x_2 \bar \xi/\xi_2$, and
{therefore Eq. \eqref{eq:DA2} becomes}:
\begin{align}
     \label{Eq:bo2}
    \int_{0}^{1/\bar{\xi}-x_{1}} d x_{2}~ D_{ij_v}^{A,2}\left(x_{1}, x_{2}, k_\perp=0\right) &=
       \sum_{\tau_1, \tau_2 = p,n} \int d\xi_1 ~\dfrac{\bar \xi}{\xi_1 }\int d\xi_2 ~
      \bar \rho_{\tau_1,\tau_2}^A (\xi_1,\xi_2) ~d_i^{\tau_1}\left(\frac{\bar{\xi} x_1}{\xi_1}\right) \int_0^{1}dy_2~  d_{j_v}^{\tau_2}(y_2)
      \\
      \nonumber
      & =\sum_{\tau_1, \tau_2 = p,n} \int d\xi_1 ~\dfrac{\bar \xi}{\xi_1 }\int d\xi_2 ~
      \bar \rho_{\tau_1,\tau_2}^A (\xi_1,\xi_2) ~d_i^{\tau_1}\left(\frac{\bar{\xi} x_1}{\xi_1}\right)  N_{j_v}^{\tau_2}~,
\end{align}
{where $N_{j_v}^{\tau_2} $ is the normalization of the valence quark PDF. In conclusion, the final result is}

\begin{align}
    \int_{0}^{1/\bar{\xi}-x_{1}} d x_{2}~ D_{ij_v}^{A,2}\left(x_{1}, x_{2}, k_\perp=0\right) &= \sum_{\tau_1, \tau_2 = p,n}  d_i^{A,\tau_1,\tau_2}\left( x_1\right)N_{j_v}^{\tau_2}~,
    \label{Eq:NSR_DPS2}
    \end{align}
where the following distribution has been defined
\begin{align}
\label{Eq:PDF_condition}
    d_i^{A,\tau_1,\tau_2}(x) \equiv \int d \xi_1 ~ \dfrac{\bar \xi}{\xi_1}\int d\xi_2 ~\bar \rho_{\tau_1,\tau_2}^A (\xi_1,\xi_2) ~d_i^{\tau_1}\left(\frac{\bar{\xi} }{\xi_1} x\right)~.
\end{align}

 By combining Eq. (\ref{NSR_DPS1}) with Eq. (\ref{Eq:NSR_DPS2}), one gets the total NSR for the nucleus 
    \begin{align}
    \label{Eq:SR_DPS1_DPS2}
     &   \int_{0}^{1/\bar{\xi}-x_{1}} d x_{2}~ D_{ij_v}^{A}\left(x_{1}, x_{2}, k_\perp=0\right) ~ = \int_{0}^{1/\bar{\xi}-x_{1}} d x_{2}~ \big[ D_{ij_v}^{A,1}\left(x_{1}, x_{2}, k_\perp=0\right) +D_{ij_v}^{A,2}\left(x_{1}, x_{2}, k_\perp=0\right) \big]
        \\
        \nonumber
        &= \sum_{\tau=n,p} \begin{cases}\left(N^\tau_{j_v}-1\right) d_{i}^{A,\tau}(x_1) & i=j \\
N_{j_v}^{\tau} d_{i}^{A,\tau}(x_1 ) & i \neq j
\\
\left(N^\tau_{j_{{v}}}+1\right) d_{i}^{A,\tau}(x_1) & i=\bar j
\end{cases} ~ + \sum_{\tau_1, \tau_2 = p,n}  d_i^{A,\tau_1,\tau_2}\left( x_1\right)N_{j_v}^{\tau_2}
   \end{align}
{Explicitly, one writes
 \begin{align}
 \label{Eq:SR_DPS1_DPS2_b}
&   \int_{0}^{1/\bar{\xi}-x_{1}} d x_{2}~ D_{ij_v}^{A}\left(x_{1}, x_{2}, k_\perp=0\right) \\
\nonumber
&= \begin{cases} (N_{j_v}^p-1)d_i^{A,p}(x_1)~ + ~(N_{j_v}^n-1)d_i^{A,n}(x_1) ~ + ~ N_{j_v}^p d_i^{A,p,p}(x_1)
\\
\nonumber
+~N_{j_v}^p d_i^{A,n,p}(x_1)~ +~ N_{j_v}^n d_i^{A,p,n}(x_1)~ + ~ N_{j_v}^n d_i^{A,n,n}(x_1) & i=j
\\
~
\\
(N_{j_v}^p+1)d_i^{A,p}(x_1) ~ + ~(N_{j_v}^n+1)d_i^{A,n}(x_1) ~ + ~ N_{j_v}^p d_i^{A,p,p}(x_1)
\\
\nonumber
+~N_{j_v}^p d_i^{A,n,p}(x_1)~ + ~ N_{j_v}^n d_i^{A,p,n}(x_1) ~ + ~ N_{j_v}^n d_i^{A,n,n}(x_1) & i=\bar j
\\
~
\\
N_{j_v}^{p} d_{i}^{A,p}(x_1 ) ~ + ~ N_{j_v}^{n} d_{i}^{A,n}(x_1) ~ + ~ N_{j_v}^p d_i^{A,p,p}(x_1)
\\
\nonumber
+~ N_{j_v}^p d_i^{A,n,p}(x_1) ~ + ~ N_{j_v}^n d_i^{A,p,n}(x_1) ~ + ~ N_{j_v}^n d_i^{A,n,n}(x_1) & i \neq j\end{cases} ~,
    \end{align}
 where
 \begin{align}
 \label{Eq:dij_def}
    d^{A,p,p}_i(x) &= (Z-1)\int d\xi_1 \dfrac{\bar \xi}{\xi_1} \bar \rho^A_{p}(\xi_1)d^p_i\left( \dfrac{\bar \xi}{\xi_1} x\right) = (Z-1) d^{A,p}_i(x)
    \\
    \label{Eq:dij_def2}
        d^{A,n,n}_i(x) &= (A-Z-1)\int d\xi_1 \dfrac{\bar \xi}{\xi_1} \bar \rho^A_{n}(\xi_1)d^n_i\left( \dfrac{\bar \xi}{\xi_1} x\right) = (A-Z-1)d^{A,n}_i(x)
 \\
 \label{Eq:dij_def3}
         d^{A,n,p}_i(x) &= Z\int d\xi_1 \dfrac{\bar \xi}{\xi_1} \bar \rho^A_{n}(\xi_1)d^n_i\left( \dfrac{\bar \xi}{\xi_1} x\right) =Z d^{A,n}_i(x)
\\
\label{Eq:dij_def4}
         d^{A,p,n}_i(x) &= (A-Z)\int d\xi_1 \dfrac{\bar \xi}{\xi_1} \bar \rho^A_{p}(\xi_1)d^p_i\left( \dfrac{\bar \xi}{\xi_1} x\right) =(A-Z)d^{A,p}_i(x)~.
\end{align}}
{Notice that in Eq. \eqref{Eq:PDF_condition}, where the nuclear PDF is defined, the nuclear two-body density $\bar \rho^A_{\tau_1,\tau_2}(\xi_1,\xi_2)$ (see Eq. (\ref{Eq:density_bar_2})) enters, while in Eqs. \eqref{Eq:dij_def}, \eqref{Eq:dij_def2},  \eqref{Eq:dij_def3}, \eqref{Eq:dij_def4} the one-body density  $\bar \rho^A_\tau(\xi)$,  Eq. (\ref{bar}),
is present.} In order to relate the {two densities}, one needs to properly define normalization properties. To this end, {use has been made of the integral properties discussed in  Appendix \ref{app:2bd}.}

By combining the terms proportional to the same PDF in Eq. \eqref{Eq:SR_DPS1_DPS2_b}, one gets
\begin{align}
    &\int_{0}^{1/\bar{\xi}-x_{1}} d x_{2}~ D_{ij_v}^{A}\left(x_{1}, x_{2}, k_\perp=0\right)
    \\
    \nonumber
&    =   \begin{cases}\big[Z N^p_{j_v} +(A-Z)N_{j_v}^n-1\big] d_{i}^{A,p}(x_1 )+ \big[(A-Z)N^n_{j_v}+ZN_{j_v}^p-1\big] d_{i}^{A,n}(x_1 ) & i=j
\\
~
\\
\big[Z N^p_{j_v} +(A-Z)N_{j_v}^n+1\big] d_{i}^{A,p}(x_1 )+ \big[(A-Z)N^n_{j_v}+ZN_{j_v}^p+1\big] d_{i}^{A,n}(x_1 ) & i=\bar j
 \\
    ~
    \\
\big [Z N_{j_v}^{p}+(A-Z)N_{j_v}^n \big]d_{i}^{A,p}(x_1 ) +\big [(A-Z)N_{j_v}^{n} +ZN_{j_v}^p\big ]d_{i}^{A,n}(x_1 ) & i \neq j\end{cases}~.
\end{align}
{Since} $Z N_{j_v}^p+(A-Z)N_{j_v}^n=N_{j_v}^A$, i.e. the number of valence quarks with flavor $j$ in the nucleus, and $d_i^{A,p}(x)+d_i^{A,n}(x)=d_i^A(x)$, i.e. the nuclear PDF, {one eventually has}
\begin{align}
        \int_{0}^{A-x_{1}} d x_{2}~ D_{ij_v}^{A}\left(x_{1}, x_{2}, k_\perp=0\right) &= \begin{cases}\left(N^A_{j_v}-1\right) d_{i}^{A}(x_1)& i=j \\
        ~
        \\
        \left(N^A_{j_v}+1\right) d_{i}^{A}(x_1)& i=\bar j
        \\
        ~
        \\
N_{j_v}^{A} d_{i}^{A}(x_1 )  & i \neq j\end{cases}~,
        \end{align}
as expected, {i.e. the  GS number sum-rule generalized to the nuclear target shown in Eq. \eqref{SR_nuclear1}}.

Let us proceed with the MSR {contribution from DPS2}. In this case, one {has} to evaluate
\begin{align}
\sum_j    \int_0^{1/\bar \xi -x_1} \hspace{-0.2cm}dx_2~x_2 D^{A,2}_{ij}(x_1,x_2,0) &=  \sum_j  \sum_{\tau_1, \tau_2=n,p} \int d\xi_1  \int d\xi_2~  \dfrac{\bar \xi^2}{\xi_1 \xi_2}  \bar \rho_{\tau_1 \tau_2}^A(\xi_1,\xi_2)
 d^{\tau_1}_{i} \hspace{-0.1cm} \left(x_1 \dfrac{\bar \xi}{\xi_1} \right)
 \\
 \nonumber
 &\times
 \int_0^{1/\bar \xi -x_1} \hspace{-0.2cm} dx_2~x_2 ~d^{\tau_2}_{j}\left(x_2 \dfrac{\bar \xi}{\xi_2} \right)~.
\end{align}
We introduce again $y_2 = x_2 \bar \xi/\xi_2$

\begin{align}
\sum_j   \int_0^{1/\bar \xi -x_1} dx_2~x_2 D^{A,2}_{ij}(x_1,x_2,0) &=    \sum_{\tau_1, \tau_2=n,p} \int d\xi_1  \int d\xi_2~  \dfrac{ \xi_2}{\xi_1}  \bar \rho_{\tau_1 \tau_2}^A(\xi_1,\xi_2)
 d^{\tau_1}_{i}\left(x_1 \dfrac{\bar \xi}{\xi_1} \right)
 \\
 \nonumber
 &\times
 \sum_j\int_0^{1} dy_2~y_2 ~d^{\tau_2}_{j}\left(y_2 \right)
 \\
 \nonumber
 &=    \sum_{\tau_1, \tau_2=n,p} \int d\xi_1  \int d\xi_2~  \dfrac{ \xi_2}{\xi_1}  \bar \rho_{\tau_1 \tau_2}^A(\xi_1,\xi_2)
 d^{\tau_1}_{i}\left(x_1 \dfrac{\bar \xi}{\xi_1} \right)~.
\label{MSR_DPS2}
\end{align}
{By using Eq. (\ref{Eq:MSRden2})} one can simplify the above expression, obtaining
\begin{align}
    \sum_j   \int_0^{1/\bar \xi -x_1} dx_2~x_2 D^{A,2}_{ij}(x_1,x_2,0) &= \sum_{\tau_1=n,p} \int \dfrac{d\xi_1}{\xi_1} ~\bar \rho^A_{\tau_1}(\xi_1)~d^{\tau_1}_{i}\left(x_1 \dfrac{\bar \xi}{\xi_1} \right)~(1-\xi_1) ~.
\end{align}

Let us conclude this Section by considering the full MSR, viz.
\begin{align}
\label{MSR_DPS2_b}
 \sum_j    \int_0^{A-x_1} dx_2~x_2 D^{A}_{ij}(x_1,x_2,0)&=
    \sum_j    \int_0^{A-x_1} dx_2~x_2 \big[D^{A,1}_{ij}(x_1,x_2,0)+ D^{A,2}_{ij}(x_1,x_2,0)\big]
    \\
    \nonumber
    &= \sum_{\tau=n,p} \int d\xi~ \bar \rho^A_\tau(\xi)d_i^\tau \left(x_1 \frac{\bar \xi}{\xi}\right)\left(1-x_1 \frac{\bar \xi}{\xi}+\frac{1}{\xi} -1\right)
     \\
    \nonumber
    &= \sum_{\tau=n,p} \int d\xi~ \dfrac{\bar \xi}{\xi}\bar \rho^A_\tau(\xi)d_i^\tau \left(x_1 \dfrac{\bar\xi}{\xi}\right) \left(\frac{1}{\bar \xi}-x_1\right) = d^A_i(x_1)\left(\dfrac{M_A}{m}-x_1 \right)
    \\
    \nonumber
    &\sim d^A_i(x_1)~(A-x_1) ~.
\end{align}
The result is in agreement with the expected {GS momentum sum-rule extended to nuclear target}  shown in Eq. (\ref{Eq:MSR}).

\section{Deuteron densities}
{The investigation of the nuclear DPDs necessarily requires a reliable and affordable relativistic framework for describing nucleons in nuclei at short distances, and consequently with large kinetic energy.}
In order to provide the first realistic,
{Poincar\'e covariant}
 calculation of nuclear DPDs,
 the deuteron target has been considered. For this two-nucleon system, a LF framework is adopted, since it  yields the possibility to  rigorously establish a suitable  Poincar\'e-covariant approach
 {(with fixed number of particles)}, where a realistic  wave function, derived from the nucleon-nucleon Av18 phenomenological potential \cite{Wiringa:1994wb}, can be exploited.  As  is well known (see Ref. \cite{Lev:2000vm} for the deuteron and \cite{DelDotto:2016vkh} for the A=3 mirror nuclei), one can take advantage of the  relativistic Hamiltonian dynamics framework (see Ref. \cite{Dirac:1949cp} for the seminal paper on these topics and Ref. \cite{KP} for a practical introduction) and the Bakamjian-Thomas \cite{Bakamjian:1953kh} construction of the Poincar\'e generators in order to embed  the highly successful phenomenology, developed within the standard nuclear physics,  in a rigorous Poincar\'e-covariant approach. A key point is represented by the construction of suitable LF spin states  starting from the canonical  (or instant-form) ones, where   the Clebsch-Gordan coefficients are used for obtaining the expected angular-momentum content.
 {In order to relate the LF spin, i.e. the  three independent-components of  the Pauli-Lubanski pseudo-vector
in the particle rest-frame that is  reached through a LF boost,  and the canonical spin (obtained by
applying a canonical boost), one has to introduce the {\em unitary} operator called   Melosh rotation, $ {\cal R}_M ({\blf p})$,
 (cf Eq. (3.105) in \cite{KP})  viz
\be
s_c^i= \left [{\cal R}_M ({\blf p}) \right ]_{ij}~s_{LF}^j~,
\ee
where ${\blf p}\equiv \{E(|{\bf p}|)+p_z, {\bf p}_\perp\}$ is
{called}  LF three-momentum (N.B. ${\bf p}$ is the three-momentum in instant form) in a generic frame.
{In what follows ${\blf p}$ will be the LF-momentum of the nucleon in the intrinsic frame of the deuteron}. The explicit expression of the Melosh rotations  is given by
\be
{\cal R}_M ({\blf p})={M+E(|{\bf p}|)+p_z -\imath
{\bfm \sigma} \cdot (\hat e_z \times
{\bf p}_{\perp}) \over \sqrt{\left ( M +E(|{\bf p}|)+p_z \right )^2 +|{\bf p}_{\perp}|^2}}~.
\ee
Then, the plane-wave in canonical and LF forms are related by
\be
|{\bf p} ;s\sigma \rangle_c =\sum_{\sigma'} ~D^{s}_{\sigma' \sigma}
\left[{\cal R}^\dagger_M ({\blf p})\right]~|{\blf p};s \sigma'
\rangle_{LF}
\nonu
|{\bf p} ;s\sigma' \rangle_{LF} =\sum_{\sigma} ~D^{s}_{\sigma \sigma'}
\left[{\cal R}_M ({\blf p})\right]~|{\blf p};s \sigma
\rangle_{c}
\ee
where
$
D^{{1 \over 2}} [{\cal R}_M ({\blf p})]_{\sigma\sigma'}=
 \chi^\dagger_{\sigma}~
{\cal R}_M ({\blf p})~\chi_{\sigma'}$, and
 the following orthonormalization rule is fulfilled

{\be
_{LF}\langle\sigma' s, {\blf p}'|{\blf p};s \sigma \rangle_{LF}
=
2 p^+ (2\pi)^3~\delta^3({\blf p}'-{\blf p})~\sum_{\mu' \mu} ~
D^{s*}_{\mu' \sigma'}
\left[{\cal R}_M ({\blf p})\right]D^{s}_{\mu \sigma}
\left[{\cal R}_M ({\blf p})\right] ~_c\langle\mu' s|s \mu\rangle_c
\nonu =
2 p^+ (2\pi)^3~\delta^3({\blf p}'-{\blf p})~\delta_{\sigma' \sigma}~.
\ee
}}

{
In instant form, the deuteron wave function is given by (N.B., for the sake of presentation  {  we have considered proton and neutron distinct in the deuteron wave function, dropping out the isospin {indeces}):
\be
\label{Eq:wfNR}
\psi^{IF}_2({\bf k}_{in, S_z})=\sum_{\mu h_1 h_2} C^{1\mu}_{1/2 h_1 1/2 h_2} \chi^{1/2}_{h_1}(p)\chi^{1/2}_{h_2}(n) \sum_{L=0,2} \sum_{m_L}~w_L(k_{in}) C^{1S_z}_{Lm_L 1\mu} Y_L^{m_L}(\theta,\phi) ~,
\ee
where  ${\bf k}_{in}$  is the intrinsic momentum, i.e. {$ {\bf k}_{in} =  ({\bf k}_1-{\bf k}_2)/2 =  {\bf k}_1$} in the frame where ${\bf k}_1+{\bf k}_2=  {\bf 0}$.
{Recall that  $w_L(k_{in})$, with $L=0,~2$ are the  radial wfs  which encode the dynamical information on the deuteron structure.}
Within the LF  approach,
 the following amplitude, shortly $_{LF}\langle \lambda_1,\lambda_2|\psi_2\rangle$, has to be used in DPS2
\begin{align}
\label{eq:deut_wf}
\psi_{\lambda_1 \lambda_2}^{LF} (\xi_1, {\bf k}_{in\perp},S_z) &= \sum_{\mu h_1, h_2}  C^{1 \mu}_{\frac{1}{2} h_1, \frac{1}{2} h_2}~
D^{{1 \over 2}}_{\lambda_1 h_1} [{\cal R}^\dagger_M ({\blf k}_1)]  D^{{1 \over 2}}_{\lambda_2 h_2 } [{\cal R}^\dagger_M ({\blf k}_2)]
\\
\nonumber
&=
 \sum_{L=0,2} \sum_{m_L}~w_L(k_{in}) C^{1S_z}_{Lm_L 1\mu} Y_L^{m_L}(\theta,\phi)
\end{align}
It should be pointed out that in the Melosh rotations the third component of ${\bf k}_1$ is given by
\be
k_{1z}= M_0(|{\bf k}_{in \perp}|,\xi_1) \left(\xi_1-{1\over 2}\right)=-k_{2z}~,\ee
where the  square free-mass of the two-nucleon system reads

\begin{align}
    M^2_0({\bf k}_{in \perp},\xi_1) ={M^2 + k^2_{in \perp}\over \xi_1 (1-\xi_1)}~.
\end{align}

{Since the nuclear LCMD {{  to be used for the convolution of}} the product of the two GPDs $H$ in the DPS2 mechanism, (cf. Eqs. (\ref{Eq:density2}) and (\ref{Eq:gen6})), { once} evaluated at ${\bf k}_\perp = {\bf 0}$ {  and integrated on $\xi_2$}, coincides to that { to be used} in the DPS1 mechanism (cf. Eqs. \eqref{Eq:density1} and \eqref{Eq:DPD1_momentum}), in this section, we provide  the calculations of all the off-forward LCMDs entering the DPS2 contribution. Moreover the LCMD entering {{Eq.}} \eqref{Eq:DPD1_momentum} is the same of that presented in Ref. \cite{Fornetti:2023gvf} and used to evaluate, e.g., the EMC effect for $^3$He \cite{Pace:2022qoj} and $^4$He \cite{Fornetti:2023gvf}. }

{In the  evaluation for the deuteron of DPS2, Eq. \eqref{Eq:DPD2full}, {one has to deal with a factorized form
{within the adopted  approach (see also Ref. \cite{Strikman:2001gz})} , viz.
\begin{align}
D^{A,2}_{ij}(x_1,x_2,{\bf k}_\perp)&= \int d\xi_1 ~ \dfrac{\bar \xi^2}{\xi_1 (1-\xi_1)}
    \int d{\bf k}_{in}~\delta\Bigl[k_{in z} -\bar k_z(|{\bf k} _{in\perp}|,\xi_1)\Bigr]~
    \\
    \nonumber
    &\times
    {\cal W}^{\mu\mu'}(\xi_1,{\bf k}_\perp,{\bf k}_{in})~ {\cal S}_{ij}^{\mu\mu'}(x_1,x_2,{\bf k}_\perp,{\blf k}_{in})
 \label{eq:dps2_fac}
 \end{align}
where the nuclear part of the DPS2 (for an unpolarized deuteron) is given by
\begin{align}
{\cal W}^{\mu\mu'}(\xi_1,{\bf k}_\perp,{\bf k}_{in})&= {1 \over 3} \sum_{S_z}\sum_{L,L'=0,2}\sum_{m_L,m_{L'}}C^{1S_z}_{Lm_L 1\mu}C^{1S_z}_{L'm_{L'} 1\mu'}
~
w_L(k_{in})~w_{L'}(k'_{in})~
\\
\nonumber
&\times
Y_L^{m_L}(\widehat{\bf k}_{in})~  \Bigl[Y_{L'}^{m_{L'}}(\widehat {\bf k}'_{in})\Bigr]^* ~,
\end{align}
 with i) ${\bf k}'_{in\perp}={\bf k}_{in\perp} +{\bf k}_{\perp}$,    ii)  $\bar k_z(| {\bf k}_{in \perp}|,\xi_1)= M_0(|{\bf k} _{in\perp}|,\xi_1)~(\xi_1-1/2)$  and iii) $k'_{in z}=M_0(|{\bf k}' _{in\perp}|,\xi_1)~(\xi_1-1/2)$.
 In Eq. \eqref{eq:dps2_fac},}
the following trace containing the partonic contribution is relevant (see Appendix \ref{app_traces} for details)
\begin{align}
\label{eq:trace}
&{\cal S}_{ij}^{\mu\mu'}(x_1,x_2,{\bf k}_\perp,{\blf k}_1)
\\
\nonumber
&=\sum_{\lambda_1\lambda_2}\sum_{\lambda'_1\lambda'_2}\frac{1}{\sqrt{2}} \langle \lambda_2| {\cal R}^\dagger_M({\blf k}_2)~
 \sigma_{\mu} ~{\cal R}_M({\blf k}_1)~(i\sigma_y )|\lambda_1\rangle~\langle \lambda_1|\hat \Phi ^i\Bigl(x_1{\bar \xi\over \xi_1},0,{\bf k}_\perp\Bigr)|\lambda'_1\rangle
 \\
 \nonumber
 &\times
 \frac{(-1)^{1+\mu'}}{\sqrt{2}}~\langle \lambda'_1| (i\sigma_y ){\cal R}^\dagger_M({\blf k}'_1)~
 \sigma_{-\mu'} ~{\cal R}_M({\blf k}'_2)|\lambda'_2\rangle~\langle \lambda'_2|\Bigl[\hat \Phi^{j }\Bigl(x_2{\bar \xi\over1- \xi_1},0,-{\bf k}_\perp\Bigr)\Bigr]^T|\lambda_2\rangle=~{(-1)^{1+\mu'}\over 2}
\\
\nonumber
&  \times Tr\Biggl\{ {\cal R}^\dagger_M({\blf k}_2)\sigma_{\mu}{\cal R}_M({\blf k}_1)~ (i\sigma_y )~\hat \Phi^i\Bigl(x_1{\bar \xi\over \xi_1},0,{\bf k}_\perp\Bigr)~(i\sigma_y )
 ~
 {\cal R}^\dagger_M({\blf k}'_1) \sigma_{-\mu'}{\cal R}_M({\blf k}'_2)~\Bigl[\hat \Phi^j\Bigl(x_2{\bar \xi\over1- \xi_1},0,-{\bf k}_\perp\Bigr)\Bigr]^T\Biggr\}
\end{align}
where the relation  $C^{1 \mu}_{\frac{1}{2} h_1, \frac{1}{2} h_2}= \langle 1/2~h_1|\sigma_\mu~i\sigma_y|h_2 1/2\rangle/\sqrt{2} $ has been used, with i) $\sigma_\mu=\hat{\bf e}_\mu \cdot {\bfm \sigma}$,  ii)  $\hat{\bf e}_0=\hat{\bf e}_z$ and iii) $\hat{\bf e}_\pm =\mp (\hat{\bf e}_x \pm  i\hat{\bf e}_y)/\sqrt{2}$.
In Eq. \eqref{eq:trace}, $\Phi^{i}\left(x,0, \mathbf k_\perp \right)$ is the LC correlator introduced in Eq. \eqref{Eq:LCGPD}.
In terms of GPDs, if one assumes  ${\bf k_\perp}=(0,k_\perp)$  {without loss of generality} ({{recall the invariance in the transverse plane, since the LF rotation around the z-axis is
kinematical}}),
the correlator can be {written} as follows:
\begin{align}
\Phi^{i}\left(x,0, \mathbf k_\perp \right) =  H_i(x,0,-{\bf k}^2_{\perp})-i \dfrac{k_\perp}{4 M}  \sigma_x   E_i(x,0,-{\bf k}^2_\perp)~.
\end{align}
Substituting in  Eq. \eqref{eq:trace},
one gets
\be
{\cal S}_{ij}^{\mu\mu'}(x_1,x_2,{\bf k}_\perp,{\blf k}_1) =~(-1)^{\mu'}{1\over 2}
\nonu
\times ~ Tr\Biggl\{ \sigma_{\mu}~{\cal R}_M({\blf k}_1)~\Bigl[\hat \Phi^i\Bigl(x_1{\bar \xi\over \xi_1},0,{\bf k}_\perp\Bigr)\Bigr]^* ~
 {\cal R}^\dagger_M({\blf k}'_1)~ \sigma_{-\mu'}~{\cal R}_M({\blf k}'_2)~\hat \Phi^j\Bigl(x_2{\bar \xi\over 1 -\xi_1},0,-{\bf k}_\perp\Bigr)~{\cal R}^\dagger_M({\blf k}_2)\Biggr\}
 \nonu
 =(-1)^{\mu'}{1\over 2}
 \nonu
\times ~ Tr\Biggl\{ \sigma_{\mu}~{\cal R}_M({\blf k}_1)~\Bigl[ H_i\Bigl(x_1{\bar \xi\over \xi_1},0, -{\bf k}^2_\perp\Bigr)+ {i \over 4M} \sigma_x  k_y~E_i\Bigl(x_1{\bar \xi\over \xi_1},0,-{\bf k}^2_\perp\Bigr)  \Bigr] ~
 {\cal R}^\dagger_M({\blf k}'_1)~ \sigma_{-\mu'}~{\cal R}_M({\blf k}'_2)
 \nonu \times ~\Bigl[H_j\Bigl(x_2{\bar \xi\over1- \xi_1},0,- {\bf k}^2_\perp\Bigr) +{i \over 4M} \sigma_x  k_y~E_j\Bigl(x_2{\bar \xi\over1- \xi_1},0,-{\bf k}^2_\perp\Bigr)\Bigr]~{\cal R}^\dagger_M({\blf k}_2)\Biggr\}
 \nonu
= H_i\Bigl(x_1{\bar \xi\over \xi_1},0, -{\bf k}^2_\perp\Bigr)H_j\Bigl(x_2{\bar \xi\over1- \xi_1},0,- {\bf k}^2_\perp\Bigr) A_{\mu \mu'}({\bf \tilde k}_1,{\bf \tilde k'}_1,{\bf \tilde k}_2,{\bf \tilde k'}_2 )
\nonu
+ ~ i ~\frac{k_y}{4 M} H_i\Bigl(x_1{\bar \xi\over \xi_1},0, -{\bf k}^2_\perp\Bigr)E_j\Bigl(x_2{\bar \xi\over1- \xi_1},0,- {\bf k}^2_\perp\Bigr) \tilde A^1_{\mu \mu'}({\bf \tilde k}_1,{\bf \tilde k'}_1,{\bf \tilde k}_2,{\bf \tilde k'}_2 )
\nonu
+~ i ~ \frac{k_y}{4 M} E_i\Bigl(x_1{\bar \xi\over \xi_1},0, -{\bf k}^2_\perp\Bigr)H_j\Bigl(x_2{\bar \xi\over1- \xi_1},0,- {\bf k}^2_\perp\Bigr)  \tilde A^2_{\mu \mu'}({\bf \tilde k}_1,{\bf \tilde k'}_1,{\bf \tilde k}_2,{\bf \tilde k'}_2)
\nonu
- ~ \frac{k^2_y}{16 M^2} E_i\Bigl(x_1{\bar \xi\over \xi_1},0, -{\bf k}^2_\perp\Bigr)E_j\Bigl(x_2{\bar \xi\over1- \xi_1},0,- {\bf k}^2_\perp\Bigr)
\tilde B_{\mu \mu'}({\bf \tilde k}_1,{\bf \tilde k'}_1,{\bf \tilde k}_2,{\bf \tilde k'}_2 )
 \label{eq:traceb}\ee
 where
 \be
 A_{\mu \mu'}({\bf \tilde k}_1,{\bf \tilde k'}_1,{\bf \tilde k}_2,{\bf \tilde k'}_2 )=(-1)^{\mu'}   ~{1\over 2} Tr\Biggl\{ \sigma_{\mu}~{\cal R}_M({\blf k}_1) ~
 {\cal R}^\dagger_M({\blf k}'_1)~ \sigma_{-\mu'}~{\cal R}_M({\blf k}'_2)~{\cal R}^\dagger_M({\blf k}_2)\Biggr\}
 \nonu
 \tilde A^1_{\mu \mu'}({\bf \tilde k}_1,{\bf \tilde k'}_1,{\bf \tilde k}_2,{\bf \tilde k'}_2 )=(-1)^{\mu'} ~{1\over 2} Tr\Biggl\{ \sigma_{\mu}~{\cal R}_M({\blf k}_1)~
 {\cal R}^\dagger_M({\blf k}'_1)~ \sigma_{-\mu'}~{\cal R}_M({\blf k}'_2)~\sigma_x  ~{\cal R}^\dagger_M({\blf k}_2)\Biggr\}
 \nonu
 \tilde A^2_{\mu \mu'}({\bf \tilde k}_1,{\bf \tilde k'}_1,{\bf \tilde k}_2,{\bf \tilde k'}_2 )=(-1)^{\mu'}~{1\over 2} Tr\Biggl\{ \sigma_{\mu}~{\cal R}_M({\blf k}_1)~ \sigma_x   ~
 {\cal R}^\dagger_M({\blf k}'_1)~ \sigma_{-\mu'}~{\cal R}_M({\blf k}'_2)~{\cal R}^\dagger_M({\blf k}_2)\Biggr\}
 \nonu
 \tilde B_{\mu \mu'}({\bf \tilde k}_1,{\bf \tilde k'}_1,{\bf \tilde k}_2,{\bf \tilde k'}_2 )=(-1)^{\mu'} ~{1\over 2} Tr\Biggl\{ \sigma_{\mu}~{\cal R}r_M({\blf k}_1)~ \sigma_x   ~
 {\cal R}^\dagger_M({\blf k}'_1)~ \sigma_{-\mu'}~{\cal R}_M({\blf k}'_2)~ \sigma_x  ~{\cal R}^\dagger_M({\blf k}_2)\Biggr\}
~.
\ee
Explicit expression of the traces can be found in Appendix \ref{app_traces}.
Notice that when $\hat \Phi$ reduces to a multiple of the  identity, e.g. for ${\bf k}_\perp={\bf k}'_{1\perp}-{ \bf k}_{1\perp}=0$,  one has
\be
\label{Eq:iden2}
{\cal R}_M({\blf k}_n)~\hat \Phi^{i}
 ~ {\cal R}^\dagger_M({\blf k}'_n)\to {\rm I}~.
\ee
and hence ${\cal S}_{ij}^{\mu\mu'}=\delta_{\mu,\mu'}~ H_i\Bigl(x_1{\bar \xi\over \xi_1},0,0\Bigr)H_j\Bigl(x_2{\bar \xi\over1- \xi_1},0,0\Bigr) $.}

Finally, the nuclear DPS2 reads
\begin{align}
\label{Eq:HHr}
    D^{A,2}_{ij}(x_1,x_2,{\bf k}_\perp)
    &= \int d\xi_1 ~ \dfrac{\bar \xi^2}{\xi_1 (1-\xi_1)}
   H_i\Bigl({\bar \xi x_1\over \xi_1},0, -{\bf k}^2_\perp\Bigr)H_j\Bigl({\bar \xi x_2\over1- \xi_1},0,- {\bf k}^2_\perp\Bigr) ~HH(\xi_1,{\bf k}_\perp)
   \\
   \label{Eq:HEr}
   &+ i \dfrac{k_y}{4M}\int d\xi_1 ~ \dfrac{\bar \xi^2}{\xi_1 (1-\xi_1)}
   H_i\Bigl({\bar \xi x_1\over \xi_1},0, -{\bf k}^2_\perp\Bigr)E_j\Bigl({\bar \xi x_2\over1- \xi_1},0,- {\bf k}^2_\perp\Bigr) ~ HE(\xi_1,{\bf k}_\perp)
   \\
   &+ i \dfrac{k_y}{4M}\int d\xi_1 ~ \dfrac{\bar \xi^2}{\xi_1 (1-\xi_1)}
   E_i\Bigl({\bar \xi x_1\over \xi_1},0, -{\bf k}^2_\perp\Bigr)H_j\Bigl({\bar \xi x_2\over1- \xi_1},0,- {\bf k}^2_\perp\Bigr) ~ EH(\xi_1,{\bf k}_\perp)
   \\
   &-\dfrac{k^2_y}{16M^2} \int d\xi_1 ~ \dfrac{\bar \xi^2}{\xi_1 (1-\xi_1)}
   E_i\Bigl({\bar \xi x_1\over \xi_1},0, -{\bf k}^2_\perp\Bigr)E_j\Bigl({\bar \xi x_2\over1- \xi_1},0,- {\bf k}^2_\perp\Bigr) ~  EE(\xi_1,{\bf k}_\perp)
\end{align}
where  the following deuteron off-forward two body LC distributions have been introduced

\begin{align}
\label{Eq:HH2}
  HH(\xi_1, {\bf k}_\perp) &=  \int d{\bf k}_{in}~\delta\Bigl[k_{in z} -\bar k_z(|{\bf k} _{in\perp}|,\xi_1)\Bigr]{\cal W}^{\mu\mu'}(\xi_1,{\bf k}_\perp,{\bf k}_{in})A_{\mu \mu'}({\bf \tilde k}_1,{\bf \tilde k'}_1,{\bf \tilde k}_2,{\bf \tilde k'}_2 )~;
  \\
  \label{Eq:HE2}
  HE(\xi_1, {\bf k}_\perp) &=  \int d{\bf k}_{in}~\delta\Bigl[k_{in z} -\bar k_z(|{\bf k} _{in\perp}|,\xi_1)\Bigr]{\cal W}^{\mu\mu'}(\xi_1,{\bf k}_\perp,{\bf k}_{in})\tilde A^1_{\mu \mu'}({\bf \tilde k}_1,{\bf \tilde k'}_1,{\bf \tilde k}_2,{\bf \tilde k'}_2)~;
  \\
    EH(\xi_1, {\bf k}_\perp) &=  \int d{\bf k}_{in}~\delta\Bigl[k_{in z} -\bar k_z(|{\bf k} _{in\perp}|,\xi_1)\Bigr]{\cal W}^{\mu\mu'}(\xi_1,{\bf k}_\perp,{\bf k}_{in})\tilde A^2_{\mu \mu'}({\bf \tilde k}_1,{\bf \tilde k'}_1,{\bf \tilde k}_2,{\bf \tilde k'}_2)~;
    \\
      EE(\xi_1, {\bf k}_\perp) &=  \int d{\bf k}_{in}~\delta\Bigl[k_{in z} -\bar k_z(|{\bf k} _{in\perp}|,\xi_1)\Bigr]{\cal W}^{\mu\mu'}(\xi_1,{\bf k}_\perp,{\bf k}_{in})\tilde B_{\mu \mu'}({\bf \tilde k}_1,{\bf \tilde k'}_1,{\bf \tilde k}_2,{\bf \tilde k'}_2)~.
\end{align}
The numerical evaluation of the above distributions will be now discussed.

In Fig. \ref{fig:DLCMD_3D} the  off-forward LCMD, $HH(\xi,{\bf k_\perp})$ (left panel) and $-i HE(\xi,{\bf k_\perp}) k_\perp/(4M)=i EH(\xi,{\bf k_\perp}) k_\perp/(4M)$  (right panel), defined through Eqs. \eqref{Eq:HH2} and \eqref{Eq:HE2}, respectively,  are shown to highlight  the full $k_\perp$ structure of the distributions.}

{As expected}, the no spin-flip contribution $HH$ is the dominant one. Both distributions exhibit their maximum absolute value for $\xi\sim0.5$, {{{since the deuteron is a two-body system, $\xi$ is the nucleon fraction of longitudinal momentum and the unpolarized LCMD is considered}}}.
The $HH$ distribution exhibits a maximum at $k_{\perp}=0$ (similar to the conventional charge form-factor).

In Fig. \ref{fig:DLCMD_3D_NR} the same quantities of Fig. \ref{fig:DLCMD_3D} have been displayed {{within the NR framework}},   see the Appendix \ref{NRLCMDs} for details. In this case, the main differences, w.r.t. the previous Poincaré covariant approach, are found for the $HE$ distribution {where  no spin-flip and spin-flip  GPDs interfere}. In fact, the {{$HE$ distribution}} now is almost zero for $\xi = 0.5$ and antisymmetric in the $\xi$ dependence.
{This peculiar effect is mainly due to the interference between S and D waves.}

In order to compare in details the $HH$ and $HE$ distributions,  these quantities {are} evaluated in different kinematic regions. In particular
in the {left}
 panel of Fig. \ref{fig:LCMD_rel_05} the $HH$ and $HE$ distributions calculated for $\xi=0.5$ are shown as functions of $k_\perp$.
{As expected}, at small momenta the no spin-flip distribution dominates. However, at higher momenta, although both quantities become relatively small, the $HE$ contribution gains significance and becomes relevant to the overall behavior (this emphasizes the role of the spin-physics at large momenta). In the right panel of Fig. \ref{fig:LCMD_rel_05} we consider the overall $k_\perp$ contribution of the nuclear DPS2 to the cross-section in order to properly establish to which extent spin-flip contributions could be relevant for future observables.

In particular we {{compare}} $k_\perp HH(0.5,k_\perp)$ {{and}} $-i k_\perp^2/(4M)HE(0.5,k_\perp)$. The extra $k_\perp$ factors w.r.t. those in  Eqs. \eqref{Eq:HHr} and \eqref{Eq:HEr} is due to the ${\bf k}_\perp$ integration appearing in the cross-section formula

\begin{align}
    \sigma_{DPS} \propto \int d^2k_\perp~ D^{A_1}_{ij}(x_1,x_2,{\bf k}_\perp) D^{A_2}_{ij}(x_1,x_2,-{\bf k}_\perp)~,
\end{align}
where $A_1$ and $A_2$ {indicate}  the kind of hadrons  or nuclei considered in the collisions.
As demonstrated by our analysis, in these kinematic conditions, the spin-flip effects are almost negligible for the eventual cross-section when $k_{\perp} < 1500~\mathrm{MeV}$. It should be emphasized that from a nuclear physics perspective, this represents a first  realistic prediction.

In Fig.  \ref{fig:LCMD_rel} the case $\xi=0.4$ is considered in order to have suitable reference values for the following comparison with the NR {{approach}}, that cannot be carried out at $\xi=0.5$ (recall that the NR $HE$ is vanishing there, as shown in the right panel of Fig. \ref{fig:DLCMD_3D_NR}). 
{Remarkably}, sizeable differences  are found  between $HH$ and $HE$ distributions, for $k_\perp<$ 1500 MeV. Also notice an order of magnitude decrease w.r.t. to  the values shown in Fig. \ref{fig:LCMD_rel_05}.

{For the sake of a detailed comparison} {of} the results obtained within the Poincaré covariant approach with those corresponding to the NR case,  the same quantities of Fig. {\ref{fig:LCMD_rel} but in the NR {{approach}}, {are shown in Fig. \ref{fig:LCMD_NR}}.
{While for the Poincar\'e-covariant calculation at $\xi=0.4$,}  the role of the $HE$ distribution is more relevant {than the $HH$ one} and cannot be neglected for $k_\perp>$ 500 MeV
{(filling also the dips  of $ k_\perp~HH$ as shown in the right panel that illustrate the relative weights in the cross-section)}, in the NR case   $HE$ is globally smaller than
$HH$,  as seen in left panel in Fig. \ref{fig:LCMD_NR},
 {and does not affect the dips, when the cross section contributions are considered.}
  These results are confirmed by properly
comparing both the $HH$ and $HE$ distributions evaluated within the Poincaré-covariant LF approach with its NR  {{counterpart}}, {as shown in Fig. \ref{fig:comp}}.
In the $HH$ case the two calculations are almost indistinguishable for $k_\perp < 400$ MeV while in the high $k_\perp$ region, relativistic effects appear. Such a result is expected from Eq. (\ref{Eq:iden2}), where for $k_\perp \sim 0$ the product of the Melosh rotations reduce to the identity. On the contrary,
in the $HE$ case,  the  product of the Melosh rotations is different from the identity also for $k_\perp \sim 0$ due to the spin flip. Therefore, relativistic effects are found also for low values of $k_\perp$. In conclusion,
this approach allows {to explore} the nuanced differences between distributions that can significantly impact observables. Notably, while  $HE(\xi,{\bf k_\perp})$  is suppressed at small $k_\perp$, $HH(\xi,{\bf k_\perp})$ and $HE(\xi,{\bf k_\perp})$ distributions exhibit diffractive minima at distinct positions. Consequently, at higher $k_\perp$ values, spin-flip effects may become non-negligible.

{Summarizing, the quantities $HH(\xi,{\bf k_\perp})$, $HE(\xi,{\bf k_\perp})$  and $EE(\xi,{\bf k_\perp})$}  represent the nuclear input needed for the calculations of the nuclear DPDs, {but only the first LCMD has been used in the calculation shown in the next  Section}.


\begin{figure*}[t]
\includegraphics[width=8.7cm]{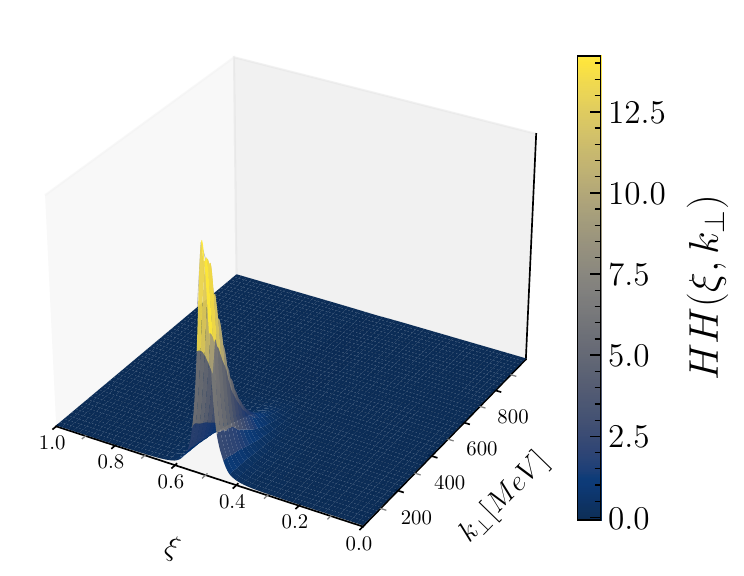}
\includegraphics[width=8.7cm]{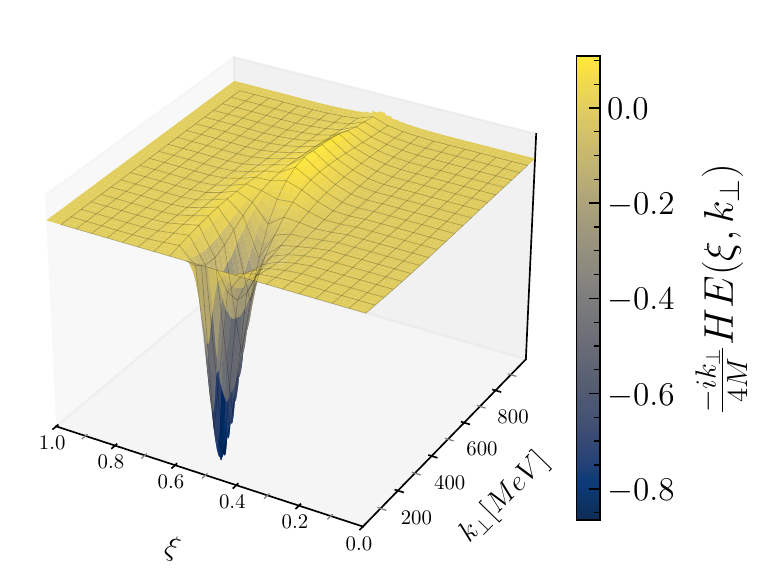}
\caption{ (Color online)  The off-forward LCMDs, $HH(\xi,{\bf k_\perp})$ (left panel) and $-i HE(\xi,{\bf k_\perp}) k_\perp/(4M)= i EH(\xi,{\bf k_\perp}) k_\perp/(4M)$ (right panel), defined in Eq. \eqref{Eq:HH2} and \eqref{Eq:HE2}, respectively.
}
\label{fig:DLCMD_3D}
\end{figure*}
\begin{figure*}[t]
\includegraphics[width=8.7cm]{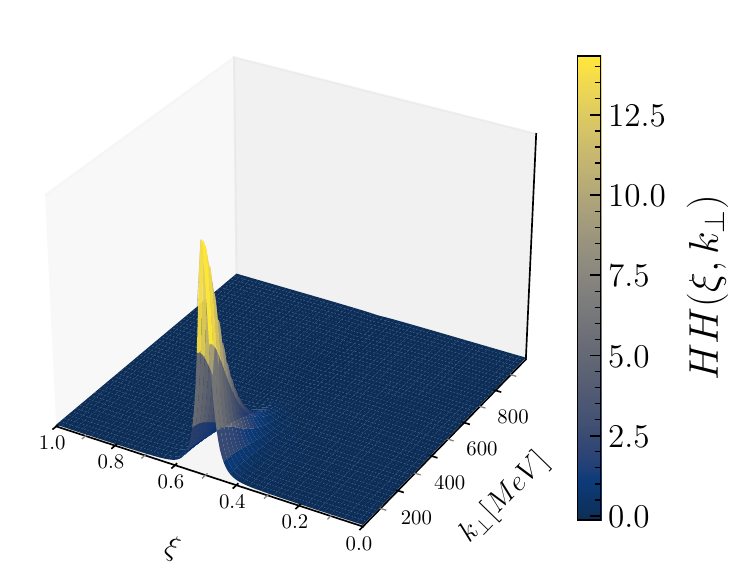}
\hskip 0.1cm
\includegraphics[width=8.7cm]{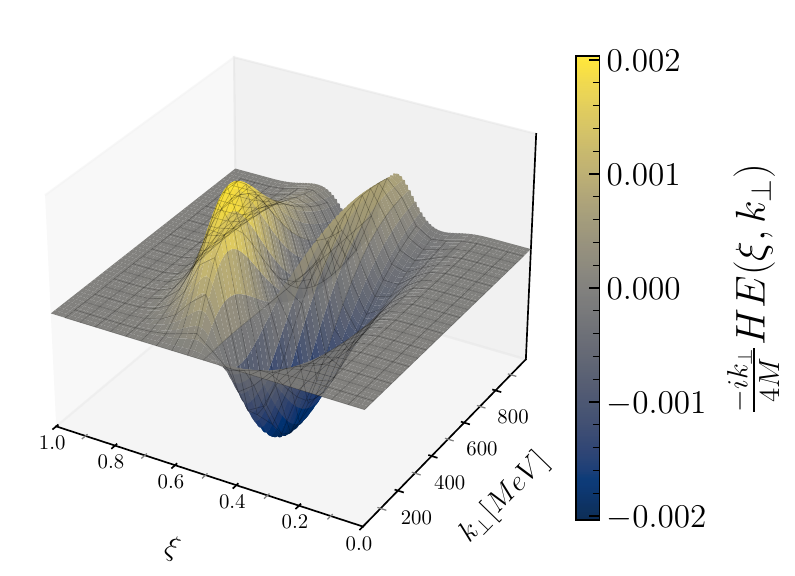}
\caption{ The same of Fig. \ref{fig:DLCMD_3D} in the NR {{framework}}.
}
\label{fig:DLCMD_3D_NR}
\end{figure*}
\begin{figure*}[htb]
\includegraphics[width=7.7cm]{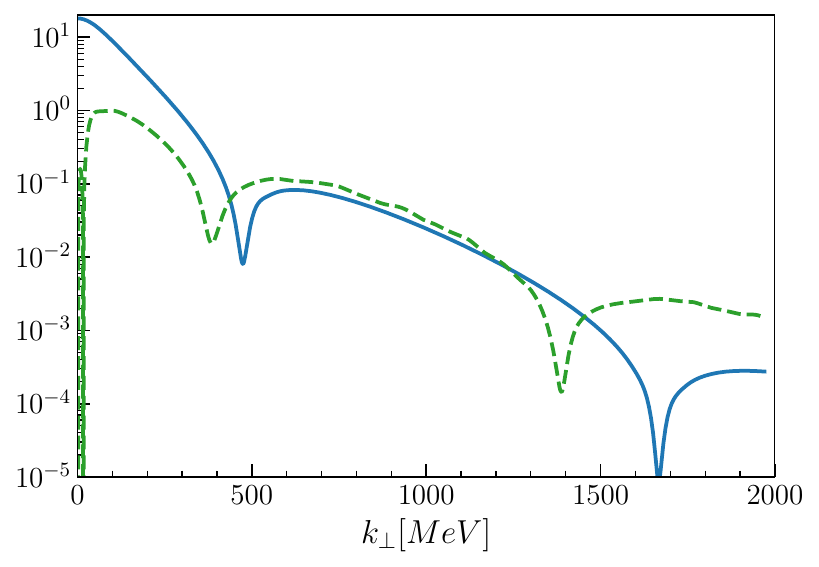}
\hskip 0.1cm
\includegraphics[width=7.7cm]{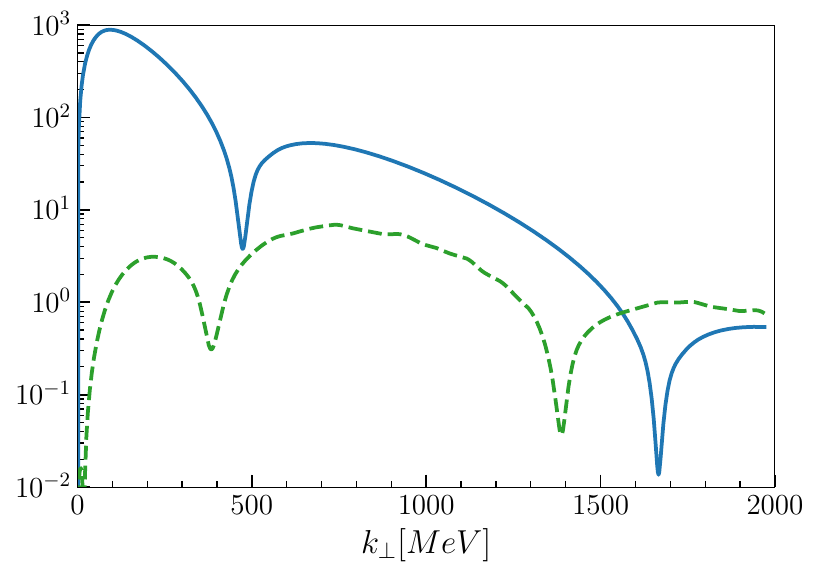}
\caption{LCMDs appearing in Eqs. (\ref{Eq:HH2}) and \eqref{Eq:HE2},
as functions $k_\perp$ and evaluated for $\xi=0.5$.
Left panel: full line for $HH(\xi,{\bf k}_\perp)$ and dashed line for $-i HE(\xi,k_\perp)$. Right panel: full line for $k_\perp HH(\xi,k_\perp)$ and dashed line  for $-i k^2_\perp/(4M) HE(\xi,{\bf k}_\perp)$.
}
\label{fig:LCMD_rel_05}
\end{figure*}

\begin{figure*}[t]
\includegraphics[width=7.7cm]{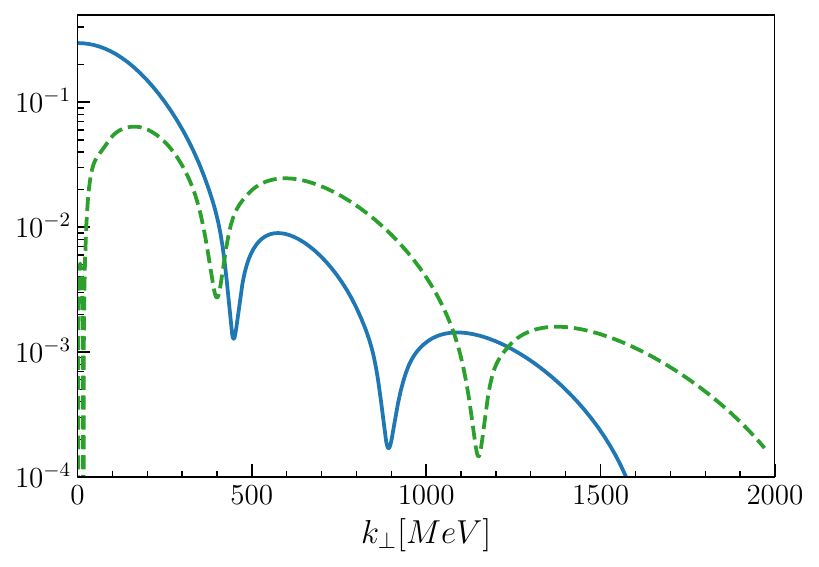}
\hskip 0.1cm
\includegraphics[width=7.7cm]{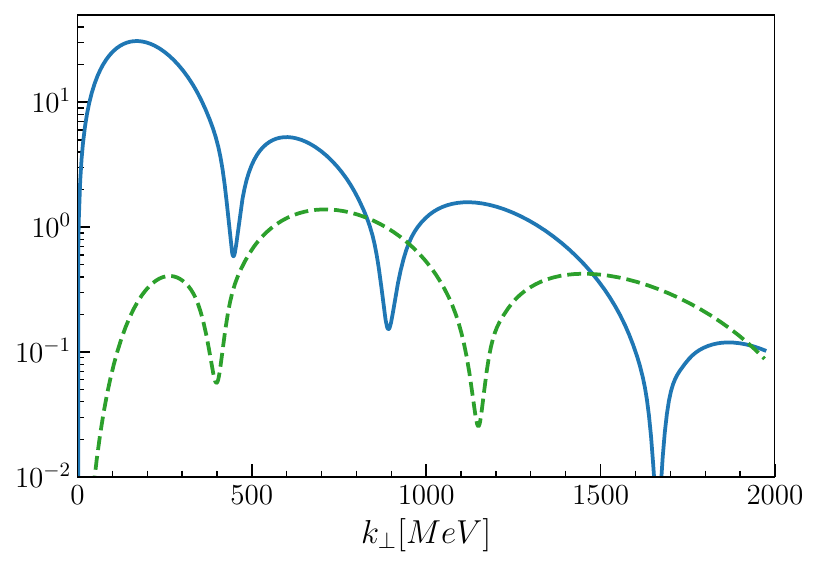}
\caption{
{The same as in Fig. \ref{fig:LCMD_rel_05}, but for $\xi=0.4$ (see text for the choice of such a $\xi$ value).}}
\label{fig:LCMD_rel}
\end{figure*}

\begin{figure*}[t]
\includegraphics[width=7.7cm]{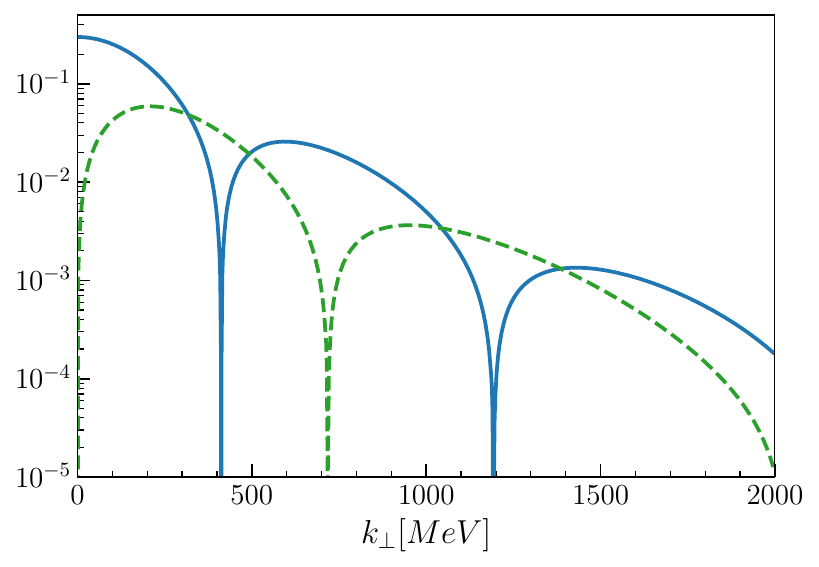}
\hskip 0.1cm
\includegraphics[width=7.7cm]{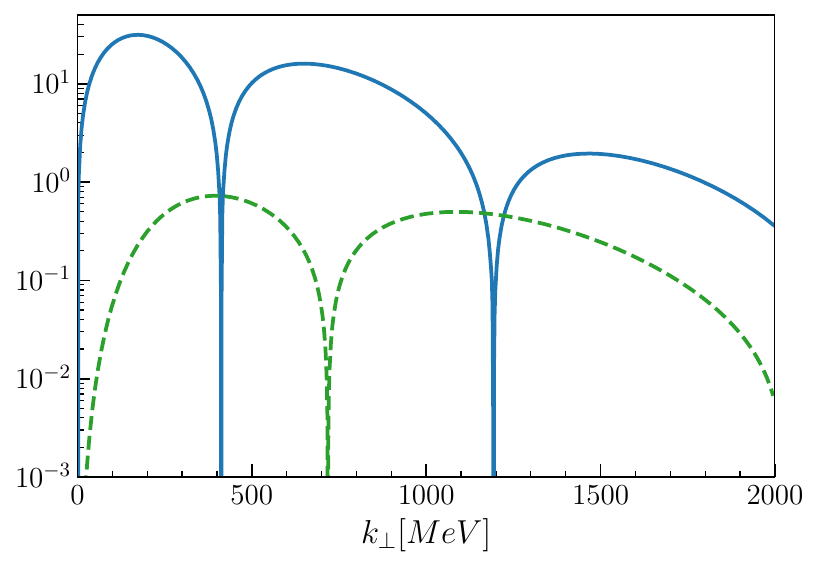}
\caption{ The same as in  Fig. \ref{fig:LCMD_rel},  but in the NR {{case}}.
}
\label{fig:LCMD_NR}
\end{figure*}

\begin{figure*}[t]
\includegraphics[width=7.7cm]{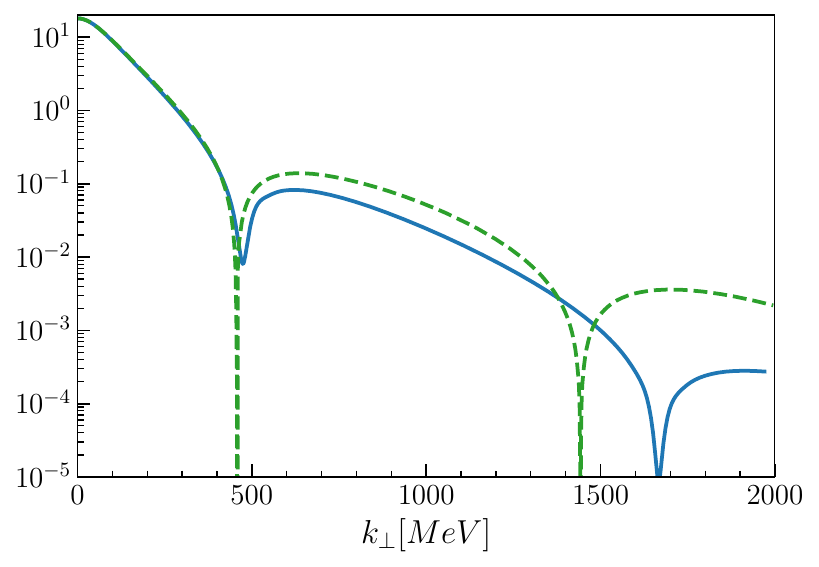}
\hskip 0.1cm
\includegraphics[width=7.7cm]{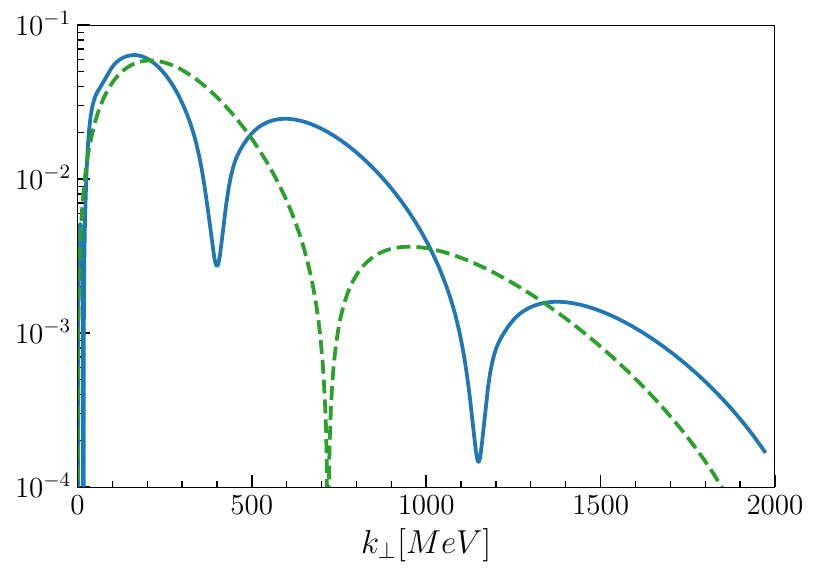}
\caption{ Left panel: comparison between the calculations of $HH(\xi=0.5, k_\perp)$ within the LF Poincaré covariant approach Eq. (\ref{Eq:HH2}) (full line) and the NR {{counterpart}} (dashed line). Right panel: the same as the left panel but for $HE(\xi=0.4, k_\perp)$, see  Eq. (\ref{Eq:HE2}).
}
\label{fig:comp}
\end{figure*}

\section{Calculation of the  DPDs}

In this Section,  the results of the calculations of deuteron DPDs corresponding to both  DPS1 and DPS2 mechanisms {are presented} {for $k_\perp=0$ in order to evaluate the distributions useful to test the nuclear GS sum rules. Moreover, the DPDs  {and GPDs} are dominated by the low $k_\perp$ region}.
In this case, for the DPS2,
only the spin conservation contribution is retained {in the LC nucleon correlator}, see Eq. (\ref{Eq:gen6}), being the dominant term. To proceed with these evaluations, one needs to adopt i) a  nucleon DPD, for the DPS1 contribution, and  ii) nucleon PDFs, entering the DPS2 mechanism for $k_\perp=0$. {Let us recall that GPDs in the forward limit, i.e. $k_\perp=0$, yield the usual PDF}.
{In the present analysis, two scenarios have been considered. In the first one, the  Harmonic Oscillator (HO) model \cite{jhepc,noice} was employed to evaluate the nucleon DPDs in Eq. (\ref{Eq:DPD1_momentum}) and the nucleon PDFs  entering the nuclear DPD, {for the DPS2 mechanism}, at $k_\perp=0$  in Eq. (\ref{Eq:gen6})}. {This model is very useful to test the nuclear GS sum rules. In fact, in this case both nucleon DPDs and PDFs  analytically fulfil the corresponding integral properties.
{All distributions evaluated within this model have been computed at the low hadronic scale. This approach is advantageous for revealing double parton correlations, which are typically suppressed following the perturbative QCD evolution procedure.}
In addition, we also consider a phenomenological approach based on the approximation of the nucleon DPDs as the product  of PDFs. This kind of ansatz has been largely used to make predictions, see e.g. Refs. \cite{Gaunt:2009re,Golec-Biernat:2014bva}. }
{In particular,}  the  Goloskov-Kroll model (GK)  \cite{Goloskokov:2007nt,Diehl:2013xca}, already adopted to evaluate the $^3$He and $^4$He GPDs in, e.g., Ref. \cite{Fucini:2021psq}, has been considered {in order} to parametrize  the nucleon
PDFs at, {e.g., $Q^2=4$ GeV$^2$ {(the evolution is included)}}.
{{For completeness, we mention that the PDFs in the GK model were obtained by fitting the {data sets of}} Refs. \cite{Martin:2009iq,Ball:2011gg,Alekhin:2012ig,Lai:2010vv,Gluck:2007ck}. In particular, the authors focused on the low and high $x$ regions. We consider this approach, instead of other possible PDF parametrizations, in view of future analyses where nucleon GPDs will be required.}}

{We build the nucleon DPDs as in Ref. \cite{Golec-Biernat:2014bva} with the correct support and with a softer behaviour for $x_1 + x_2 \to 1$  compared to the standard factorization ansatz},
{i.e.}
\begin{align}
    \nonumber
    D^\tau_{ij}(x_1,x_2,{\bf k_\perp}) = & \frac{\lambda^{\tau}_{i j}}{4}\Bigg[ \frac{1}{1-x_2} \left[ d_i^\tau \left(\frac{x_1}{1-x_2}\right)d_j^\tau(x_2) + d_j^\tau\left(\frac{x_1}{1-x_2}\right)d_i^\tau(x_2) \right]+
    \\
    & +
\frac{1}{1-x_1} \left[ d_j^\tau(x_1) d_i^\tau \left(\frac{x_2}{1-x_1}\right) + d_i^\tau(x_1) d_j^\tau \left(\frac{x_2}{1-x_1}\right) \right] \Bigg] \theta(1-x_1 - x_2) g(k_\perp)~,
\label{Eq:fact}
\end{align}
where $g(k_\perp)$ is {an} effective form factor \cite{rapid} {and  $\lambda^\tau_{ij} = 1$,  except for  $\lambda^p_{d_v d_v} = \lambda^n_{u_v u_v} = 0$}. {Recall that }{in this first analysis, as an example, we consider the case of $k_\perp=0$.}
{The theta function in the above equation is included by hand in order to preserve the support conditions for which the distribution is zero for $x_1+x_1>1$.}
Let us remark that the normalization of the DPDs leads to
 $g(0)=1$ \cite{rapid}, {and} {that} within this approach, commonly used for phenomenological and experimental analyses of DPS at the LHC, double parton correlation are neglected.
{A relevant difference between the two mechanisms, independently of the model which is used, is that  for the DPS1 distribution one has the constraint $x_1/\xi + x_2/\xi \le M_A / M $ (see Eq. \eqref{DA1}) and, since the nuclear LCMD resembles a $\delta$  function, $\delta(\xi - M/M_A)$, this constraint becomes $x_1 + x_2 \le 1$, obviously only  approximately.}

In Fig. \ref{fig:DPS12_GK} the  nuclear DPDs, weighted by $x_1x_2$,  for the $ud$ flavors and corresponding to both the DPS1  (left panel)
and DPS2 (right panel) mechanisms are shown for the GK model. As one can see, the absence of double parton correlations in the nucleon DPDs  {(cf. Eq. \eqref{Eq:fact})} leads to distributions that similarly populate the phase space. In fact,
at $k_\perp=0$ {both DPS1 and DPS2 distributions depend on the product of nucleon PDFs (recall that the product of nucleon GPDs reduces to the product of PDFs, for $k_\perp=0$). {{ The effect of the constraint $x_1 + x_2 \le 1$ is clearly visible for DPS1.}}
{Interestingly}, one  notice that the DPS2 contribution is the dominant one.

In Fig. \ref{fig:DPS12_HO} the same quantities shown in Fig. \ref{fig:DPS12_GK} have been plotted for the HO model. In this case,  nucleon DPDs are {remarkably} not factorized in terms of PDFs,
{since} double parton correlations are included at the low energy scale of the model, {where there is the valence-quark dominance}.{
One should notice that PDFs and DPDs evaluated within this model have the corresponding maxima in the valence region.
In particular, the PDF maximum is located at $x\sim 0.145$ (see Ref. \cite{noice} for details), therefore also the double distribution related to the DPS2 mechanism is peaked at $x_1 \sim x_2 \sim 0.145$. Instead, due to double parton correlations, the DPD for the DPS1 contribution has maximum values for $x_1 \sim x_2 \sim 0.32$. One should notice that in these figures, these values are shifted to higher $x_i$ since we display the DPD multiplied by $x_1 x_2$. In addition,
 the following features are addressed: $i)$ support effects are evident in particular in the DPS1 case where, due to the peaked {nucleon} LCMDs, the relative DPD goes to zero as in the free nucleon case, i.e. $x_1+x_2 \leq 1$; $ii)$ since the GK model provides nucleon PDFs at high energy scale, the corresponding nuclear PDFs and DPDs are dominated by the lower $x_i$ region w.r.t. the HO calculations, at the hadronic scale, and therefore boundary effects are less evident and $iii)$ due to all these features the two models populate the phase space quite differently.}

{Finally, one can study in detail the magnitude of nuclear effects induced by the nuclear potential through the comparison of the computed nuclear DPSs with the free ones (i.e when no interactions between nucleons are assumed),   defined by
\begin{align}
\label{Eq:free}
    D^{free,1}_{ij}(x_1,x_2,\mathbf{0}_\perp) = &Z D^{p}_{ij}(x_1,x_2,\mathbf{0}_\perp) + (A-Z)D^{n}_{ij}(x_1,x_2,\mathbf{0}_\perp) \\
    \nonumber D^{free,2}_{ij}(x_1,x_2,\mathbf{0}_\perp) = &{Z(Z-1) d^{p}_{i}(x_1) d^p_j(x_2) + Z(A-Z) \big[d^{p}_{i}(x_1) d^n_j(x_2) + d^{n}_{i}(x_1) d^p_j(x_2) \big]}+ \\
    & \nonumber (A-Z)(A-Z-1) d^{n}_{i}(x_1) d^n_j(x_2)~.
\end{align}
}
{
In particular, one can 
{build} an EMC-like ratio for each contribution, as well for the {{sum of the two contributions}} as
\begin{align}
\label{Eq:EMC1}
    R^{A,n}_{ij}(x_1,x_2) = & \frac{D^{A,n}_{ij}(x_1,x_2,\mathbf{0}_\perp)}{D^{free,n}_{ij}(x_1,x_2,\mathbf{0}_\perp)}  \quad \quad \quad n=1,2
\end{align}

In Fig. \ref{fig:Ratios} the ratios {$R_{ud}^{2,1}$ and $R_{ud}^{2,2}$}, computed for the deuteron with the HO model, are shown for $x_1 = 0.3$ and $x_1 = 0.5$ {as a function of $x_2$}}.
All ratios display the typical behaviour of the nuclear EMC-like effect, usually defined for structure functions \cite{Griffioen:2015hxa}.
In perspective, in order to qualitatively establish to what extent DPS processes could provide
useful insights on the {short-range correlations} origin of the EMC effect, we
{also} evaluate  the ratio involving free and nuclear PDFs, by using the same model as in Fig. \ref{fig:Ratios}. In particular, Fig. \ref{fig:Ratio_pdf}  shows the following {{ratio for the $u$ quark}}:  }

\begin{align}
\label{Eq:ratio2}
    \bar R_{i}^A(x) = \dfrac{d^A_i(x)}{d^p_i(x)+d_i^n(x)}~.
\end{align}

By comparing Figs. \ref{fig:Ratios} and \ref{fig:Ratio_pdf}, it becomes evident that the ratio  {$R_{ud}^{2,2}$, corresponding to the DPS2 mechanism} and defined in Eq. (\ref{Eq:EMC1}) {(dot-dashed  lines)}, which depends on the product of PDFs, follows the same {pattern} as the ratio in Eq. (\ref{Eq:ratio2}), related to the conventional deuteron EMC effect \cite{Griffioen:2015hxa} {(modulo the actual  position of the dip, that interestingly depends upon the value of $x_1$)}. In contrast, $R^{2,1}$ (cf. Eq. \eqref{Eq:EMC1}) exhibits a markedly different behavior.
Since the  $^2$H  LCMD  is peaked around $\xi \sim \bar{\xi}$, the nucleon DPDs entering Eq. \eqref{Eq:DPD1_momentum} vanish for $x_1+x_2 \sim 1$. Consequently, Fermi motion effects manifest at smaller values of $x_2$ compared to those observed in the DPS2 contribution. More significantly, the depletion of this ratio,  {which manifests at $x_2\sim 1-x_1$, } is more pronounced than the one obtained from either the DPS2 contribution or the conventional EMC effect.
The principal conclusion of this analysis is that the DPS1 contribution, which involves DPDs, exhibits greater sensitivity to nuclear effects than single parton scattering processes such as DIS. Therefore, future experimental investigations of the DPS1 mechanism could provide novel fundamental insights into the {short-range correlations and the} origin of the EMC effect. In Ref. \cite{preparation2}, a preliminary exploratory study of a \textit{double EMC effect} {(related to DPS1)} for light nuclei is presented.

\begin{figure*}[t]
\includegraphics[width=8.7cm]{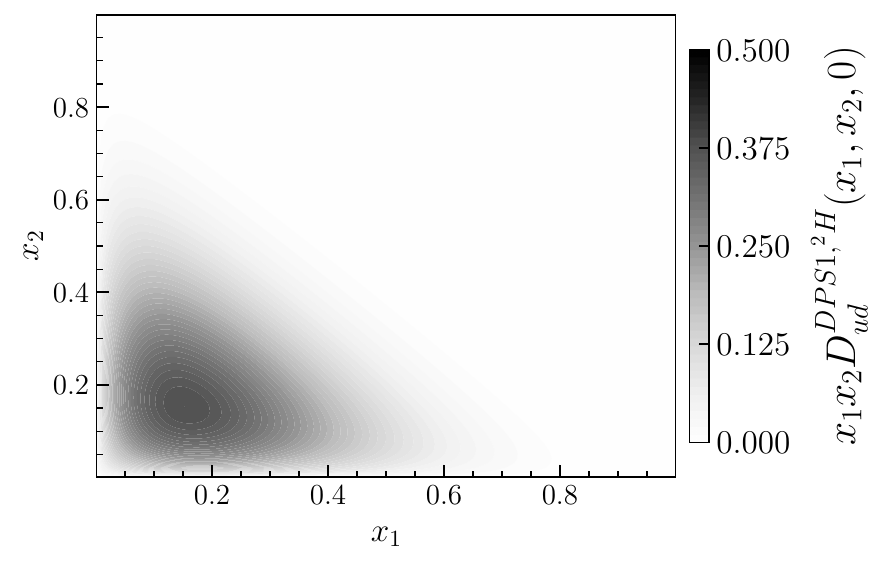}
\hskip 0.1cm
\includegraphics[width=8.7cm]{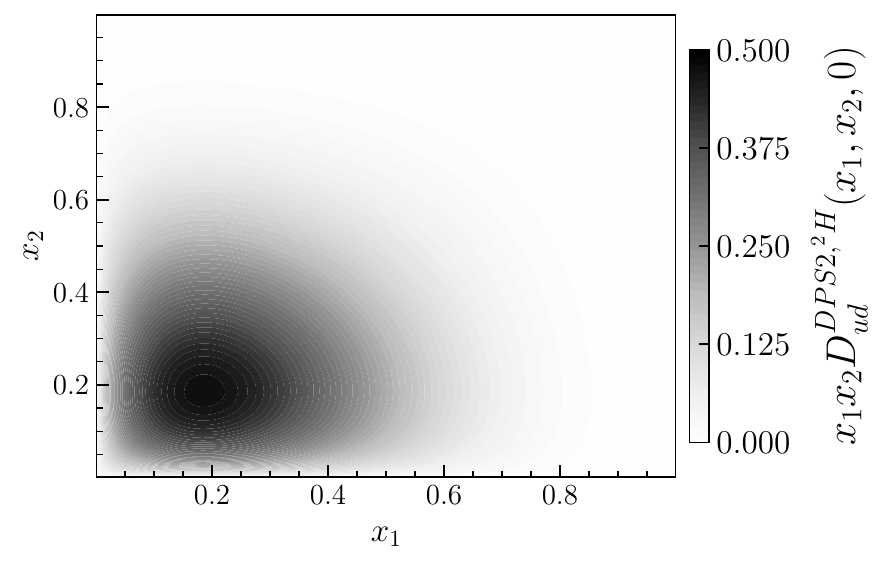}
\caption{Deuteron DPDs, weighted by $x_1x_2$,  evaluated for $i=u$ and $j=d$ within the GK model.
Left Panel:  DPS1 mechanism. The corresponding distribution have been calculated through Eq. \eqref{Eq:DPD1_momentum} for $k_\perp = 0$.
Right Panel: the same as in the left panel, but for the  DPS2 mechanism, calculated through Eq. \eqref{Eq:gen6} for $k_\perp = 0$.
}
\label{fig:DPS12_GK}
\end{figure*}

\begin{figure*}[t]
\includegraphics[width=8.7cm]{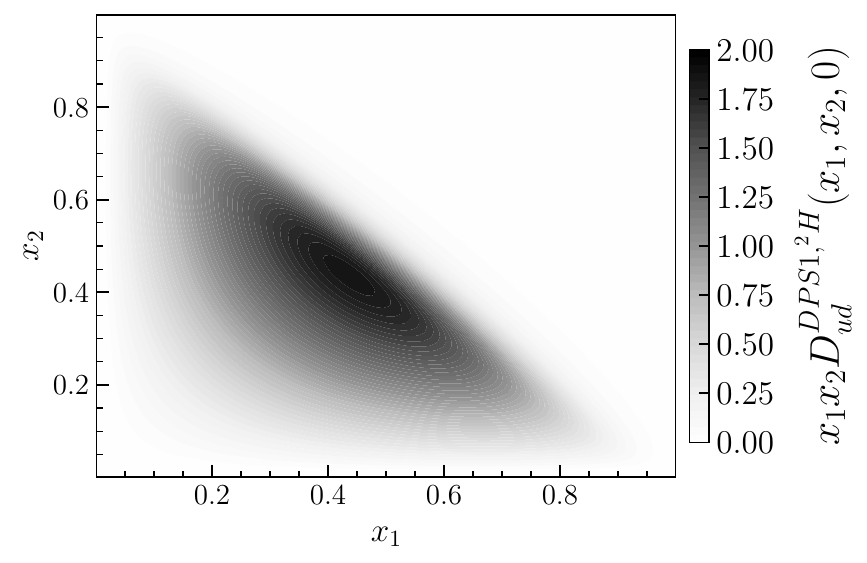}
\hskip 0.1cm
\includegraphics[width=8.7cm]{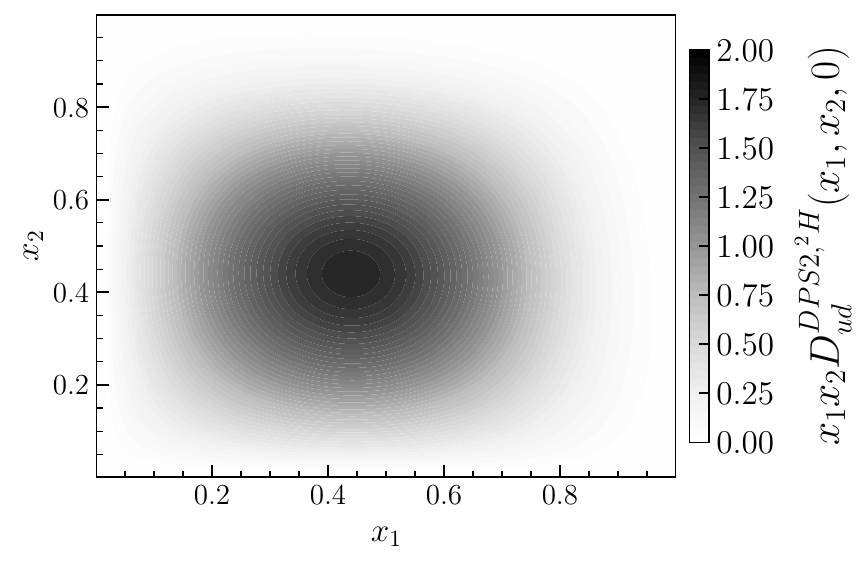}
\caption{The same as in  Fig. \ref{fig:DPS12_GK}, but  for the HO model
{(see text for the discussion)}.}
\label{fig:DPS12_HO}
\end{figure*}

\begin{figure*}[t]
\includegraphics[width=8.7cm]{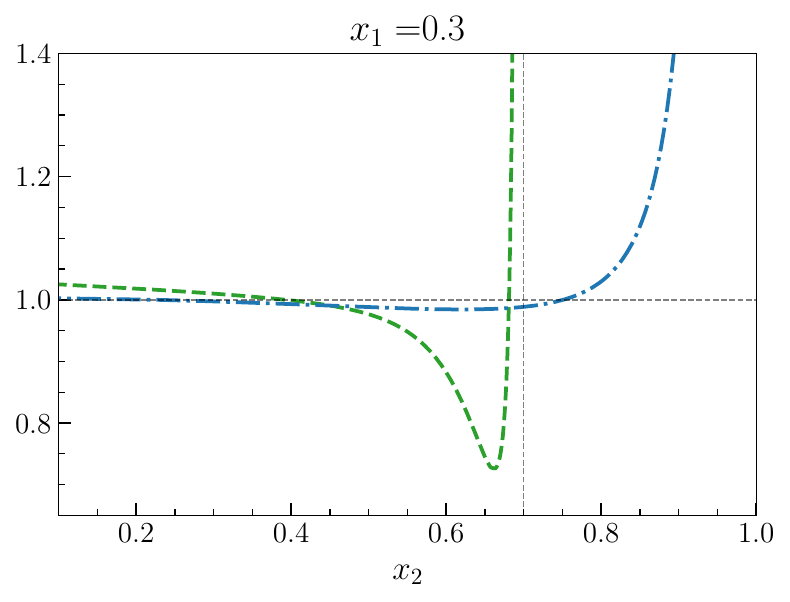}
\hskip 0.1cm
\includegraphics[width=8.7cm]{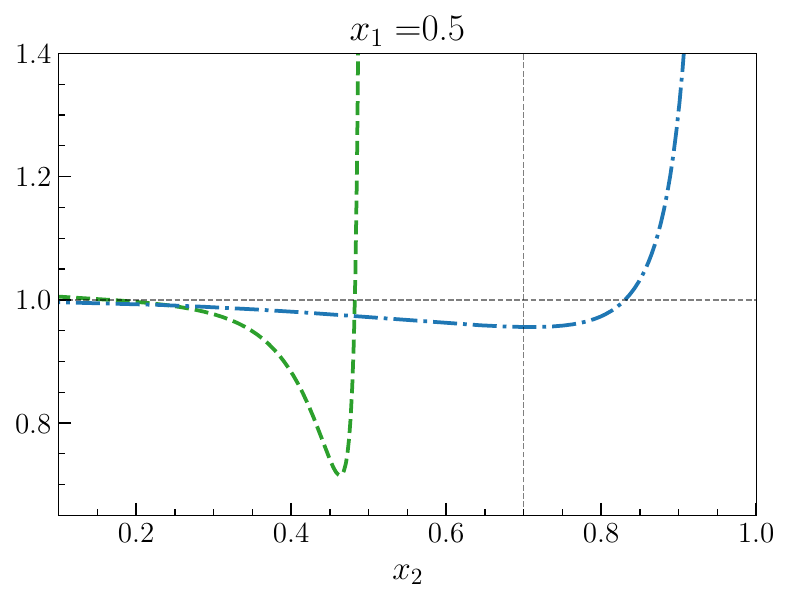}
\caption{
EMC-like ratios for the deuteron. Dashed lines: 
DPS1 contributions, i.e. $n=1$ in Eq. \eqref{Eq:EMC1}. Dot-dashed lines:
DPS2 contributions, i.e $n=2$ in Eq. \eqref{Eq:EMC1}.
Here we considered the $u$ and $d$ flavors and the HO model.
 Left panel:  $x_1 = 0.3$. Right panel:
 $x_1 = 0.5$.
 }
\label{fig:Ratios}
\end{figure*}

\begin{figure*}[t]
\includegraphics[width=7.7cm]{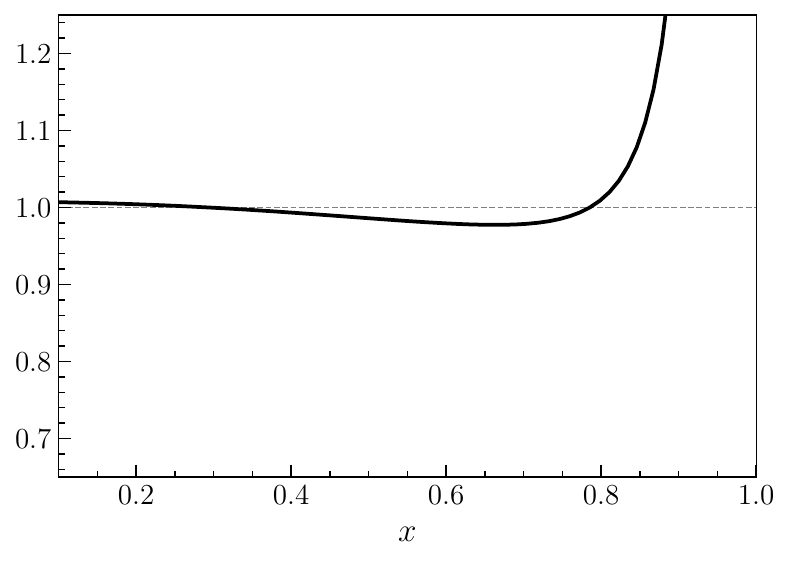}
\caption{
{ The
EMC-like ratio Eq. (\ref{Eq:ratio2}) for the deuteron and the $u$ quark, evaluated within the HO model.
 }}
\label{fig:Ratio_pdf}
\end{figure*}

\subsection{Evaluation of the nuclear GS sum rules for the deuteron}
{ The numerical results for  the GS {nuclear} sum rules, i.e.  the number sum rule, Eq. \eqref{SR_nuclear1}, and the momentum one,  Eq. \eqref{Eq:MSR}, are discussed in what follows for the deuteron case.
Importantly, the evaluation of   these integral properties requires  the nucleon DPDs within a suitable model. In fact,  the factorization ansatz, such as the one in Eq. (\ref{Eq:fact}), fails to satisfy all the nucleon  GS sum rules. Therefore, to test Eqs. \eqref{SR_nuclear1} and \eqref{Eq:MSR}, the Harmonic Oscillator  model \cite{jhepc,noice} was employed for  the nucleon DPDs.
In this case, since the nucleon DPDs are evaluated from the corresponding wf, the probabilistic interpretation of DPDs is preserved and the relative sum rules are directly fulfilled.

In  Fig. \ref{fig:sumrule} ,   the NSR,  rhs of Eq. \eqref{SR_nuclear1},  (full-line in the left panel) and the MSR,  rhs of Eq. \eqref{Eq:MSR}, (full-line in the right panel) are shown for the $uu$ combination. These results are compared with the  contributions of the PSR corresponding to  i) the DPS1 mechanism,
{shown by} dotted lines   and ii) the  DPS2 mechanism, dashed lines. In the left panel, the DSP1 contribution to NSR is evaluated by the lhs of Eq. \eqref{NSR_DPS1}  and the DPS2 by the  lhs of Eq.  \eqref{Eq:NSR_DPS2}, respectively. In the right panel, the DSP1 contribution to MSR is obtained from the lhs of Eq.
\eqref{MSR_DPS1}, and the DPS2 from Eq. \eqref{MSR_DPS2}, respectively. Notably, the sum of the two contributions, dash-dotted lines,
{are hardly distinguishable from}   the exact result, given by the full lines.}
{One should notice that the DPS1 contribution to the NSR is almost half that of the DPS2. However, in the MSR case the two contributions are comparable.}

\begin{figure*}[t]
\includegraphics[width=7.7cm]{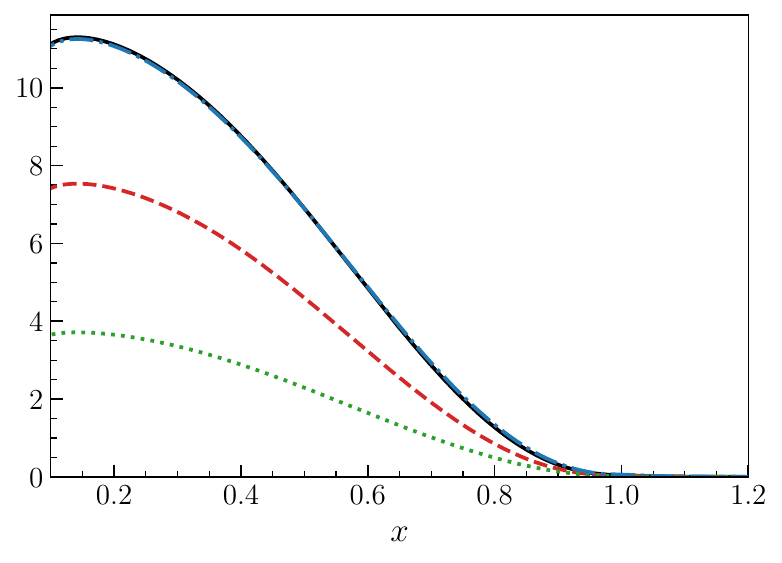}
\hskip 0.1cm
\includegraphics[width=7.7cm]{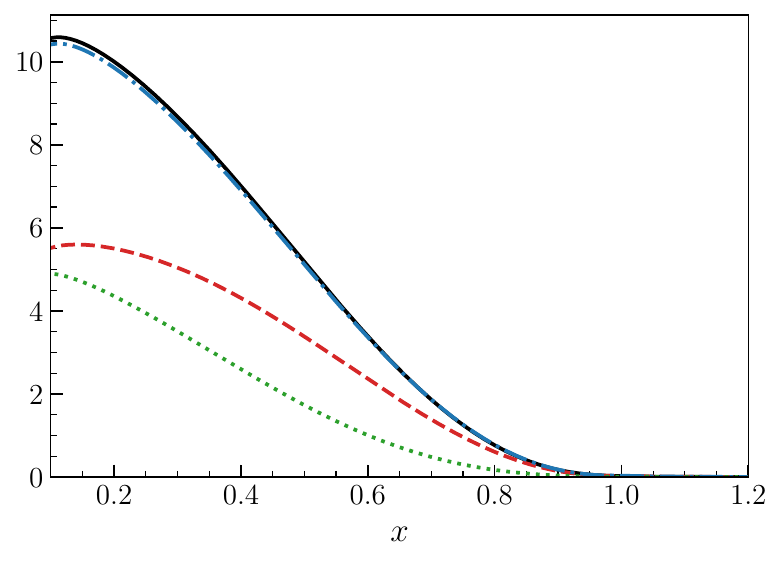}
\caption{ (Color online). Left Panel: NSR for the deuteron {with} $i_v = j_v = u_v$. The full line is the rhs of Eq. \eqref{SR_nuclear1}. The dotted line is the NSR for the DPS1 mechanism, calculated through the lhs of Eq. \eqref{NSR_DPS1}. The dashed line is the NSR for the DPS2 mechanism, calculated through the lhs of Eq. \eqref{Eq:NSR_DPS2}. The dash-dotted line is the lhs  of eq. \eqref{SR_nuclear1}.
Right Panel: MSR for the deuteron {with} $i_v = j_v = u_v$. The full line is the rhs of Eq. \eqref{Eq:MSR}. The dotted line is the MSR for the DPS1 mechanism, calculated through the lhs of Eq. \eqref{MSR_DPS1}. The dashed line is the MSR for the DPS2 mechanism, calculated through the lhs of Eq. \eqref{MSR_DPS2}. The dash-dotted line is the lhs  of Eq. \eqref{Eq:MSR}
}

\label{fig:sumrule}
\end{figure*}

\section{Conclusions}
This investigation introduces a novel approach to describe the reaction mechanism of double parton scattering  off nuclear targets. By retaining only nucleonic degrees of freedom, we formally demonstrate{, within an impulse approximation framework,} that the nuclear double parton distribution can be decomposed into two distinct contributions: i) the $DPS1$
 mechanism,  where the two active partons belong to the {\em same} nucleon, and ii) the $DPS2$
  mechanism where the two partons belong to {\em different} nucleons.
In the DPS1 case, the nuclear distribution results from the convolution of  the light-cone momentum distribution with the free nucleon double-parton distribution, while
in the DPS2 contribution, one has the convolution between an off-forward momentum nuclear distribution with  two free nucleon light-cone correlators, parametrized through GPDs.
Notably, this analysis highlights the possible roles of both GPDs $H$ and $E$ in evaluating the nuclear DPD at leading twist.
Since the total double-parton scattering  cross-section encompasses both DPS1 and DPS2 contributions,
it is fundamental {to quantitatively assess their relative weight}. To this aim, we consider the integral properties of DPDs by {introducing for the first time}  partial sum rules for nuclear DPDs.
This approach has been developed by generalizing the Gaunt-Stirling sum rules to the nuclear context.

To rigorously validate our theoretical framework, we also  defined and analysed the  properties of the two-body light-front nuclear density {necessary for fully evaluating}  the nuclear DPDs.
In particular, our investigation presents the first detailed calculations of the deuteron double-parton distributions for both DPS1 and DPS2 mechanisms using a robust, phenomenologically grounded approach. Specifically, we {adopted a} nuclear wave functions derived from a realistic nucleon-nucleon phenomenological potential
{and embedded in a rigorous Poincar\'e covariant formalism}.
 We {then} calculated the off-forward light-cone momentum distributions corresponding to spin-conserving and single spin-flip {processes}. {In particular,} we showed that for the deuteron the first contribution is the  dominant one. To complete the analysis, the nuclear sum rules have been successfully tested and we found that the DPS1 contribution to the NSR is almost half the DPS2 one, while the two contributions to the MSR are comparable.
Finally, the nuclear DPDs, evaluated at $k_\perp = 0$ and corresponding to both mechanisms, have been compared in a complete realistic framework.
{EMC-like ratios involving nuclear and free DPDs were introduced to address the potential role of DPS in understanding this effect.}

{In conclusion, it should be emphasized that the current} analysis provides a foundational methodology for future nuclear double-parton scattering  studies, {{{and enables}} more precise accounting of nuclear effects } in theoretical calculations of DPDs.

\section*{Acknowledgements}
M. Rinaldi thanks for the financial support received under the STRONG-2020 project of the European Union’s
Horizon 2020 research and innovation programme: Fund
no 824093.

\appendix

\section{One-body LC momentum number  densisty }
\label{app:1bd}

Here we introduce the  baryon number and momentum sum rules for the {nuclear} one-body LC momentum number density.
Let us {{recall}} that, from Eq. (\ref{Eq:density1})
\be
\bar \rho_\tau(\xi_1)= A \rho_\tau(\xi_1)=
A \sum_{\tau_2} \int_0^{1-\xi_1} d\xi_2~\int  d^2 k_{1,\perp} \int  d^2k_{2,\perp} |\psi(\xi_1,\xi_2, \mathbf k_{1,\perp},\mathbf k_{2,\perp},\tau,\tau_2)|^2
\ee

From the normalization of the LCMD Eq. \eqref{eq:normrho}, the baryon number sum rule reads as follows
\begin{align}
\label{Eq:numden}
    \sum_\tau \int d\xi ~\bar \rho_\tau (\xi) = A
\end{align}
 therefore
\begin{align}
    \int d\xi ~\bar \rho_\tau (\xi) = N_\tau~,
\end{align}
being $N_\tau$ the number of nucleons with isospin $\tau$, {i.e.}
\be
\int_0^1 d\xi_1~\bar \rho_{p}(\xi_1)=Z~, \quad {\rm and} \quad
\int_0^1 d\xi_1~\bar \rho_{n}(\xi_1)=(A-Z)~.\ee
In addition, the momentum sum rule {reads}
\begin{align}
\label{Eq:momcons}
    \langle \xi \rangle_p+\langle \xi \rangle_n = \sum_{\tau} \int d\xi~ \xi\bar \rho^A_{\tau}(\xi) =1~,
\end{align}
where
\begin{align}
\label{Eq:momcons2}
    \langle \xi \rangle_\tau = \int d\xi~\xi \bar \rho^A_\tau(\xi)~.
\end{align}

 \section{Two-body LC momentum number  density}
\label{app:2bd}
In this Appendix,  the number and momentum sum rules for the two-body LC momentum number density are introduced.

Let us recall the definition of  $\bar \rho_{\tau_1 \tau_2}^A (\xi_1,\xi_2)$ given in Eq. \eqref{Eq:density_bar_2}
\begin{align}
\label{eq:density_bar_2_app}
    \bar \rho_{\tau_1 \tau_2}^A (\xi_1,\xi_2) \equiv \bar \rho_{\tau_1 \tau_2}^A (\xi_1,\xi_2, {0})=A (A-1) \int  d^2 k_{1,\perp} \int  d^2k_{2,\perp} |\psi(\xi_1,\xi_2, \mathbf k_{1,\perp},\mathbf k_{2,\perp},\tau_1,\tau_2)|^2~.
\end{align}
Thanks to the normalization of the nuclear wave-function, Eq. \eqref{den2}, one gets
\be
\sum_{\tau,\tau_2}\int_0^1 d\xi_1 \int_0^{1-\xi_1} d\xi_2~\bar \rho_{\tau \tau_2}^A (\xi_1,\xi_2)=A(A-1)~.
\ee
By comparing Eqs. \eqref{Eq:density1} and \eqref{Eq:density2b}, one obtains the usual relation between many-body densities, i.e.:
\begin{align}
    \rho_\tau(\xi) = \sum_{\tau_2} \int d\xi_2~\rho_{\tau \tau_2}(\xi_1,\xi_2)~,
\end{align}
where we {{remember}} that $\rho_{\tau \tau_2}(\xi_1,\xi_2)=\rho_{\tau \tau_2}(\xi_1,\xi_2,{\bf k}_\perp = {\bf 0})$. The above condition can be easily generalized to LC momentum number density, {by using Eq. \eqref{Eq:numden}} one gets

\begin{align}
 \sum_{\tau_2} \int d\xi_2~\bar \rho_{\tau \tau_2}(\xi_1,\xi_2) = (A-1)\bar \rho_{\tau}(\xi_1) ~.
\end{align}
One can show  that the GS sum rules \cite{Gaunt:2009re} applied to the two-body LC momentum number density  fulfills the {following} constraints
\begin{align}
\label{Eq:GSN0}
  \int_0^{1-\xi_1} d\xi_2 \bar \rho_{\tau_1 \tau_2}(\xi_1,\xi_2) &=
 \begin{cases} N_{\tau_2} \bar \rho_{\tau_1}(\xi_1) & \text { for } \tau_1 \neq \tau_2 \\
\left(N_{\tau_2}-1\right) \bar \rho_{\tau_1}(\xi_1) &\text { for } \tau_2=\tau_1\end{cases} ~.
\end{align}
The above
{expressions can rewritten} as follows

\begin{align}
\label{Eq:GSN}
  \int_0^{1-\xi_1} d\xi_2 \bar \rho_{\tau_1 \tau_2}(\xi_1,\xi_2) &=
 ( N_{\tau_2} -\delta_{\tau_2\tau_1})~\bar \rho_{\tau_1}(\xi_1)~,
\end{align}
{hence,}

\begin{align}
\label{Eq:normd2}
&\sum_{\tau_2} \int_0^{1-\xi_1} d\xi_2~\bar \rho_{\tau \tau_2}^A (\xi_1,\xi_2) =
\sum_{\tau_2}~(N_{\tau_2} -\delta_{\tau_2 \tau_1} )~\bar \rho_{\tau_1}(\xi_1) = (A-1)\bar \rho_{\tau_1}(\xi_1)~.
\end{align}

Therefore, for a generic $A$ nucleus the {partial} sum rules Eq. \eqref{Eq:GSN} lead to
\begin{align}
\label{Eq:rel}
        \int  d\xi_2~ \bar \rho^A_{pp}(\xi_1,\xi_2) &= (Z-1)\bar \rho^A_p(\xi_1)
    \\
        \int  d\xi_2~ \bar \rho^A_{pn}(\xi_1,\xi_2) &= (A-Z) \bar \rho^A_p(\xi_1)
        \\
        \int  d\xi_2~ \bar \rho^A_{np}(\xi_1,\xi_2) &= Z\bar \rho^A_n(\xi_1)
\\
        \int  d\xi_2~ \bar \rho^A_{nn}(\xi_1,\xi_2) &= (A-Z-1)\bar \rho^A_n(\xi_1)~.
        \label{Eq:rel2}
\end{align}
Moreover, {from} the integral property shown in Eq. \eqref{Eq:normd2} one can obtain

\begin{align}
    \int d\xi_1 d\xi_2~ \bar \rho^A_{pp}(\xi_1,\xi_2) &= Z(Z-1)
    \\
        \int d\xi_1 d\xi_2~ \bar \rho^A_{pn}(\xi_1,\xi_2) &= Z(A-Z)
        \\
        \int d\xi_1 d\xi_2~ \bar \rho^A_{np}(\xi_1,\xi_2) &= Z(A-Z)
\\
        \int d\xi_1 d\xi_2~ \bar \rho^A_{nn}(\xi_1,\xi_2) &= (A-Z)(A-Z-1)~.
\end{align}
{{Thanks to the analogy in the   interpretation of the two-body LC momentum  number density  and DPDs,}}
{one can also {introduce} the {{following}} momentum sum-rule   by following the GS {definition \cite{Gaunt:2009re}, viz.}
\begin{align}
\sum_{\tau_2} \int d\xi_2~\xi_2 \bar \rho^A_{\tau_1 \tau_2}(\xi_1,\xi_2)   = (1-\xi_1) \bar \rho^A_{\tau_1}(\xi_1)~.
\label{Eq:MSRden2}
\end{align}
{ The validity of this relation can be easily checked.} In fact, let us {consider the integral}
\begin{align}
    {\cal I}=\sum_{\tau_1} \int d \xi_1~ \sum_{\tau_2} \int d\xi_2~\xi_2 \bar \rho^A_{\tau_1 \tau_2}(\xi_1,\xi_2)~,
\end{align}
then, by using Eqs. \eqref{Eq:MSRden2},  one  gets
\begin{align}
    {\cal I} &= \sum_{\tau_1} \int d\xi_1~ (1-\xi_1)\bar \rho^A_{\tau_1}(\xi_1) =  \sum_{\tau_1} \int d\xi_1~\bar \rho^A_{\tau_1}(\xi_1)-
    \sum_{\tau_1} \int d\xi_1~ \xi_1\bar \rho^A_{\tau_1}(\xi_1)
   = A-1~,
\end{align}
where use has been made of Eqs. {\eqref{Eq:numden}} and \eqref{Eq:momcons}.
However, such a result can be   re-obtained by using only the momentum sum-rule related to the one-body LC momentum number density, i.e.
\be
    {\cal I}=  \int d\xi_2~\xi_2  \underbrace{\int d\xi_1~ \bar \rho^A_{pp}(\xi_1,\xi_2)}_{I_{pp}}~ + \int d\xi_2~\xi_2  \underbrace{\int d\xi_1~ \bar \rho^A_{pn}(\xi_1,\xi_2)}_{I_{pn}}
    \nonu + ~
    \int d\xi_2~\xi_2  \underbrace{\int d\xi_1~ \bar \rho^A_{np}(\xi_1,\xi_2)}_{I_{np}}+\int d\xi_2~\xi_2  \underbrace{\int d\xi_1~ \bar \rho^A_{nn}(\xi_1,\xi_2)}_{I_{nn}}~.
\ee

From  Eqs. \eqref{Eq:rel}-\eqref{Eq:rel2}, one writes

\begin{align}
    I_{pp}&= (Z-1) \bar \rho^A_{p}(\xi_2)
    \\
    \nonumber
    I_{pn}&= Z \bar \rho^A_{n}(\xi_2)
        \\
    \nonumber
    I_{np}&= (A-Z) \bar \rho^A_{p}(\xi_2)
        \\
    \nonumber
    I_{nn}&= (A-Z-1) \bar \rho^A_{n}(\xi_2)~,
\end{align}
and
therefore one has
\begin{align}
     {\cal I} &= (Z-1) \langle \xi \rangle_p+ (A-Z)\langle \xi \rangle_p+(A-Z-1)\langle \xi \rangle_n+ Z \langle \xi \rangle_n~,
 \end{align}
 where $\langle \xi \rangle_\tau$ has been defined in Eq. \eqref{Eq:momcons2}.
By using the momentum conservation written in  Eq. \eqref{Eq:momcons},
\begin{align}
    \langle \xi \rangle_p+\langle \xi \rangle_n = \sum_{\tau} \int d\xi~ \xi\bar \rho^A_{\tau}(\xi) =1~,
\end{align}
 one eventually re-obtains
\begin{align}
    {\cal I} = A-1~.
\end{align}

\section{Evaluating the traces in ${\cal S}^{\mu\mu'}_{ij}$}
\label{app_traces}
In this Appendix, the evaluation of the four traces present in Eq. \eqref{eq:traceb} is discussed in detail. Notice that in what follows the LF momentum is indicated by ${\blf p}$ and amounts to ${\blf p}\equiv \{p^+, p_x,p_y\} $ with, importantly, $p^+=E(|{\bf p}|)+p_z$.

In the LF deuteron wave function,  one has to deal with the following  combination  { of Melosh rotations: $\sum_{h_1, h_2} D^{{1 \over 2}} [{\cal R}^\dagger_M ({\blf k}_1)]_{\lambda_1 h_1}  D^{{1 \over 2}} [{\cal R}^\dagger_M ({\blf k}_2)]_{\lambda_2 h_2}~ C^{1 \mu}_{\frac{1}{2} h_1, \frac{1}{2} h_2}$}. It can be recast in a new expression by using the relation $C^{1 \mu}_{\frac{1}{2} h_1, \frac{1}{2} h_2}= \langle 1/2 ~h_1|\sigma_\mu~i\sigma_y|1/2 ~h_2\rangle /\sqrt{2} $, with i) $\sigma_\mu=\hat{\bf e}_\mu \cdot {\bfm \sigma}$,  ii) $\hat{\bf e}_0=\hat{\bf e}_z$ and iii) $\hat{\bf e}_\pm =\mp (\hat{\bf e}_x \pm  i\hat{\bf e}_y)/\sqrt{2}$.  One gets:
\be
\sum_{h_1, h_2} D^{{1 \over 2}} [{\cal R}^\dagger_M ({\blf k}_1)]_{\lambda_1 h_1}  D^{{1 \over 2}} [{\cal R}^\dagger_M ({\blf k}_2)]_{\lambda_2 h_2}~ C^{1 \mu}_{\frac{1}{2} h_1, \frac{1}{2} h_2}
\nonu
= \frac{1}{\sqrt{2}}\sum_{h_1, h_2} \langle \lambda_2| {\cal R}^\dagger_M({\blf k}_2)|h_2\rangle \langle h_2|
 \sigma_{\mu} (i\sigma_y )|h_1\rangle\langle h_1|{\cal R}^*_M({\blf k}_1)|\lambda_1\rangle
\nonu
= \frac{1}{\sqrt{2}} \langle \lambda_2| {\cal R}^\dagger_M({\blf k}_2)~
 \sigma_{\mu} (i\sigma_y )~{\cal R}^*_M({\blf k}_1)|\lambda_1\rangle
\nonu
= \frac{1}{\sqrt{2}}\langle \lambda_2| \frac{M + k^+_2 +i {\bfm \sigma} \cdot (\hat e_z \times
{\bf k}_{2\perp}) }{\sqrt{(M + k^+_2)^2 + k_{2\perp}^2}}   ~  \sigma_{\mu}  (i\sigma_y) ~\frac{M + k^+_1 +i {\bfm \sigma}^* \cdot (\hat e_z \times
{\bf k}_{1\perp})}{\sqrt{(M + k^+_1)^2 + k_{1\perp}^2}} |\lambda_1\rangle
\nonu
=\frac{1}{\sqrt{2}}\langle \lambda_2| \frac{M + k^+_2 +i {\bfm \sigma} \cdot (\hat e_z \times
{\bf k}_{2\perp}) }{\sqrt{(M + k^+_2)^2 + k_{2\perp}^2}}   ~  \sigma_{\mu}   ~\frac{M + k^+_1 - i {\bfm \sigma} \cdot (\hat e_z \times
{\bf k}_{1\perp})}{\sqrt{(M + k^+_1)^2 + k_{1\perp}^2}} ~(i\sigma_y)|\lambda_1\rangle
\nonu
=\frac{1}{\sqrt{2}}\langle \lambda_2| {\cal R}^\dagger_M ({\blf k}_2)~  \sigma_{\mu}   ~{\cal R}_M ({\blf k}_1)~(i\sigma_y)|\lambda_1\rangle
\label{eq:deu_wf_spin}~,
\ee
where $(i\sigma_y)\sigma^{T}_k(i\sigma_y)=\sigma_k$ has been  also used.
For the conjugated wf one has
\be
\sum_{h'_1, h'_2} D^{{1 \over 2}} [{\cal R}_M ({\blf k}'_1)]_{h'_1\lambda'_1 }  D^{{1 \over 2}} [{\cal R}_M ({\blf k}'_2)]_{h'_2\lambda'_2 }~ C^{1 \mu'}_{\frac{1}{2} h'_1, \frac{1}{2} h'_2}
\nonu
=\sum_{h'_1, h'_2} D^{{1 \over 2}} [{\cal R}^T_M ({\blf k}'_1)]_{\lambda'_1h'_1 }  C^{1 \mu'}_{\frac{1}{2} h'_1, \frac{1}{2} h'_2}  D^{{1 \over 2}} [{\cal R}_M ({\blf k}'_2)]_{h'_2\lambda'_2}
\nonu
= \frac{1}{\sqrt{2}}\sum_{h'_1, h'_2} \langle \lambda'_1| {\cal R}^TM({\blf k}'_1)| h'_1\rangle \langle h'_1|
 \sigma_{\mu'} (i\sigma_y )|h'_2\rangle\langle h'_2|{\cal R}_M({\blf k}'_2)|\lambda'_2\rangle
\nonu
=-(-1)^{\mu'} \frac{1}{\sqrt{2}}~\langle \lambda'_1|  {\cal R}^T_M({\blf k}'_1)~(i\sigma_y )~
 \sigma_{-\mu'}~{\cal R}_M({\blf k}'_2)|\lambda'_2\rangle
 \nonu=-(-1)^{\mu'}\frac{1}{\sqrt{2}}~\langle \lambda'_1| (i\sigma_y )~ {\cal R}^\dagger_M({\blf k}'_1)~
 \sigma_{-\mu'}~{\cal R}_M({\blf k}'_2)|\lambda'_2\rangle~,
\label{eq:deu_wf_con_spin2}
\ee
where in the last line one has $\sigma_{\mu'}\sigma_y=-(-1)^{\mu'}\sigma_y \sigma_{-\mu'}$.
Recall that ${\cal R}_M^T\ne {\cal R}_M^*$, since the Melosh rotations are unitary but not Hermitian.

The relevant quantity to be evaluated is
\be
{\cal S}_{ij}^{\mu\mu'}=\sum_{\lambda_1\lambda_2}\sum_{\lambda'_1\lambda'_2}\frac{1}{\sqrt{2}} \langle \lambda_2| {\cal R}^\dagger_M({\blf k}_2)~
 \sigma_{\mu} ~{\cal R}_M({\blf k}_1)~(i\sigma_y )|\lambda_1\rangle~\langle \lambda_1|\hat \Phi ^i(x_1,0,{\bf k}_{\perp})|\lambda'_1
 \rangle
 \nonu \times ~
 \frac{(-1)^{1+\mu'}}{\sqrt{2}}~\langle \lambda'_1| (i\sigma_y ){\cal R}^\dagger_M({\blf k}'_1)~
 \sigma_{-\mu'} ~{\cal R}_M({\blf k}'_2)|\lambda'_2\rangle~\langle \lambda'_2|\Bigl[\hat \Phi^{j }(x_2,0,-{\bf k}_\perp)\Bigr]^T|\lambda_2\rangle
 \nonu
 =~-(-1)^{\mu'}~{1\over 2}
 \nonu \times ~
 Tr\Biggl\{ {\cal R}^\dagger_M({\blf k}_2)~\sigma_{\mu}~{\cal R}_M({\blf k}_1)~ (i\sigma_y )~\hat \Phi^i(x_1,0,{\bf k}_\perp)~ (i\sigma_y )
 ~
 {\cal R}^\dagger_M({\blf k}'_1)~ \sigma_{-\mu'}~{\cal R}_M({\blf k}'_2)~\Bigl[\hat \Phi^j(x_2,0,-{\bf k}_\perp)\Bigr]^T\Biggr\}~.
\nonu
\label{eq:app_trace}\ee
The  LC correlator is defined as follows (recall that $\sigma^{+j} = i \gamma^+ \gamma^j$ and only the large component of the nucleon spinor is acting)
\be
 \hat \Phi^{i}\left(x,0,{\bf k}_{\perp}\right)=H_i(x,0,-{\bf k}^2_{\perp}) -
\dfrac{i}{4 M}( \hat e_z\cdot {\bfm \sigma}\times {\bf k}_\perp  )  E_i(x,0,-{\bf k}^2_\perp)~,
\ee
with ${\bf k}_\perp={\bf k}'_{1\perp}-{\bf k}_{1\perp}= -({\bf k}'_{2\perp}-{\bf k}_{2\perp} )$. Moreover,
by  taking ${\bf k}_\perp\equiv \{0,k_y\}$, without loss of generality, one has
\be
\hspace{-.8cm} \hat \Phi^i (x,0,\pm{\bf k}_\perp)= H_i(x,0,- {\bf k}^2_\perp) \mp{i \over 4M} \sigma_x  k_y~E_i(x,0,-{\bf k}^2_\perp)
=\Bigl[\hat \Phi^i(x,0,\pm{\bf k}_\perp)\Bigr]^T
\\ &&
\hspace{-.8cm} (i\sigma_y )~\hat \Phi^i(x,0,{\pm\bf k}_\perp)~ (i\sigma_y )= -\Bigl[H_i(x,0,- {\bf k}^2_\perp)\pm {i \over 4M} \sigma_x  k_y~E_i(x,0,-{\bf k}^2_\perp)\Bigr]= - \Bigl[\hat \Phi^i(x,0,\pm{\bf k}_\perp)\Bigr]^*.
\ee
Therefore
\be
{\cal S}_{ij}^{\mu\mu'}=~(-1)^{\mu'}{1\over 2}
\nonu \times ~
Tr\Biggl\{ \sigma_{\mu}~{\cal R}_M({\blf k}_1)~\Bigl[\hat \Phi^i(x_1,0,{\bf k}_\perp)\Bigr]^* ~
 {\cal R}^\dagger_M({\blf k}'_1)~ \sigma_{-\mu'}~{\cal R}_M({\blf k}'_2)~\hat \Phi^j(x_2,0,-{\bf k}_\perp)~{\cal R}^\dagger_M({\blf k}_2)\Biggr\}=
 \nonu
 =(-1)^{\mu'}{1\over 2} Tr\Biggl\{ \sigma_{\mu}~{\cal R}_M({\blf k}_1)~\Bigl[ H_i( x_1,0,-{\bf k}^2_\perp)+ {i \over 4M} \sigma_x  k_y~E_i(x_1,0,-{\bf k}^2_\perp)  \Bigr] ~
 {\cal R}^\dagger_M({\blf k}'_1)~ \sigma_{-\mu'}~{\cal R}_M({\blf k}'_2)
 \nonu \times~\Bigl[H_j(x_2,0,- {\bf k}_\perp) +{i \over 4M} \sigma_x  k_y~E_j(x_2,0,-{\bf k}_\perp)\Bigr]~{\cal R}^\dagger_M({\blf k}_2)\Biggr\}
 \nonu
 =~ H_i(x_1,0, -{\bf k}^2_\perp)H_j(x_2,0,- {\bf k}^2_\perp)  ~{(-1)^{\mu'}\over 2} Tr\Biggl\{ \sigma_{\mu}~{\cal R}_M({\blf k}_1) ~
 {\cal R}^\dagger_M({\blf k}'_1)~ \sigma_{-\mu'}~{\cal R}_M({\blf k}'_2)~{\cal R}^\dagger_M({\blf k}_2)\Biggr\}
 \nonu
 +i { k_y\over 4M}H_i(x_1,0, -{\bf k}^2_\perp) ~ E_j(x_2,0,-{\bf k}^2_\perp)~{(-1)^{\mu'}\over 2} Tr\Biggl\{ \sigma_{\mu}~{\cal R}_M({\blf k}_1)~
 {\cal R}^\dagger_M({\blf k}'_1)~ \sigma_{-\mu'}~{\cal R}_M({\blf k}'_2)~\sigma_x  ~{\cal R}^\dagger_M({\blf k}_2)\Biggr\}
 \nonu
 +i {k_y\over 4M} ~E_i(x_1,0,-{\bf k}^2_\perp) H_j(x_2,0,- {\bf k}^2_\perp) ~{(-1)^{\mu'}\over 2} Tr\Biggl\{ \sigma_{\mu}~{\cal R}_M({\blf k}_1)~ \sigma_x   ~
 {\cal R}^\dagger_M({\blf k}'_1)~ \sigma_{-\mu'}~{\cal R}_M({\blf k}'_2)~{\cal R}^\dagger_M({\blf k}_2)\Biggr\}
 \nonu
 -{k^2_y\over 16M^2}~E_i(x_1,0,-{\bf k}^2_\perp)  ~E_j(x_2,0,-{\bf k}^2_\perp) ~{(-1)^{\mu'}\over 2}
 \nonu
 \times ~ Tr\Biggl\{ \sigma_{\mu}~{\cal R}_M({\blf k}_1)~ \sigma_x   ~
 {\cal R}^\dagger_M({\blf k}'_1)~ \sigma_{-\mu'}~{\cal R}_M({\blf k}'_2)~ \sigma_x  ~{\cal R}^\dagger_M({\blf k}_2)\Biggr\}~.
\ee
When $\hat \Phi$ reduces to a multiple of the  identity, e.g. for ${\bf k}_\perp={\bf k}'_{1\perp}-{ \bf k}_{1\perp}=0$,  one gets
\be
\label{Eq:iden}
{\cal R}_M({\blf k}_n)~\hat \Phi^{i}
 ~ {\cal R}^\dagger_M({\blf k}'_n)\to {\rm I}~,
\ee
and the trace becomes $\delta_{\mu,\mu'}$. In fact, one should recall that  (cf Varshalovich et al p. 48) $$\sigma_\mu=\hat {\bf e}_\mu \cdot {\bfm \sigma}~~,$$ leading to
\be
 (-1)^{\mu'}{1\over 2} Tr \{\sigma_{\mu} ~ \sigma_{-\mu'}\}= (-1)^{\mu'}\Bigl[\hat {\bf e}_\mu \Bigr]_i  \Bigl[\hat {\bf e}_{-\mu '}\Bigr]_j ~{1\over 2} Tr \{\sigma_{i} ~ \sigma_{j}\}=\hat {\bf e}_\mu \cdot \hat {\bf e}^{\mu'}= \delta_{\mu,\mu'}~.
\ee

\subsection{Simplifying $S^{\mu\mu'}_{ij}$}
First, let us considering the following products involving  Melosh rotations,  recalling $k^+_n= E(|{\bf k}_n|) +k_{nz}$,
\be
{\cal R}_M({\blf k}'_n)  ~{\cal R}^\dagger_M({\blf k}_n)=
{\Bigl[ k^{\prime +}_n +M-i {\bfm \sigma}\cdot (\hat e_z \times {\bf k}'_{n\perp})\Bigr]~\Bigl[ k^{ +}_n+M +i {\bfm \sigma}\cdot (\hat e_z \times {\bf k}_{n\perp})\Bigr] \over  \sqrt{ | {\bf k}_{n\perp} |^ 2 +(k^+_n +M)^2 } ~
  \sqrt{ | {\bf k}^\prime_{n \perp}|^ 2 +(k^{\prime +}_n +M)^2 }}
 \nonu ={{\cal A}_1({\blf k}'_n,{\blf k}_n)
  +i {\bfm \sigma}\cdot  {\bfm {\cal B}}_1({\blf k}'_n,{\blf k}_n)\over  \sqrt{ | {\bf k}_{n\perp} |^ 2 +(k^+_n +M)^2 } ~
  \sqrt{ | {\bf k}^\prime_{n \perp}|^ 2 +(k^{\prime +}_n +M)^2  }}
 ~, \ee
  and \be
  {\cal R}_M({\blf k}'_n)~ \sigma_x  ~{\cal R}^\dagger_M({\blf k}_n)=
{\Bigl[ k^{\prime +}_n+M -i {\bfm \sigma}\cdot (\hat e_z \times {\bf k}'_{n\perp})\Bigr]~\sigma_x~\Bigl[ k^{ +}_n +M+i {\bfm \sigma}\cdot (\hat e_z \times {\bf k}_{n\perp})\Bigr] \over  \sqrt{ | {\bf k}_{n\perp} |^ 2 +(k^+_n +M)^2 } ~
  \sqrt{ | {\bf k}^\prime_{n \perp}|^ 2 +(k^{\prime +}_n +M)^2 }}
  \nonu=i{{\cal A}_2({\blf k}'_n,{\blf k}_n)+i{\bfm \sigma}\cdot {\bfm {\cal B}}_2({\blf k}'_n,{\blf k}_n)
   \over  \sqrt{ | {\bf k}_{n\perp} |^ 2 +(k^+_n +M)^2 } ~
  \sqrt{ | {\bf k}^\prime_{n \perp}|^ 2 +(k^{\prime +}_n +M)^2 } }~.\ee
  In the above equations, one has
  \be
  {\cal A}_1({\blf k}'_n,{\blf k}_n)=(k^{\prime +}_n+M)~(k^{ +}_n+M) +  {\bf k}'_{n\perp}\cdot   {\bf k}_{n\perp}
  \\ &&
  {\bfm {\cal B}}_1({\blf k}'_n,{\blf k}_n)=
  -\hat e_x [(k^{\prime +}_n+M) k_{ny}-(k^+_n+M) k'_{ny}]+\hat e_y [(k^{\prime +}_n+M) k_{nx}
  \nonu
  - ~ (k^+_n+M) k'_{nx}]
  + \hat  e_z (k^{\prime}_{nx} k_{ny}-k_{nx} k'_{ny})~,
  \ee
and
\begin{align}
{\cal A}_2({\blf k}'_n,{\blf k}_n)&=(k^{ +}_n+M) k'_{ny}-(k^{\prime  +}_n+M) k_{ny}
\\
\nonumber
{\bfm {\cal B}}_2({\blf k}'_n,{\blf k}_n)&=-\hat e_x [(k^{\prime +}_n+M) (k^{ +}_n+M) +{ k}^\prime_{ny} { k}_{ny}-{ k}'_{nx} { k}_{nx}]
  +\hat e_y ({ k}'_{ny} { k}_{nx}+{ k}'_{nx} { k}_{ny} )
  \\
  \nonumber
  &+\hat e_z [(k^{\prime +}_n+M) k_{nx}+ (k^{ +}_n+M) k'_{nx}]~.
\end{align}
For the two-nucleon case, notice that for ${\bf k}_\perp =0$, i.e. ${\bf k}'_{1\perp}={\bf k}_{1\perp}$, one has $ k_{1z}=M_0 (\xi_1 -1/2)= k'_{1z}$ and gets
\be
{\cal A}_1({\blf k}_1,{\blf k}_1)=(k^{ +}_1+M)^2~ +  |{\bf k}_{1\perp}|^2 , \quad {\bfm {\cal B}}_1({\blf k}_1,{\blf k}_1)=0
\nonu
{\cal A}_2({\blf k}_1,{\blf k}_1)=0 ,
\nonu
{\bfm {\cal B}}_2({\blf k}_1,{\blf k}_1)=
 -\hat e_x \Bigl[(k^{ +}_1+M)^2+k_{1y}^2 -k_{1x}^2\Bigr]
  +2\hat e_y { k}_{1y} { k}_{1x}+2\hat e_z (k^{ +}_1+M) k_{1x}~,
\nonu
\label{eq:calAB_kperp0}\ee
and analogously for ${\blf k}_2$. Then:
\be
{\cal S}_{ij}^{\mu\mu'}= {H_i( x_1,0,-{\bf k}^2_\perp)H_j(x_2,0,- {\bf k}^2_\perp)~T^{\mu\mu'}_{11}\over \Bigl[  | {\bf k}_{n\perp} |^ 2 +(k^+_n +M)^2 \Bigr] ~
  \Bigl[ | {\bf k}^\prime_{n \perp}|^ 2 +(k^{\prime +}_n +M)^2 \Bigr] }
 \nonu
 -{ k_y\over 4M}{H_i(x_1,0, -{\bf k}^2_\perp) ~ E_j(x_2,0,-{\bf k}^2_\perp)~T^{\mu\mu'}_{12}\over  \Bigl[  | {\bf k}_{n\perp} |^ 2 +(k^+_n +M)^2 \Bigr] ~
  \Bigl[ | {\bf k}^\prime_{n \perp}|^ 2 +(k^{\prime +}_n +M)^2 \Bigr]}
 \nonu
 - {k_y\over 4M}  ~{E_i(x_1,0,-{\bf k}^2_\perp) H_j(x_2,0,- {\bf k}^2_\perp)~T^{\mu\mu'}_{21}\over  \Bigl[  | {\bf k}_{n\perp} |^ 2 +(k^+_n +M)^2 \Bigr] ~
  \Bigl[ | {\bf k}^\prime_{n \perp}|^ 2 +(k^{\prime +}_n +M)^2 \Bigr]}
 \nonu
 +{k^2_y\over 16M^2}~{E_i(x_1,0,-{\bf k}^2_\perp)  ~E_j(x_2,0,-{\bf k}^2_\perp)~T^{\mu\mu'}_{22}\over  \Bigl[  | {\bf k}_{n\perp} |^ 2 +(k^+_n +M)^2 \Bigr] ~
  \Bigl[ | {\bf k}^\prime_{n \perp}|^ 2 +(k^{\prime +}_n +M)^2 \Bigr] }
~.
\label{eq:s_mmup}\ee
The four traces $T^{\mu\mu'}_{rq}$  amount to
\be
T^{\mu\mu'}_{11}=(-1)^{\mu'}\sum_{n\ell}
[\hat e_\mu]_n [\hat e_{-\mu'}]_\ell
\nonu \times
~{1\over 2} Tr\Biggl\{ \sigma_{n}~\Bigl[ {\cal A}_1({\blf k}_1,{\blf k}'_1)
  +i {\bfm \sigma}\cdot  {\bfm {\cal B}}_1({\blf k}_1,{\blf k}'_1)\Bigr]~ \sigma_{\ell}~\Bigl[ {\cal A}_1({\blf k}'_2,{\blf k}_2)
  +i {\bfm \sigma}\cdot  {\bfm {\cal B}}_1({\blf k}'_2,{\blf k}_2)\Bigr]\Biggr\}
  \nonu
 T^{\mu\mu'}_{12}=(-1)^{\mu'}\sum_{n\ell}
[\hat e_\mu]_n [\hat e_{-\mu'}]_\ell
\nonu \times
~{1\over 2} Tr\Biggl\{ \sigma_{n}~\Bigl[ {\cal A}_1({\blf k}_1,{\blf k}'_1)
  +i {\bfm \sigma}\cdot  {\bfm {\cal B}}_1({\blf k}_1,{\blf k}'_1)\Bigr]~ \sigma_{\ell}~\Bigl[ {\cal A}_2({\blf k}'_2,{\blf k}_2)
  +i {\bfm \sigma}\cdot  {\bfm {\cal B}}_2({\blf k}'_2,{\blf k}_2)\Bigr]\Biggr\}
  \nonu
  T^{\mu\mu'}_{21}=(-1)^{\mu'}\sum_{n\ell}
[\hat e_\mu]_n [\hat e_{-\mu'}]_\ell
\nonu \times
~{1\over 2} Tr\Biggl\{ \sigma_{n}~\Bigl[ {\cal A}_2({\blf k}_1,{\blf k}'_1)
  +i {\bfm \sigma}\cdot  {\bfm {\cal B}}_2({\blf k}_1,{\blf k}'_1)\Bigr]~ \sigma_{\ell}~\Bigl[ {\cal A}_1({\blf k}'_2,{\blf k}_2)
  +i {\bfm \sigma}\cdot  {\bfm {\cal B}}_1({\blf k}'_2,{\blf k}_2)\Bigr]\Biggr\}
  \nonu
 T^{\mu\mu'}_{22}=(-1)^{\mu'}\sum_{n\ell}
[\hat e_\mu]_n [\hat e_{-\mu'}]_\ell
\nonu \times
~{1\over 2} Tr\Biggl\{ \sigma_{n}~\Bigl[ {\cal A}_2({\blf k}_1,{\blf k}'_1)
  +i {\bfm \sigma}\cdot  {\bfm {\cal B}}_2({\blf k}_1,{\blf k}'_1)\Bigr]~ \sigma_{\ell}~\Bigl[ {\cal A}_2({\blf k}'_2,{\blf k}_2)
  +i {\bfm \sigma}\cdot  {\bfm {\cal B}}_2({\blf k}'_2,{\blf k}_2)\Bigr]\Biggr\} ~.
\nonu\ee
In general
\begin{align}
\nonumber
&T^{\mu\mu'}_{r q}=(-1)^{\mu'}\sum_{n\ell}[\hat e_\mu]_n [\hat e_{-\mu'}]_\ell~{1\over 2} Tr\Biggl\{ \sigma_{n}~\Bigl[ {\cal A}_r({\blf k}_1,{\blf k}'_1)
  +i {\bfm \sigma}\cdot  {\bfm {\cal B}}_r({\blf k}_1,{\blf k}'_1)\Bigr]~ \sigma_{\ell}~\Bigl[ {\cal A}_q({\blf k}'_2,{\blf k}_2)
  + i {\bfm \sigma}\cdot  {\bfm {\cal B}}_q({\blf k}'_2,{\blf k}_2)\Bigr]\Biggr\}
\\
& =(-1)^{\mu'}\sum_{n\ell}[\hat e_\mu]_n [\hat e_{-\mu'}]_\ell~\Biggl\{{\cal A}_r({\blf k}_1,{\blf k}'_1)  {\cal A}_q({\blf k}'_2,{\blf k}_2) \delta_{n,\ell}
 -{\cal A}_r({\blf k}_1,{\blf k}'_1)\sum_u ~\varepsilon_{n \ell u} ~\Bigl[{\bfm {\cal B}}_q({\blf k}'_2,{\blf k}_2)\Bigr]_u
 \\
 \nonumber
 &-
 {\cal A}_q({\blf k}'_2,{\blf k}_2)\sum_u ~\varepsilon_{n  u \ell} ~\Bigl[{\bfm {\cal B}}_r({\blf k}_1,{\blf k}'_1)\Bigr]_u
 -  \sum_{uv}\Bigl[{\bfm {\cal B}}_r({\blf k}_1,{\blf k}'_1)\Bigr]_u \Bigl[{\bfm {\cal B}}_q({\blf k}'_2,{\blf k}_2)\Bigr]_v ~\Bigl(\delta_{n,u}\delta_{\ell,v} -\delta_{n,\ell}\delta_{u,v}+\delta_{n,v}\delta_{\ell,u} \Bigr)\Biggr\}~,
 \end{align}
 where standard traces of two, three and four Pauli matrices have been used, viz.
\be
\hspace{-1.cm} {1\over 2} Tr\Bigl[ \sigma_n \sigma_\ell\Bigr]=\delta_{n,\ell}~, \quad {1\over 2} Tr\Bigl[ \sigma_n \sigma_\ell \sigma_u \Bigr]=i\varepsilon_{n\ell u}~, \quad
{1\over 2} Tr\Bigl[ \sigma_n \sigma_u \sigma_\ell \sigma_v \Bigr]=
\delta_{n,u}\delta_{\ell,v} -\delta_{n,\ell}\delta_{u,v}+\delta_{n,v}\delta_{\ell,u}
~.\ee
Finally, one gets
 \be
 T^{\mu\mu'}_{r q}=
 \delta_{\mu,\mu'}\Bigl[ {\cal A}_r({\blf k}_1,{\blf k}'_1)  {\cal A}_q({\blf k}'_2,{\blf k}_2) +{\bfm {\cal B}}_r({\blf k}_1,{\blf k}'_1)\cdot {\bfm {\cal B}}_q({\blf k}'_2,{\blf k}_2)\Bigr]
 \nonu
 - i (-1)^{\mu'}\sqrt{2} C^{1\lambda}_{1\mu1-\mu'}~\Bigl\{{\cal A}_r({\blf k}_1,{\blf k}'_1) ~\Bigl[{\bfm {\cal B}}_q({\blf k}'_2,{\blf k}_2)\Bigr]_\lambda-
 {\cal A}_q({\blf k}'_2,{\blf k}_2) ~\Bigl[{\bfm {\cal B}}_r({\blf k}_1,{\blf k}'_1)\Bigr]_\lambda \Bigr\}
 \nonu
 - (-1)^{\mu'}\Bigl\{\Bigl[{\bfm {\cal B}}_r({\blf k}_1,{\blf k}'_1)\Bigr]_\mu \Bigl[{\bfm {\cal B}}_q({\blf k}'_2,{\blf k}_2)\Bigr]_{-\mu'} +\Bigl[{\bfm {\cal B}}_r({\blf k}_1,{\blf k}'_1)\Bigr]_{-\mu'} \Bigl[{\bfm {\cal B}}_q({\blf k}'_2,{\blf k}_2)\Bigr]_{\mu}
 \Bigr\}~
~,\ee
where Greek indexes indicate spherical harmonics coordinates, with
\be
(-1)^{\mu'}\sum_n[\hat e_\mu]_n [\hat e_{-\mu'}]_n= (-1)^{\mu'}\hat e_\mu\cdot \hat e_{-\mu'}=  \hat e_\mu\cdot \hat e^{\mu'}= \delta_{\mu,\mu'}~, \nonu (-1)^{\mu'}\sum_{n\ell}\varepsilon_{n\ell u}[\hat e_\mu]_n [\hat e_{-\mu'}]_\ell= (-1)^{\mu'}[\hat e_\mu\times \hat e_{-\mu'}]_u= i (-1)^{\mu'}\sqrt{2} C^{1\lambda}_{1\mu1-\mu'} [\hat e_\lambda]_u ~.
\ee
From the above expression, one writes
\be
T^{\mu\mu'}_{21 }({\blf k}_1,{\blf k}'_1;{\blf k}'_2,{\blf k}_2)=
(-1)^{\mu+\mu'}~T^{\mu'\mu}_{12 }({\blf k}'_2,{\blf k}_2;{\blf k}_1,{\blf k}'_1)
\ee
It is easily seen that for ${\bf k}_\perp=0$, i.e. ${\bf k}'_{1\perp}={\bf k}_{1\perp}$ and ${\bf k}'_{2\perp}={\bf k}_{2\perp}$, by using Eq. \eqref{eq:s_mmup}, one has
only
\be
{\cal S}^{\mu\mu'}_{ij}= \delta_{\mu,\mu'} ~H_i( x_1,0,0)H_j(x_2,0,0)
\ee

\section{Deuteron LCMDs in the NR {\bf{framework}}}
\label{NRLCMDs}

 { In this Appendix it is shown  {{how to evaluate the deuteron light-cone momentum distributions in the NR framework.}}}
{
{We remark that in the NR framework Melosh rotation are not present and}}
the deuteron wave-function in momentum space, 
for a given spin third-component $M$ reads
\begin{align}
\label{Eq:wf0}
    \psi^M_2({\bf k}_{in}) = u (k_{in}) \chi_1^M Y^0_0(\theta,\phi)+ w(k_{in}) \sum_{m_L} C^{1M}_{2m_L~1M-m_L} Y_2^{m_L}(\theta,\phi) \chi_1^{M-m_L}~,
\end{align}
where  ${\bf k}_{in}$ is the intrinsic momentum, i.e. {$ {\bf k}_{in} =  ({\bf k}_1-{\bf k}_2)/2 =  {\bf k}_1$} in the frame where ${\bf k}_1+{\bf k}_2=  {\bf 0}$. Moreover, $\theta$ and $\phi$ are the angles defining the {direction of the} vector ${\bf k}_{in}$.
{{The radial wf's are obtained solving the Schroedinger equation with the Av18 potential.}} The following normalization condition holds:

\begin{align}
    \int dk_{in}~k_{in}^2 \big[ u(k_{in})^2+w(k_{in})^2 \big]=1.
\end{align}
 One should notice that Eq. {\eqref{Eq:wf0}} can be written within the same spin-dependence convention   {of Eq. (\ref{Eq:A_state})}:

\begin{align}
    \psi^M_2({\bf k}_{in}) &= \sum_{\lambda_1,\lambda_2} \psi^M_2({\bf k}_{in},\lambda_1,\lambda_2)~{\chi_{1/2}^{\lambda_1}(p) \chi_{1/2}^{\lambda_2}(n) } = \sum_{\lambda_1,\lambda_2} \chi_{1/2}^{\lambda_1}(p) \chi_{1/2}^{\lambda_2}(n)
    \\
    \nonumber
    &\times \Big[  u (k_{in})  C^{1M}_{\frac{1}{2} \lambda_1 \frac{1}{2}\lambda_2  }   Y^0_0(\theta,\phi)+ w(k_{in}) \sum_{m_L} C^{1M}_{2m_L~1M-m_L} Y_2^{m_L}(\theta,\phi) C^{1M-m_L}_{\frac{1}{2} \lambda_1 \frac{1}{2}\lambda_2  }   \Big]~.
\end{align}

Within this formalism, one can evaluate both the nuclear DPDs also including spin-flip effects in the DPS2 contribution (see Eq. (\ref{Eq:DPD2full})).  In order to properly {calculate} Eq. (\ref{Eq:DPD2full}) for the deuteron, one has to study the following quantity,{ $ P^\tau_{ij}(x_1,x_2,\xi_1,{\bf k}_\perp)$, } {where i) the {deuteron} isospin dependence has been understood   and {ii) an average} over the deuteron spin polarizations is {performed. Hence, one has}
\begin{align}
    P^\tau_{ij}(x_1,x_2,\xi_1,{\bf k}_\perp)&=  \sum_{M=-1}^1 \sum_{\lambda_1,\lambda_2} \sum_{\lambda'_1,\lambda'_2} g^M(\xi_1, {\bf k}_{\perp},\lambda_1,\lambda_2,\lambda'_1,\lambda'_2)
    \\
    \nonumber
&\times    \Phi^{\tau,i}_{\lambda_1,\lambda'_1}\left(x_1 \dfrac{\bar \xi}{\xi_1},0, \mathbf k_\perp \right) \Phi^{\tau,j}_{\lambda_2,\lambda'_2}\left(x_2 \dfrac{\bar \xi}{1-\xi_1},0, -\mathbf k_\perp \right)~.
\end{align}
where
\begin{align}
 g^M(\xi_1, {\bf k}_\perp,\lambda_1,\lambda_2,\lambda'_1,\lambda'_2) =\int d{{\bf k}_{in}}
    \psi_2^M({\bf k}_{in},\lambda_1,\lambda_2)\psi_2^{M \dagger}({\bf k'}_{in},\lambda'_1,\lambda'_2) \delta \left( \xi_1-{\dfrac{k_{in}^+}{ P^+}} \right),
\end{align}
with \ {${ \bf k'}_{in}={\bf k}_{in}+ {\bf k_\perp}$} { (cf. the variables definition in Ref. \cite{Diehl:2011yj})}.

In terms of GPDs, if one assumes  ${\bf k_\perp}=(0,k_\perp)$  {without loss of generality} ({\bf{remember}} {the role of the LF} transverse-rotation invariance), the correlator can be {written} as follows:
\begin{align}
\Phi^{i}_{\lambda,\lambda'}\left(x,0, \mathbf k_\perp \right) = \delta_{\lambda,\lambda'} H_i(x,0,-{\bf k}^2_{\perp})-i \dfrac{k_\perp}{4 M}  \delta_{\lambda,-\lambda'}   E_i(x,0,-{\bf k}^2_\perp)~.
\end{align}
Therefore, one gets
\begin{align}
\label{eq:p1}
      P^\tau_{ij}(x_1,x_2,\xi_1,{\bf k}_\perp) &= \underbrace{  \sum_{M=-1}^1 \sum_{\lambda_1,\lambda_2}
g^M(\xi_1, {\bf k}_{\perp},\lambda_1,\lambda_2,\lambda_1,\lambda_2) }_{HH(\xi_1,{\bf k}_\perp)}H^\tau_i\left(x_1 \dfrac{\bar \xi}{\xi_1},0,-{\bf k}^2_\perp\right)H^\tau_j\left(x_2 \dfrac{\bar \xi}{1-\xi_1},0,-{\bf k}^2_\perp\right)&
      \\
\nonumber
&
\hspace{-1.cm} +  i \dfrac{k_\perp}{4 M}\underbrace{ \sum_{M=-1}^1\sum_{\lambda_1,\lambda_2}
g^M(\xi_1, {\bf k}_{\perp},\lambda_1,\lambda_2,\lambda_1,-\lambda_2)
}_{HE(\xi_1,{\bf k}_\perp)}H^\tau_i\left(x_1 \dfrac{\bar \xi}{\xi_1},0,-{\bf k}^2_\perp\right)E^\tau_j\left(x_2 \dfrac{\bar \xi}{1-\xi_1},0,-{\bf k}^2_\perp\right)&
\\
\nonumber
&
\hspace{-1.cm} +i \dfrac{k_\perp}{4 M}\underbrace{\sum_{M=-1}^1 \sum_{\lambda_1,\lambda_2}
g^M(\xi_1, {\bf k}_{\perp},\lambda_1,\lambda_2,-\lambda_1,\lambda_2)
}_{EH(\xi_1,{\bf k}_\perp)}E^\tau_i\left(x_1 \dfrac{\bar \xi}{\xi_1},0,-{\bf k}^2_\perp\right)H^\tau_j\left(x_2 \dfrac{\bar \xi}{1-\xi_1},0,-{\bf k}^2_\perp\right)&
 \\
\nonumber
 &\hspace{-1.cm} - \dfrac{k_\perp^2}{16 M^2}\underbrace{\sum_{M=-1}^1 \sum_{\lambda_1,\lambda_2}
 g^M(\xi_1, {\bf k}_{\perp},\lambda_1,\lambda_2,-\lambda_1,-\lambda_2)
 }_{EE(\xi_1,{\bf k}_\perp)}E^\tau_i\left(x_1 \dfrac{\bar \xi}{\xi_1},0,-{\bf k}^2_\perp\right)E^\tau_j\left(x_2 \dfrac{\bar \xi}{1-\xi_1},0,-{\bf k}^2_\perp\right)&
\end{align}

Once all the off-forward {LCMD} $HH(\xi,{\bf k_\perp}),~HE(\xi,{\bf k_\perp})=-EH(\xi,{\bf k_\perp})$ and $EE(\xi,{\bf k_\perp})$ are evaluated, the complete set of nuclear ingredients required for calculating {both  DPS1 and  DPS2 contributions to the nuclear DPS are available. Indeed} the one-body {LCMD}, entering the DPD corresponding to the DPS1 mechanism, Eq. (\ref{Eq:density1}), is:

\begin{align}
\rho_\tau^2(\xi)= \sum_{M=-1}^1 \sum_{\lambda_1,\lambda_2} g^M(\xi,{\bf 0}_\perp, \lambda_1,\lambda_2,\lambda_1,\lambda_2)   = HH(\xi,{\bf 0}_\perp)
\end{align}

\bibliographystyle{unsrt}
\bibliography{bib.bib}

\end{document}